\documentclass[fleqn,usenatbib]{mnras}

\usepackage{newtxtext,newtxmath}

\usepackage[T1]{fontenc}

\DeclareRobustCommand{\VAN}[3]{#2}
\let\VANthebibliography\thebibliography
\def\thebibliography{\DeclareRobustCommand{\VAN}[3]{##3}\VANthebibliography}

\usepackage{graphicx}	
\usepackage{amsmath}	
\usepackage{caption}

\usepackage{makecell}
\usepackage{svg}
\usepackage{comment}
\usepackage{pdflscape}
\usepackage{hyperref}
\usepackage{orcidlink}

\title[ALMA EGO-10]{The ALMA EGO-10 Survey of Massive Protoclusters: Correlation of 1.3\,mm Continuum Source Clustering with Evolutionary State}

\author[M.C Logue et al.]{Michael C. Logue\textsuperscript{\,\orcidlink{0009-0005-7600-7190}},$^{1}$\thanks{E-mail: mcl21@st-andrews.ac.uk} 
Claudia J. Cyganowski\textsuperscript{\,\orcidlink{0000-0001-6725-1734}},$^{1}$\thanks{E-mail: cc243@st-andrews.ac.uk} 
Crystal L. Brogan\textsuperscript{\,\orcidlink{0000-0002-6558-7653}},$^{2}$
Todd R. Hunter\textsuperscript{\,\orcidlink{0000-0001-6492-0090}},$^{2}$
\newauthor
Alessio Traficante\textsuperscript{\,\orcidlink{0000-0003-1665-6402}},$^{3}$
Allison P. M. Towner\textsuperscript{\,\orcidlink{0000-0001-5933-824X}}\,$^{4}$ and Stella S. R. Offner\textsuperscript{\,\orcidlink{0000-0003-1252-9916}}\,$^{5}$\\
$^{1}$Scottish Universities Physics Alliance (SUPA), School of Physics and Astronomy, University of St Andrews, North Haugh, St Andrews, Fife KY16 9SS, UK\\
$^{2}$National Radio Astronomy Observatory, 520 Edgemont Road, Charlottesville, VA 22903, USA\\
$^{3}$INAF-IAPS, Via Fosso del Cavaliere, 100, 00133 Rome, Italy\\
$^{4}$University of Arizona Department of Astronomy and Steward Observatory, 933 North Cherry Ave., Tucson, AZ 85721, USA\\
$^{5}$Department of Astronomy, University of Texas at Austin, TX 78712, USA
}

\date{Accepted XXX. Received YYY; in original form ZZZ}

\pubyear{\the\year{}}

\begin{document}
\label{firstpage}
\pagerange{\pageref{firstpage}--\pageref{lastpage}}
\maketitle

\begin{abstract}
Massive stars characteristically form in clustered 
environments. Characterising young massive `protoclusters' 
is therefore crucial to constraining the mechanism(s) of
massive star formation, and of the assembly of stellar 
clusters. We present 1.3\,mm continuum results from the ALMA
EGO-10 imaging survey, targeting ten \textit{Spitzer} 
GLIMPSE Extended Green Objects (EGOs) -- massive protostars 
with active outflows traced by extended 4.5\,$\mu$m
emission. Our sensitive 
1\farcm6$\times$1\farcm6 mosaics reveal
rich protoclusters associated with all targets. With a
mean spatial resolution 2200$\times$1600\,AU, we identify
570 cores -- between 13 and 135 per field. We
quantify protocluster structure with the $\mathcal{Q}$-parameter, finding structural diversity with 0.5 $\lesssim\mathcal{Q}\lesssim$ 0.9. The sample is notable for
the wealth of complementary high-resolution multiwavelength
data available. Correlating our cores with these observations, we find only 2\%, 5\% and 4\%
of cores host 6.7\,GHz CH$_3$OH masers, 22\,GHz
H$_2$O masers and cm-$\lambda$ continuum sources, respectively. 
The massive protostars traced by 6.7\,GHz masers typically reside near protocluster centres (median offset 0.045\,pc), and all at $d<$ 3\,kpc are found in clustered locales, with $>$10 cores within 10,000\,AU.
Using VLA cm-$\lambda$ continuum observations, we construct a new
evolutionary indicator: the ratio of protocluster
cm-$\lambda$ continuum luminosity to the mass of the associated
ATLASGAL clump ($L_\text{cm}/M_\text{AGAL}$). This ratio correlates positively 
with $\mathcal{Q}$, with the correlation driven primarily by the cm-$\lambda$
continuum emission from MYSOs. This suggests dynamic
protocluster structure, evolving from subclustered to
centrally condensed, consistent with the global collapse in hierarchical, clump-fed models of massive
star formation. 

\end{abstract}

\begin{keywords}
stars:formation -- stars:protostars -- submillimetre: ISM
\end{keywords}



\section{Introduction} \label{sec:intro}
Massive stars ($M_\text{ZAMS}>$ 8\,$M_\odot$, $L\gtrsim$ 10$^3\,$$L_\odot$) dominate the physical and chemical processes of the interstellar medium on a galactic scale \citep{Oey2009}, but the specific mechanism(s) of their formation have yet to be determined \cite[see reviews by][]{Tan2014, Motte2018, Beuther2025}. Observationally constraining proposed formation scenarios requires the detailed characterisation of large samples of massive star-forming regions and their constituent massive young stellar objects (MYSOs) at early evolutionary stages. Early indications from optical and NIR observations that massive stars formed in clustered environments \citep{Palla2000, Lada2003, deWit2005} were supported by observations of (sub)millimetre-wavelength multiplicity with the Submillimeter Array (SMA) and the  Plateau de Bure Interferometer (PdBI) \citep[e.g.][]{Hunter2006, Beuther2007, Cyganowski2007, Zhang2007}.
More recently, Atacama Large Millimeter/submillimeter Array (ALMA) mosaics have revealed unprecedented richness in massive star-forming
regions, often with $\gtrsim$tens of sources within $\lesssim$1\,pc \cite[e.g.][]{Cyganowski2017, Sanhueza2019, Cheng2018, Kong2019, Sadaghiani2020, Budaiev2024, Morrii2024, Louvet2024, Xu2024}. Given the ubiquity of clustered MYSO environs, any self-consistent massive star formation scenario must incorporate massive protoclusters (protoclusters containing massive (proto)stars), explaining their origin, structure and dynamics, and the effect of the mutual interaction of several (M)YSOs in relatively close proximity \cite[e.g.][]{Bonnell2004, Smith2009}. 

A fundamental parameter of any (proto)stellar cluster is the structure of the (proto)stellar distribution. This is perhaps most commonly quantified with the dimensionless $\mathcal{Q}$-parameter \citep{Cartwright2004}, constructed from the Minimum Spanning Tree (MST) of the source spatial distribution. The MST is defined as the set of straight lines, termed `edges', connecting all points with the minimum total length, without forming any closed loops. $\mathcal{Q}$ is then defined as:
\begin{equation}
\mathcal{Q} = \frac{\bar{m}}{\bar{s}} = \frac{m/\biggr(\sqrt{\frac{N_\text{cores} \pi R_{\text{clus}}^2}{(N_\text{cores}-1)}}\biggr)}{s/R_\text{clus}}
	\label{eq:q}
\end{equation}

\noindent where $\bar{m}$ is the mean edge length of the MST ($m$), normalised to account for the number of points in the distribution, and $\bar{s}$ is the normalised correlation length - the mean inter-point separation ($s$), normalised by $R_\text{clus}$. $R_\text{clus}$ is the cluster radius defined by \cite{Cartwright2004}: the distance from the cluster centre, defined as the mean position of all sources, to the source furthest from this position. $N_\text{cores}$ is the total number of points (cores) in the distribution. \citet{Cartwright2004} demonstrated $\mathcal{Q}$ can distinguish synthetically generated clusters with fractal, subclustered source distributions ($\mathcal{Q}<$ 0.8; smaller values indicate more subclustering) from those with centrally-condensed distributions and corresponding radial density gradients ($\mathcal{Q}>$ 0.8; larger values indicate more centrally-condensed structure), with the boundary value $\mathcal{Q=}$ 0.8 indicating a uniform source distribution. Applying the technique to observational data, these authors found the $\mathcal{Q}$-parameters of a set of relatively nearby clusters to span a range 0.47 $<\mathcal{Q}<$ 0.98. Since this initial application, the $\mathcal{Q}$-parameter has been extensively used to quantify cluster structure, including in more evolved clusters and low-mass star-forming regions \citep{Kumar2007, Schmeja2006, Gieles2008, Bastian2009, Sanchez2009, Alfaro2011, Gregorio-Hetem2015, Dib2019, Hetem2019}. The aforementioned large source counts achieved via (sub)millimetre-wavelength interferometric observations have also enabled this analysis in massive protoclusters. The first application, to 
SMA observations of NGC 6334 I(N) \citep{Hunter2014}, has recently been succeeded by similar, but more sensitive, studies with ALMA, yielding values 0.3 $\lesssim\mathcal{Q}\lesssim$ 1 \citep{Sanhueza2019, ONeill2021, Sadaghiani2020, Xu2024, Morrii2024, Kinman2025, Schisano2025, Zhang2025}.

Given the potential sensitivity of (proto)cluster structure to the initial formation conditions and subsequent dynamical evolution \cite[e.g.][and references therein]{Alfaro2011, Farias2024}, quantifying the structure of young massive protoclusters may provide crucial insight into the (proto)cluster assembly process and by extension, the mechanism(s) of massive star formation. Simulations of massive protocluster formation have demonstrated that whilst the (proto)stellar distribution is initially subclustered, the global gravitational contraction of the star-forming cloud leads to the hierarchical merging of subclusters, giving a more centrally-condensed (proto)stellar distribution \citep{Bonnell2003, Vazquez-Semadeni2017, Guszejnov2022}. Such structure evolution has been quantified by tracking the $\mathcal{Q}$-parameter of the simulation sink particle distribution, where sink particles correspond to individual (proto)stars \citep{Maschberger2010, Gavagnin2017, Ballone2020, Laverde-Villarreal2025}. These works all find the evolution of massive protocluster structure from initially subclustered to more centrally condensed, traced by increasing $\mathcal{Q}$.

Observational evidence of this phenomenon has emerged from recent ALMA surveys that have allowed correlations between protocluster structure and indicators of evolutionary state to be investigated within statistically significant samples. The \citet{Traficante2023} Star formation in QUiescent And Luminous Objects (SQUALO) survey found decreasing core separations as a function of the luminosity-to-mass ratio ($L/M$) of clumps - a well-established indicator of evolutionary state \cite[e.g.][]{Molinari2008, Molinari2016}. This result is supported by the similar findings of the ALMA Survey of Star Formation and Evolution in Massive Protoclusters with Blue Profiles (ASSEMBLE) pilot survey of \citet{Xu2024}. These authors specifically quantified protocluster structure with the $\mathcal{Q}$-parameter of the dense core spatial distribution in their ALMA mosaics, finding a positive correlation between $\mathcal{Q}$ and $L/M$ within the ASSEMBLE pilot sample. 
To investigate the $\mathcal{Q}$--$L/M$ relationship across a wider range of clump $L/M$, \citet{Xu2024} applied the same analysis to the pilot sample of the ALMA Survey of 70\,$\mu$m Dark High-mass Clumps in Early Stages \citep[ASHES;][]{Sanhueza2019}. 
This revealed an increase in $\mathcal{Q}$ from $\sim$0.4 to $\sim$0.9 across the full $L/M$ range occupied by these samples (ASHES: 0.05 $< L/M<$ 1.7\,$L_\odot/M_\odot$; ASSEMBLE: \mbox{11 $<L/M<$ 79\,$L_\odot/M_\odot$}; see \citealt{Xu2024} Figure~11). 

Recently, \citet{Schisano2025} also identified a weak $\mathcal{Q}$--$L/M$ correlation in the ALMA evolutionary study of high-mass protocluster formation in the Galaxy (ALMAGAL) sample. 
In general, these observational results are consistent with those 
from simulations, in which the global gravitational contraction of massive star-forming clumps is traced by the structure of the core spatial distribution, which evolves from subclustered to centrally condensed over time. 

SQUALO, ASSEMBLE, ASHES and ALMAGAL are among several recent ALMA surveys targeting samples of massive protoclusters. Table~\ref{tab:summ} summarises key parameters for such surveys, placing our ALMA EGO-10 survey in context. 
The ALMA EGO-10 1.3\,mm images are relatively large \mbox{($\sim$1\farcm6 $\times$ 1\farcm6)}, sensitive (rms noise $\sim$0.17\,mJy beam$^{-1}$), high-resolution ($\sim$0\farcs66) mosaics, and are thus well-suited to resolving the dense core population across the entirety of the target clumps. Fig.~\ref{fig:mm} presents three-colour \emph{Spitzer} images of our fields, showing the ALMA mosaic coverage of these targets. 
As detailed in Table~\ref{tab:summ}, the EGO-10 sample occupies a relatively narrow range in $L/M$ (3 $< L/M<$ 21 $L_\odot/M_\odot$), intermediate between the early stage clumps of the ASHES pilot sample, and samples reaching $L/M\gtrsim$ 100, which likely include more evolved regions; \citet{Urquhart2018}, for example, find compact \ion{H}{ii} regions become common at $L/M\gtrsim$ 40\,$L_\odot/M_\odot$.

\begin{figure*}
   \includegraphics[width=0.95\textwidth]{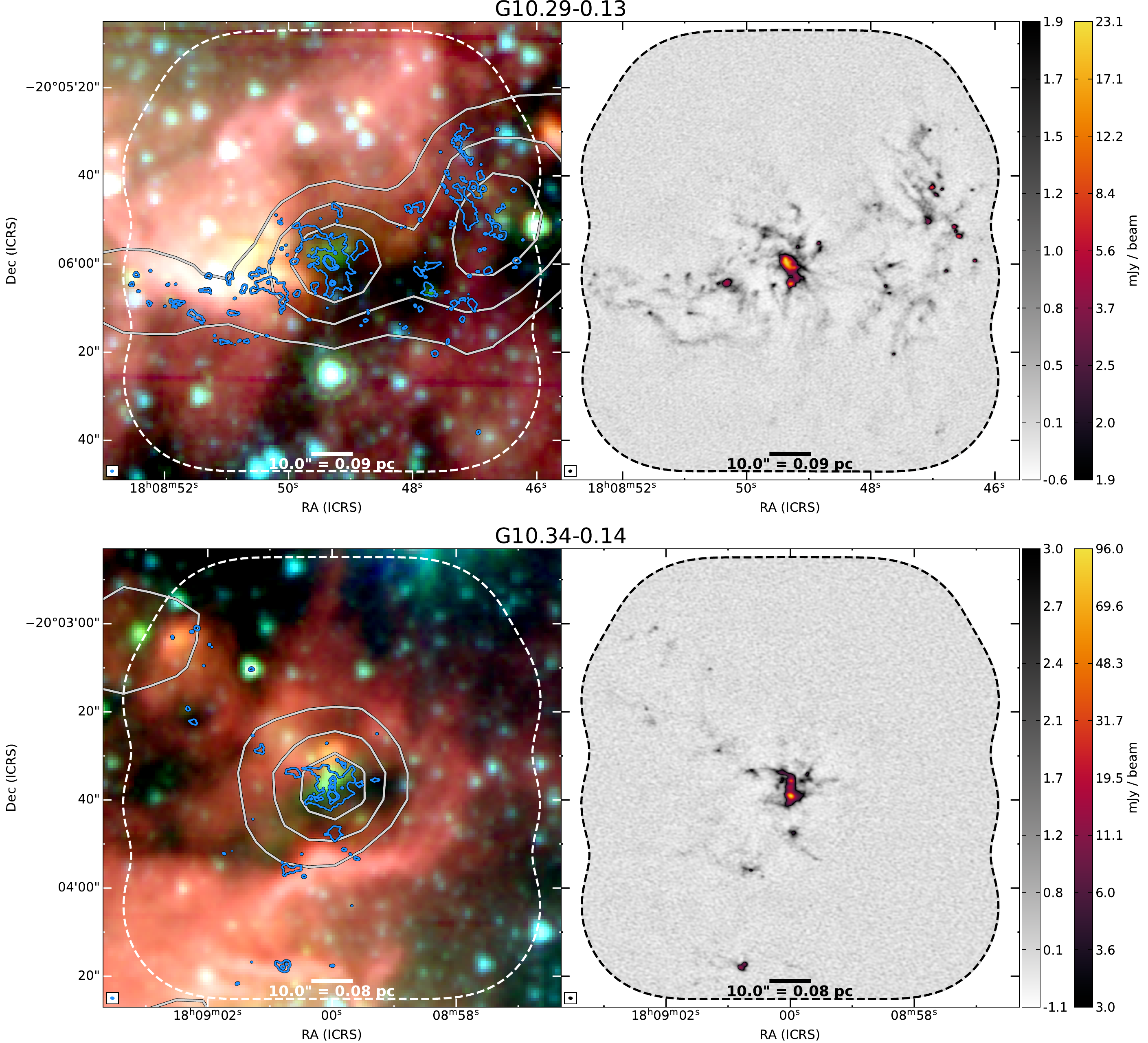}
    
\caption{\textit{Left}: \textit{Spitzer} GLIMPSE IRAC images (RGB: 8, 4.5, 3.6\,$\mu$m) of the EGO-10 fields, overlaid with ALMA 1.3\,mm continuum contours (blue; 12\,m-only images, prior to correction for the primary beam response, levels [5, 50, 100] $\times \sigma_\text{rms}$, where $\sigma_\text{rms}$ is listed in Table~\ref{tab:img}). ATLASGAL 870\,$\mu$m emission \citep{Schuller2009} is shown in silver contours (levels [0.25, 0.5, 0.75] $\times$ $I_{\text{peak}}$, the peak intensity of the central ATLASGAL clump).  
\textit{Right}: ALMA 1.3\,mm continuum  as in \textit{left}, using a double colour bar. The greyscale shows emission from $-$7 to 20 $\times\sigma_\text{rms}$ on a power law stretch with exponent $=$ 1.7. The colourscale shows the emission from 20 $\times\sigma_\text{rms}$ to the value of the image peak intensity, also on a power law stretch, with exponent $=$ 0.4.
Each panel shows the 10\% response level of the ALMA mosaic (dashed contour), the synthesised beam (bottom left), and a 10\arcsec\/ scalebar.} 
\label{fig:mm}
\end{figure*}

\begin{figure*}
\ContinuedFloat
\includegraphics[width=0.95\textwidth]{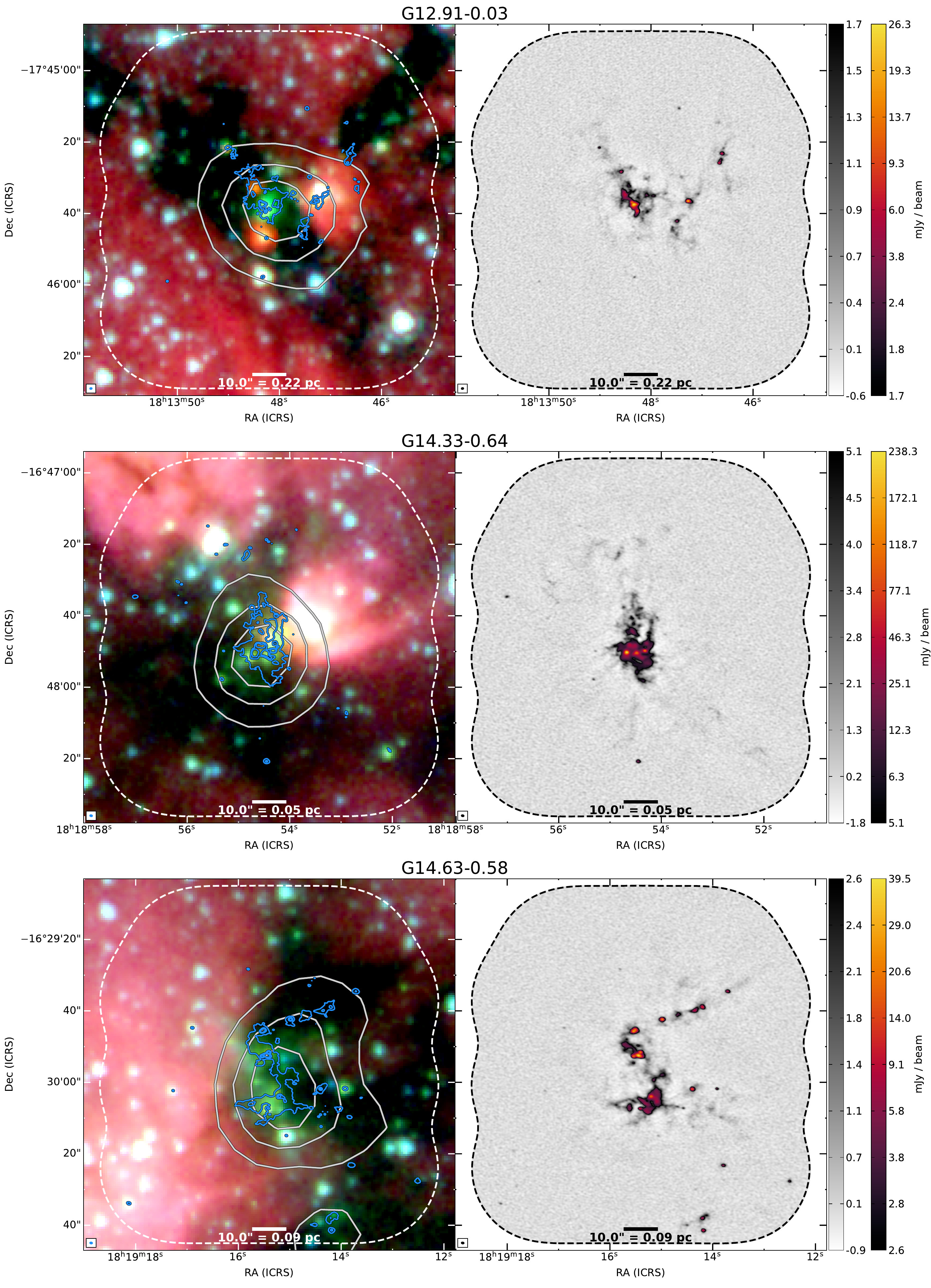}
    
\caption{(Continued.)}
\end{figure*}

\begin{figure*}
\ContinuedFloat
   \includegraphics[width=0.95\textwidth]{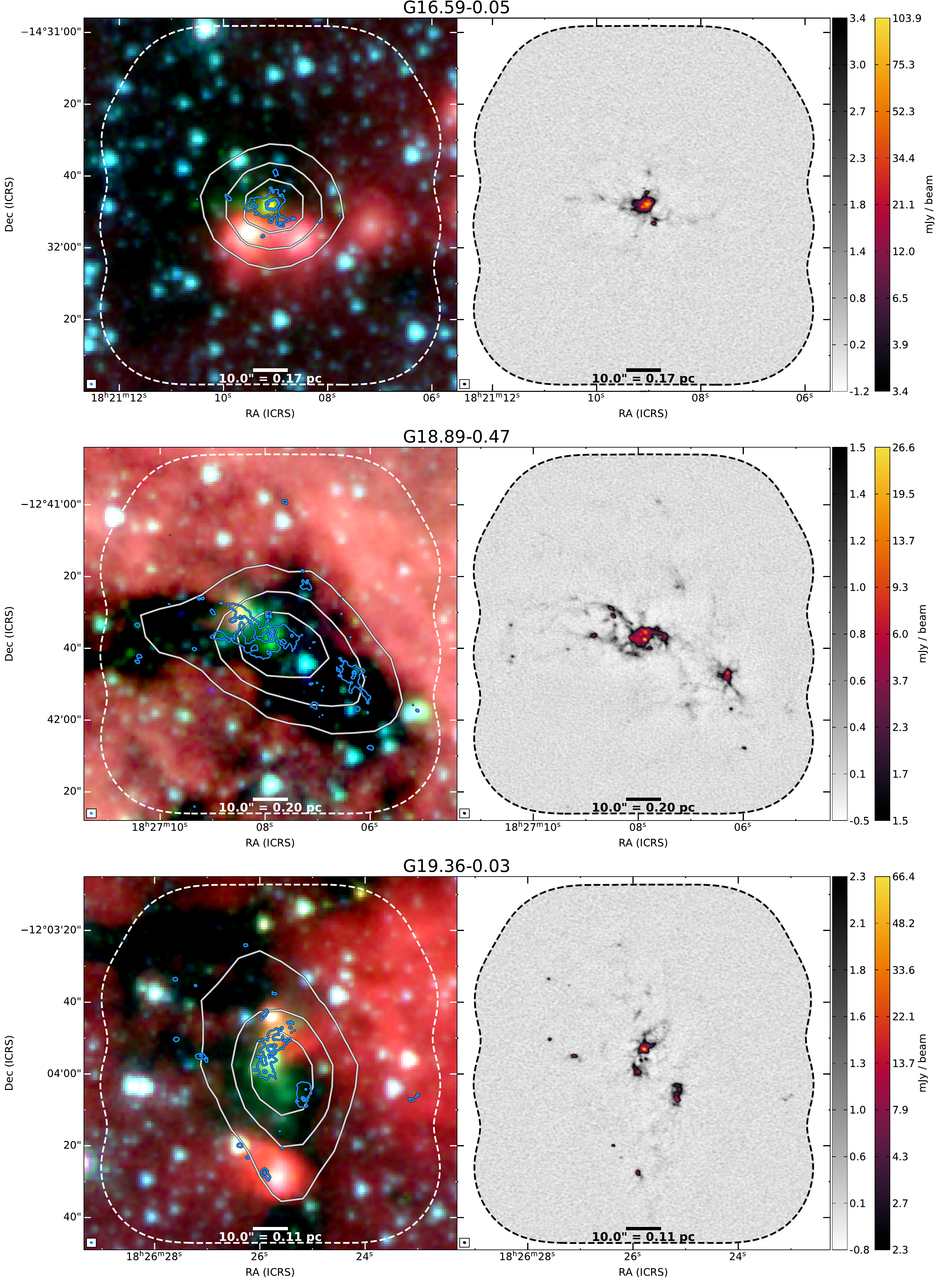}
    
\caption{(Continued.)}
\end{figure*}

\begin{figure*}
\ContinuedFloat
   \includegraphics[width=0.95\textwidth]{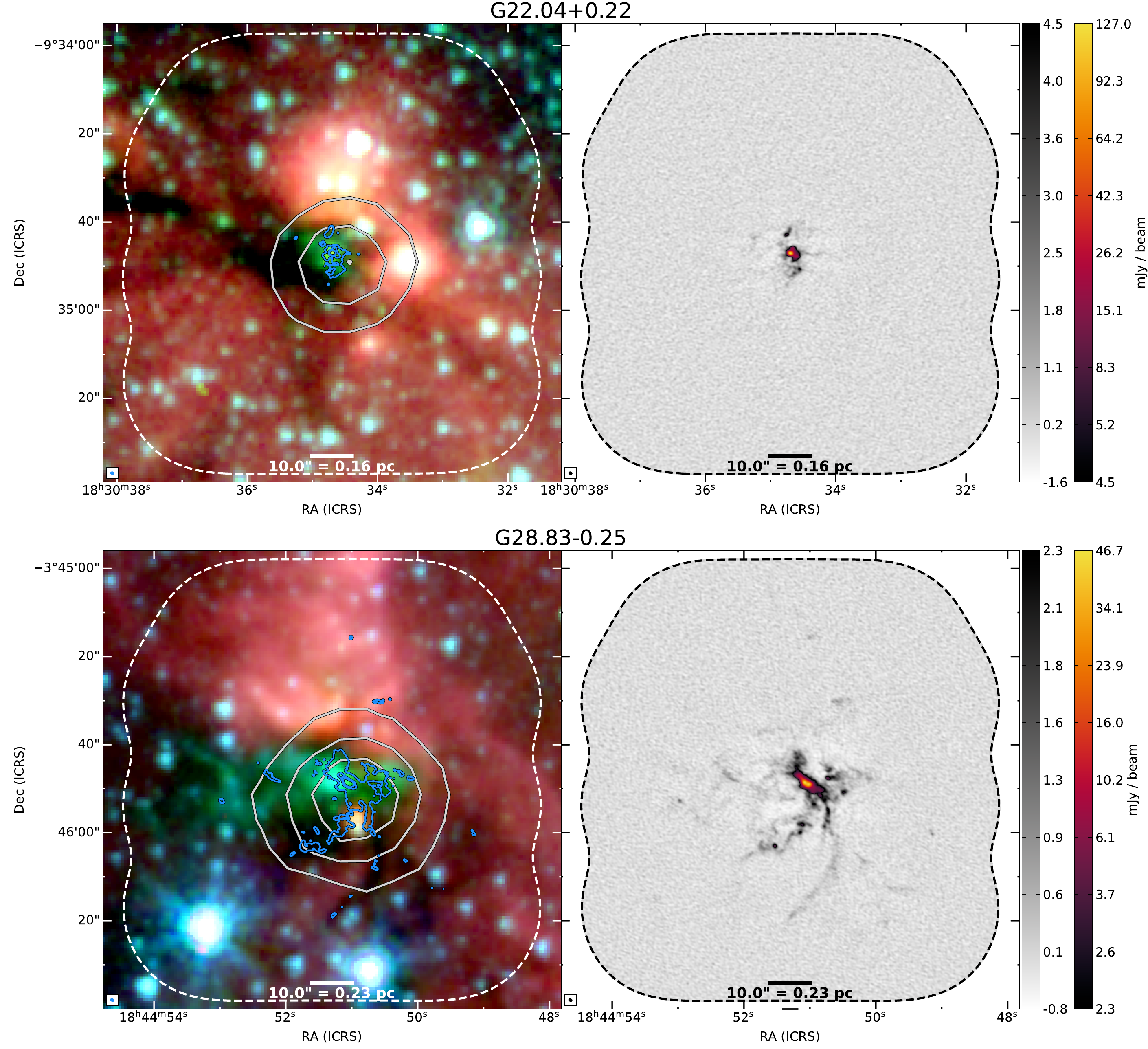}
    
\caption{(Continued.)}
\end{figure*}

\begin{table*}
\caption{Overview of ALMA surveys targeting massive protoclusters.}
\begin{center}
\begin{tabular}{lccccccccc}
\hline
\footnotesize
Survey & N$^\text{a}$  & $\lambda$$^\text{b}$ & $d$$^\text{c}$ & Angular Resolution$^\text{d}$ & Sensitivity$^\text{e}$ & Mosaic?$^\text{f}$ & $L/M$$^\text{g}$\\
& & & (kpc) & ($''$) & (mJy / beam) & (Y/N) & ($L_\odot/M_\odot$) \\
\hline
EGO-10 (this work) & 10 & 1.3\,mm & 1.13 -- 4.8 & 0.66 & 0.17 & Y ($\sim$ 1\farcm6 $\times$ 1\farcm6) & 3 -- 21 \\
ALMA-IMF$^{1, 2, 3}$ & 15 & 1.3\,mm & 2 -- 5.5 & 0.31 -- 0.87 & 0.25 & Y ($\sim$ 1 -- 2$'$ $\times$ 1 -- 2$'$) & 12 -- 108 \\
ALMA-IRDC$^4$ & 10 & 3\,mm & 1.8 -- 5.7 & 2.9 & 0.08 & Y ($\sim$ 1$'$ $\times\sim$ 1 -- 3$'$) & -- \\
INFANT$^5$ & 8 & 1.3\,mm & 3.8 -- 5.4 & 0.64 & 0.10 & Y ($\sim$ 1 -- 2$'$ $\times$ 1 -- 2$'$) & 2 -- 12 \\
ASHES (pilots)$^6$ & 12 & 1.3\,mm & 2.9 -- 5.4 & 1.2 & 0.1 & Y ($\sim$1$'$ $\times$ 1$'$) & 0.05 -- 1.7 \\
ASSEMBLE (pilots)$^7$ & 11 & 870\,$\mu$m & 2.5 -- 7.54 & 0.77 & 1.2 & Y ($\sim$ 0\farcm75 $\times$ 0\farcm75) & 11 -- 79 \\
ALMAGAL$^{8, 9, 10}$ & 1013 & 1.3\,mm & 2 -- 8 & 0.4, 0.2 & 0.1 & N & 0.05 -- 450 \\
QUARKS$^{11, 12}$ & 139 & 1.3\,mm & 1.2 -- 12.9 & 0.3 & 0.1 -- several & N & 4 -- 460  \\
ATOMS$^{13, 14}$ & 146 & 3\,mm & 0.4 -- 13 & 1.5 & 0.4 & N & 3 -- 400\\
DIHCA$^{15, 16}$ & 30 & 1.3\,mm & 1.3 -- 5.26 & 0.30 & 0.25 & N & 0.7 -- 450 \\
SQUALO$^{17}$ & 13 & 1.3\,mm & 2 -- 5.5 & 1.2 & 4.1 & N & 0.1 -- 107\\
TEMPO$^{18}$ & 38 & 1.3\,mm & 1.8 -- 6.3 & 0.8 & 0.26 & N & 0.4 -- 700 \\

\hline
\end{tabular}
\end{center}
\begin{flushleft}

        \small{
        References: ALMA-IMF: $^1$\citet{Motte2022}, $^2$\citet{DellOva2024}, $^3$\citet{Louvet2024}; ALMA-IRDC: $^4$\citet{Barnes2021}; INFANT: $^5$\citet{Cheng2024}; ASHES (pilots): $^6$\citet{Sanhueza2019}; ASSEMBLE (pilots): $^7$\citet{Xu2024}; ALMAGAL; $^8$\citet{Molinari2025}, $^9$\citet{Sanchez-Monge2025},  $^{10}$\citet{Coletta2025};  QUARKS: $^{11}$\citet{Liu2024}, $^{12}$\citet{Xu2024_Q}; ATOMS: $^{13}$\citet{Liu2020}, $^{14}$\citet{Liu2021}; DIHCA: $^{15}$\citet{Olguin2021}, $^{16}$\citet{Ishihara2024}; SQUALO: $^{17}$\citet{Traficante2023}; TEMPO: $^{18}$\citet{Avison2023}.\\

        $^\text{a}$ Sample size.\\
        $^\text{b}$ Wavelength at which continuum source extraction was performed.\\
        $^\text{c}$ Distance range of the sample, as reported by the authors (see Table~\ref{tab:tbl1} for EGO-10).\\
        $^\text{d}$ Average angular resolution reported by the authors. For EGO-10 we report the average geometric mean of the synthesised beams of the 12\,m-only images (see Section~\ref{sec:obs}). In ALMA-IMF and ALMAGAL, regions were imaged at different angular resolution depending on distance to achieve uniform spatial resolution. We report the range of values for ALMA-IMF and the values for the (`near', `far') subsamples for ALMAGAL.\\
        
        $^\text{e}$ Mean image rms noise, as reported by the authors. For EGO-10 we report the mean rms noise of the primary beam-corrected 12\,m-only images (see Section~\ref{sec:obs}).\\
        
        $^\text{f}$ Approximate mosaic dimensions, as reported by the authors or estimated from published images, are given in parentheses. For EGO-10 we report the size at the 10\% response level. For surveys in which mosaics are not uniform in size and/or are elongated (ALMA-IMF, ALMA-IRDC, INFANT), we list approximate ranges for the major and minor mosaic dimensions. 
        
        $^\text{g}$ Luminosity-to-mass ratio range of sample, where reported by the authors, evaluated from spectral energy distribution (SED) fitting with far-infrared (FIR) data (see Table~\ref{tab:tbl1} for EGO-10).}
        
    \end{flushleft}
\label{tab:summ}
\end{table*}

\subsection{The EGO-10 Sample}
\label{sec:ego}
In this work, we present an analysis of the 1.3\,mm continuum emission observed with ALMA toward ten Extended Green Objects (EGOs) - the EGO-10 sample. EGOs are extended 4.5\,$\mu$m emission features 
first 
catalogued by \cite{Cyganowski2008} in \textit{Spitzer} Galactic Legacy Infrared Mid-Plane Survey Extraordinaire \cite[GLIMPSE;][]{Benjamin2003, Churchwell2009} images, which are thought to trace shocked molecular gas in protostellar outflows \citep[e.g.][]{Smith2005,DeBuizer2010,Ray2023}. This interpretation is supported by EGOs' association with collisionally-pumped masers \citep{Cyganowski2009, Chen2011, Cyganowski2013} and with SiO emission \citep{Cyganowski2009}, which indicates the corresponding outflows are active \citep[e.g.][]{PineaudesForets1997}. Crucially, GLIMPSE EGOs are also associated with 6.7\,GHz Class II CH$_3$OH masers \citep{Cyganowski2008, Cyganowski2009}, which are exclusively associated with massive protostars \citep{Minier2003, Breen2013}. Initial Very Large Array (VLA) centimetre continuum studies showed these EGOs are associated with weak, compact centimetre continuum emission, indicating the driving MYSOs are at an early evolutionary stage, prior to the disruption of the natal cloud by ionising radiation \citep{Cyganowski2011_VLA}. GLIMPSE EGOs are therefore outflows, specifically driven by actively-accreting \emph{massive} protostars, that signpost a specific, short-lived evolutionary stage in which the MYSO is just beginning to excite cm-$\lambda$ continuum emission.

To better understand this crucial stage of MYSO evolution, we have undertaken extensive high-resolution multiwavelength follow-up observations of EGO subsamples. 
The \citet{Towner2017} VLA survey revealed EGO-associated CH$_3$OH emission (thermal and 25\,GHz masers) and weak 1.3\,cm continuum emission associated with 6.7\,GHz CH$_3$OH masers. Follow-up Stratospheric Observatory for Infrared Astronomy (SOFIA) observations \citep{Towner2019} suggested there are $\sim$1 to a few MYSOs per EGO, with EGOs in transition from the IR-quiet to IR-bright stages of MYSO evolution \cite[see e.g.][]{Motte2018}. Of particular relevance to this work is the \citet{Towner2021} VLA survey - a deep (sensitivity \mbox{$7-$14\,$\mu$Jy beam$^{-1}$}), high-resolution (\mbox{$\lesssim$0\farcs5}) census of the maser and cm-$\lambda$ continuum emission in a sample of nine EGOs (the EGO-9 sample, a subset of the EGO-10, see below). This survey confirmed moderate MYSO multiplicity, with one to three 6.7\,GHz CH$_3$OH masers detected per field. Ubiquitous weak, compact cm-$\lambda$ continuum emission was also detected, with properties largely consistent with thermal free-free emission from ionised jets, alongside some examples of nascent radiatively-ionised (\ion{H}{ii}) regions and thermal emission from warm dust in dense cores. 

Table~\ref{tab:tbl1} presents the fundamental properties of the EGO-10 regions. 
All were covered by the \citet{Towner2017} and \citet{Towner2019} surveys, and all but G16.59$-$0.05 by \citet{Towner2021} - and for this field, comparable data exist in the literature (see Section~\ref{sec:cm_assoc}). A notable strength of the EGO-10 sample is therefore the wealth of complementary sensitive, high-resolution, multiwavelength data available. In this work, we leverage this multiwavelength synergy by combining these data with our new, sensitive, wide-field 1.3\,mm ALMA mosaics to correlate millimetre cores with cm-$\lambda$ tracers of active star formation and investigate the evolution of structure in the early phases of massive (proto)cluster formation.

In Section~\ref{sec:obs} we detail our observations and imaging procedure. In Section~\ref{sec:source} we present the 1.3\,mm continuum images and describe our source identification approach and the resulting 1.3\,mm core catalogue. In Section~\ref{sec:struc} we quantify protocluster structure, including with the $\mathcal{Q}$-parameter, and in Section~\ref{sec:cm_assoc} we investigate the association between the 1.3\,mm cores and cm-$\lambda$ tracers of active star formation. In Section~\ref{sec:evol} we combine the results of Section~\ref{sec:struc} with the cm-$\lambda$ data in Section~\ref{sec:cm_assoc}, to investigate the relationship between protocluster structure and evolutionary state. A summary of our results and the conclusions of this work are presented in Section~\ref{sec:concl}.

\begin{table*}
\caption{Fundamental properties of the EGO-10 regions.}
\begin{center}
\begin{tabular}{lcccccccc}
\hline
EGO$^\text{a}$  & AGAL$^\text{b}$ & \multicolumn{2}{c}{Position (ICRS)$^\text{c}$} & $V_\text{LSR}$$^\text{d}$ & Distance$^\text{e}$ & $T_{\text{NH}_3}$$^\text{d}$ & $L_\text{AGAL}/M_\text{AGAL}$$^\text{f}$\\
\cline{3-4}
 & & R.A. (h m s) & Dec. ($^\circ$ $^\prime$ $^{\prime\prime}$) & (km s$^{-1}$) & (kpc) & (K) & ($L_\odot/M_\odot$)\\
\hline
G10.29$-$0.13 & \makecell[c]{010.288-00.124\\010.284-00.114} & 18:08:49.3 & -20:05:57.0 & 14 & 1.9 (0.3) & 21.19 & 7\\
\hline
G10.34$-$0.14 & 010.342-00.142 & 18:09:00.0 & -20:03:35.0 & 12 & 1.6 (0.2) & 28.23 & 20\\
G12.91$-$0.03 & 012.904-00.031 & 18:13:48.2 & -17:45:39.0 & 57 & 4.5 (0.7) & 23.56 & 3\\
G14.33$-$0.64 & 014.331-00.644 & 18:18:54.4 & -16:47:46.0 & 23 & 1.13$^{_{+0.14}}_{^{-0.11}}$ & 25.26 & 8\\
\hline
G14.63$-$0.58 & \makecell[c]{014.632-00.577\\014.621-00.579} & 18:19:15.4 & -16:29:55.0 & 19 & 1.83$^{_{+0.08}}_{^{-0.07}}$ & 20.76 & 3\\
\hline
G16.59$-$0.05 & 016.586-00.051 & 18:21:09.1 & -14:31:48.0 & 60 & 3.58$^{_{+0.32}}_{^{-0.27}}$ & 20.51 & 14\\
G18.89$-$0.47 & 018.888-00.474 & 18:27:07.9 & -12:41:36.0 & 66 & 4.2 (0.6) & 28.24 & 11\\
G19.36$-$0.03 & 019.362-00.031 & 18:26:25.8 & -12:03:57.0 & 27 & 2.2 (0.3) & 24.90 & 4\\
G22.04$+$0.22 &  022.038+00.222 & 18:30:34.7 & -09:34:47. & 51 & 3.4 (0.5) & 26.71 & 7\\
G28.83$-$0.25 & 028.831-00.252 & 18:44:51.3 & -03:45:48.0 & 87 & 4.8 (0.7) & 28.27 & 17\\
\hline
\end{tabular}
\end{center}
\begin{flushleft}

        \small{
        $^\text{a}$ Name of the target EGO from \citet{Cyganowski2008}. \\
        $^\text{b}$ ATLASGAL Complete Source Catalogue \cite[CSC;][]{Contreras2013, Urquhart2014} designation of the 870\,$\mu$m clump(s) associated with the field. Where the peaks of two clumps lie within the 10\% response level of the ALMA mosaic (Fig.~\ref{fig:mm}), both are listed, with the central EGO-hosting clump first. \\
        $^\text{c}$ ALMA source position, corresponding to the centre of the mosaic coverage.\\
        $^\text{d}$ From single-dish NH$_3$ observations of \citet{Cyganowski2013} \cite[see also][]{Towner2019}.\\
        $^\text{e}$ Protocluster distance and associated uncertainty \cite[][and references therein]{Towner2019, Towner2021}.\\
        $^\text{f}$ Luminosity-to-mass ratio of the associated ATLASGAL 870\,$\mu$m clump(s) calculated from the (log) values reported in \citet{Urquhart2018}.  For G10.29$-$0.13 and G14.63$-$0.58, we report $L_\text{AGAL}/M_\text{AGAL}$ as the ratio of the sum of clump luminosities to the sum of clump masses.\\      
        }
    \end{flushleft}
\label{tab:tbl1}
\end{table*}

\section{Observations}
\label{sec:obs}
Our 1.3\,mm (Band 6) ALMA EGO-10 observations were taken in 2017--18 as part of projects 2016.1.00747.S and 2017.1.00983.S (PI C.\ Brogan). These ALMA projects included both 12\,m array and Atacama Compact Array (ACA) observations.  

Each field was observed using a mosaic of 27 pointings of the 12\,m array to cover the extent of the ATLASGAL 870\,$\mu$m clump associated with the target EGO (see Fig.~\ref{fig:mm}). The ALMA source positions, which correspond to the centres of the mosaic coverage, are listed in Table~\ref{tab:tbl1}. The dimensions of the resulting mosaics are $\sim$ 95\arcsec$ \times$ 95\arcsec\/ $\approx$ 1\farcm6 $\times$ 1\farcm6, within the 10\% response level.

The 12\,m array observations were taken in configuration C40-3; key parameters of these observations are listed in Table~\ref{tab:obs}.
For our observations, the ALMA 12\,m array correlator was configured to cover 6 spectral windows (spws): 5 narrower spws targeting particular spectral line(s), and one wide (1.875\,GHz bandwidth) spw, centred at $\sim$234.2\,GHz and observed with a channel width of 0.488\,MHz. Two of the narrow spws, tuned to cover C$^{18}$O (2-1) at 219.560358\,GHz and $^{13}$CO (2-1) at 220.398684\,GHz, have bandwidths of 468.75\,MHz and channel widths of 0.244\,MHz.  A further two narrow spws, tuned to cover SiO (5-4) at 217.104919\,GHz and H$_2$CO 3$_{\rm 0,3}$-2$_{\rm 0,2}$ at 218.222192\,GHz, have bandwidths of 937.50\,MHz and channel widths of 0.977\,MHz.  The final narrow spw, tuned to cover N$_2$D$^{+}$ (3-2) at 231.321665\,GHz, has a bandwidth of 937.50\,MHz and channel width of 0.244\,MHz.  

The data were calibrated with CASA \citep{CASATEAM2022} using version 6.2.1.7 of the ALMA calibration pipeline, which includes spectral renormalization to correct for the presence of strong line emission in ALMA autocorrelation spectra \citep[see][]{Hunter2023b}.  
For each target region, a pseudo-continuum dataset was constructed from line-free channels identified using custom runs of the Python function \texttt{findContinuum} \citep{Hunter2023b}; iterative phase-only self-calibration was then performed for each continuum dataset.  The aggregate continuum bandwidth of the pseudocontinuum data for each region is listed in Table~\ref{tab:obs}. In this paper, we consider only the 1.3\,mm continuum data; future papers will present the line emission. 

As the focus of this paper is on identifying dense cores, we opted to make images using only the 12\,m array data to isolate emission on core scales. 
Two image sets were made to address our source extraction goals (see Section~\ref{sec:source}): the `12\,m-only images' and the `40\,k$\lambda$ images'. Both image sets were made via interactive clean using CASA's \texttt{tclean} command, using multi-term multi frequency synthesis (\texttt{mtmfs}) and a Brigg's robust parameter of 0. The 40\,k$\lambda$ images were made by considering only baselines $>$40\,k$\lambda$, implemented via \texttt{tclean}'s \texttt{uvrange} parameter. The rms noise of each image was measured in an emission-free region, with the same region used for all images of a given field. Image parameters are given in Table~\ref{tab:img}. 
All images shown are 12\,m-only images, unless otherwise stated.

\begin{table*}
\caption{Parameters of EGO-10 12\,m array 1.3\,mm observations.}
\begin{tabular}{lccccccc}
\hline
EGO & Project & Obs. Dates & No. Antennas & \multicolumn{3}{c}{Calibrators} & Cont. BW \\
 & & & & Flux & Bandpass & Phase & (GHz) \\
\hline
G10.29$-$0.13 & 2017.1.00983.S & 2018 Apr 20 - May 5 & 43-44 & J1924-2914 & J1924-2914 & J1832-2039 & 1.70 \\
G10.34$-$0.14 & 2016.1.00747.S & 2017 Apr 27-29 & 39-40 & Titan & J1924-2914 & J1832-2039 & 1.11 \\ 
G12.91$-$0.03 &  2017.1.00983.S  & 2018 Apr 20 - May 5 & 43-44 & J1924-2914 & J1924-2914 & J1832-2039 & 1.85 \\
G14.33$-$0.64 & 2016.1.00747.S & 2017 Apr 27-29 & 39-40 & Titan & J1924-2914 & J1832-2039 & 0.38 \\
G14.63$-$0.58 & 2016.1.00747.S  & 2017 Apr 27-29 & 39-40 & Titan & J1924-2914 & J1832-2039 & 2.45 \\
G16.59$-$0.05 & 2016.1.00747.S & 2017 Apr 28-May 1 & 38-40 & J1733-1304, J1924-2914, Titan & J1924-2914 & J1832-2039 & 0.53 \\
G18.89$-$0.47 & 2017.1.00983.S & 2018 Apr 8-11 & 43-45 & J1751+0939 & J1751+0939 & J1832-1035 & 3.06\\
G19.36$-$0.03 & 2016.1.00747.S & 2017 Apr 28-May 1 & 38-40 & J1733-1304, J1924-2914, Titan & J1924-2914 & J1832-2039 & 1.59\\
G22.04$+$0.22 &  2016.1.00747.S & 2017 Apr 28-May 1 & 38-40 & J1733-1304, J1924-2914, Titan & J1924-2914 & J1832-2039 &  0.37\\
G28.83$-$0.25 &  2017.1.00983.S & 2018 Apr 8-11 & 43-45  & J1751+0939 & J1751+0939 & J1832-1035 & 1.16\\
\hline
\end{tabular}
\label{tab:obs}
\end{table*}

\begin{table*}
\caption{Properties of 1.3 mm continuum images$^\text{a}$.}
\begin{tabular}{lccccccccc}
\hline
\hline
 & \multicolumn{5}{c}{12\,m-only Images} & & \multicolumn{3}{c}{40\,k$\lambda$ Images}\\
\cline{2-6}
\cline{8-10}
EGO & Synthesised Beam & MRS$^\text{b}$ & $\sigma_\text{rms}$$^\text{c}$ & $\sigma_\text{rms, pb}$$^\text{d}$ & $M_{5\sigma}$$^\text{e}$ & & Synthesised Beam & $\sigma_\text{rms}$$^\text{c}$ & $\sigma_\text{rms, pb}$$^\text{d}$\\
 & ($''\times''$) $[^\circ]$ & ($''$) & (mJy beam$^{-1}$) & (mJy beam$^{-1}$) & ($M_\odot$) & & ($''\times''$) $[^\circ]$ & (mJy beam$^{-1}$) & (mJy beam$^{-1}$)\\
\hline
G10.29$-$0.13 & 0.72$\times$0.55[-81.71] & 8.19 & 0.093 & 0.105 & 0.04 &  & 0.70$\times$0.53[-81.6] & 0.089 & 0.102 \\
G10.34$-$0.14 & 0.82$\times$0.58[81.85] & 6.58 & 0.152 & 0.187 & 0.03 &  & 0.80$\times$0.57[81.75] & 0.158 & 0.194 \\
G12.91$-$0.03 & 0.72$\times$0.55[-81.08] & 8.19 & 0.083 & 0.104 & 0.17 &  & 0.69$\times$0.53[-80.94] & 0.087 & 0.110 \\
G14.33$-$0.64 & 0.83$\times$0.58[80.79] & 6.58 & 0.253 & 0.313 & 0.03 &  & 0.80$\times$0.57[80.88] & 0.258 & 0.320 \\
G14.63$-$0.58 & 0.84$\times$0.59[80.29] & 6.58 & 0.132 & 0.165 & 0.05 &  & 0.81$\times$0.57[80.29] & 0.123 & 0.157 \\
G16.59$-$0.05 & 0.75$\times$0.56[81.85] & 6.31 & 0.168 & 0.208 & 0.26 &  & 0.73$\times$0.55[82.03] & 0.173 & 0.214 \\
G18.89$-$0.47 & 0.76$\times$0.54[64.5] & 8.32 & 0.077 & 0.105 & 0.12 &  & 0.74$\times$0.52[64.55] & 0.078 & 0.106 \\
G19.36$-$0.03 & 0.75$\times$0.57[80.31] & 6.31 & 0.117 & 0.148 & 0.05 &  & 0.74$\times$0.55[80.42] & 0.119 & 0.151 \\
G22.04$+$0.22 & 0.76$\times$0.57[80.21] & 6.31 & 0.225 & 0.267 & 0.22 &  & 0.75$\times$0.56[80.16] & 0.233 & 0.276 \\
G28.83$-$0.25 & 0.82$\times$0.56[61.61] & 8.32 & 0.116 & 0.141 & 0.21 &  & 0.79$\times$0.54[61.68] & 0.124 & 0.151 \\
\hline
Mean & 0.78$\times$0.57 & 7.2 & 0.14 & 0.17 & 0.12 & & 0.76$\times$0.55 & 0.14 & 0.18 \\
\hline
\end{tabular}
\begin{flushleft}
        \small{$^\text{a}$ $T_B(K)\approx$ 5e-2 $\times$ $I_\text{1.3mm}$ (mJy beam$^{-1}$) $\times$ $\frac{0.42}{\theta_\text{max} \times \theta_\text{min}}$, where $\theta_\text{max}$ and $\theta_\text{min}$ are the dimensions of the synthesised beam in arcseconds (0.42 arcsec$^{2}$ is the mean value of $\theta_\text{max} \times \theta_\text{min}$ for the 12\,m-only images).\\
        $^\text{b}$ Maximum Recoverable Scale, the largest angular scale on which the image is sensitive to smooth emission, estimated using the CASA Analysis Utilities \citep{Hunter2023_au} task \texttt{au.estimateMRS} from the fifth percentile shortest baseline.\\
        $^\text{c}$ rms noise in the non-primary beam-corrected image, measured in CASA by drawing a polygon in an emission-free region within the 50\% response level of the ALMA mosaic. The same emission-free region is used to measure the noise in both the 12\,m-only and 40\,k$\lambda$ images.\\
        $^\text{d}$ rms noise in the image after correction for the primary beam response, measured within the same emission-free region as $\sigma_\text{rms}$.\\ 
        $^\text{e}$ The 5\,$\sigma_\text{rms, pb}$ point-mass sensitivity, estimated by inputting $\sigma_\text{rms, pb}$ to Equation 1 of \citet{Cyganowski2017} with an assumed gas-to-dust mass ratio of $R=$ 100, dust temperature equal to the $T_{\text{NH}_3}$ of the clump (see Table~\ref{tab:tbl1}) and an assumed opacity of $\kappa_\nu=$ 1\,cm$^2$\,g$^{-1}$. 
        }
        \\
    \end{flushleft}
\label{tab:img}
\end{table*}

\section{1.3 mm Continuum: Source Identification and Core Catalogue} 
\label{sec:source}

The ALMA EGO-10 1.3\,mm continuum mosaics are presented in Fig.~\ref{fig:mm}. In all fields, the target EGO - traced by the GLIMPSE 4.5\,$\mu$m emission - is coincident with strong 1.3\,mm continuum. A striking feature of Fig.~\ref{fig:mm} is the morphological diversity of the sample. In some regions, numerous compact millimetre continuum sources trace filamentary structure that extends across the mosaicked area, while in others the 1.3\,mm emission is largely concentrated near the target EGO.

Source identification in (sub)millimetre interferometric images of massive star-forming regions is in general non-trivial, requiring a procedure capable of distinguishing compact emission features against complex backgrounds, often in the presence of non-uniform noise. A number of algorithms have been developed for this purpose over the last few decades, for example: \textit{Gaussclumps} \citep{Stutzki1990}, \textit{Clumpfind} \citep{Williams1994}, \textit{Astrodendro} \citep{Rosolowsky2008}, \textit{CuTEx} \citep{Molinari2011}, \textit{Hyper} \citep{Traficante2015} and \textit{getsf} \citep{Menshchikov2021}. In this section we describe our source identification approach, which combines the \textit{Astrodendro} and \textit{Hyper} algorithms to produce a source catalogue that we find corresponds well to the core-scale structure based on careful visual inspection of the 1.3\,mm continuum images. In general, we identify sources in images that have not been corrected for the primary beam response - to prevent spurious detections near the edge of the mosaics. Source photometry is then performed in images that have been primary beam-corrected \cite[e.g.][]{Sanhueza2019, Svoboda2019, Barnes2021, Avison2023, Louvet2024, Zhang2025}.

\subsection{Astrodendro}
\label{sec:dendro}
Dendrograms \citep{Rosolowsky2008} remain one of the most commonly used source identification tools at far-infrared and (sub)millimetre wavelengths. The procedure takes an input image and assigns the emission to a set of nested hierarchical structures defined by the image isocontours. The `leaves' of the dendrogram are the highest-level structures in the hierarchy, with no discernible substructure given the user-defined input parameters. Three input parameters are required: the flux threshold below which pixels are ignored (\texttt{min$\_$flux}), the minimum isocontour separation required to define an independent structure (\texttt{min$\_$delta}) and the minimum number of pixels in an independent structure
(\texttt{min$\_$npix}). 
When applied to images of molecular clouds and star-forming regions, dendrogram leaves often correspond to small-scale emission features embedded in larger-scale, lower-intensity backgrounds - and are thus generally interpreted as dense cores \cite[e.g.][]{Cheng2018, Liu2018, Williams2018, Kong2019, Li2019, Sanhueza2019, Svoboda2019, Li2020, Barnes2021, Liu2021, Avison2023, Budaiev2024, Li2024, Zhang2025}.

We used the well-documented Python implementation of dendrograms, \textit{Astrodendro}\footnote{\url{https://dendrograms.readthedocs.io/en/stable/}}. We initially explored running \textit{Astrodendro} on our non-primary beam-corrected 12\,m-only images, using standard input parameters \cite[\texttt{min$\_$flux} $=$ 4 $\sigma_{\text{rms}}$, \texttt{min$\_$delta} $=$ $\sigma_{\text{rms}}$, \texttt{min$\_$npix} $=$ $N_{\text{beam}}$/2, as in e.g.][]{Liu2018, Kong2019}. This attempt, however, gave numerous irregular leaves that appeared to trace fragments of filamentary structure rather than cores. 
To minimise the identification of filament fragments as leaves, we instead ran \textit{Astrodendro} on our 40\,k$\lambda$ images (see Section~\ref{sec:obs}). The minimum baseline cut-off of 40\,k$\lambda$ corresponds to an angular scale $\sim$5\arcsec\/ and physical scale 0.075\,pc $\approx$ 15,000\,AU at the mean distance of the sample (2.9\,kpc). Extended filamentary emission is thus filtered out in the 40\,k$\lambda$ images, which remain sensitive to emission on core scales. 

\begin{figure*}
   \includegraphics[width=\textwidth]{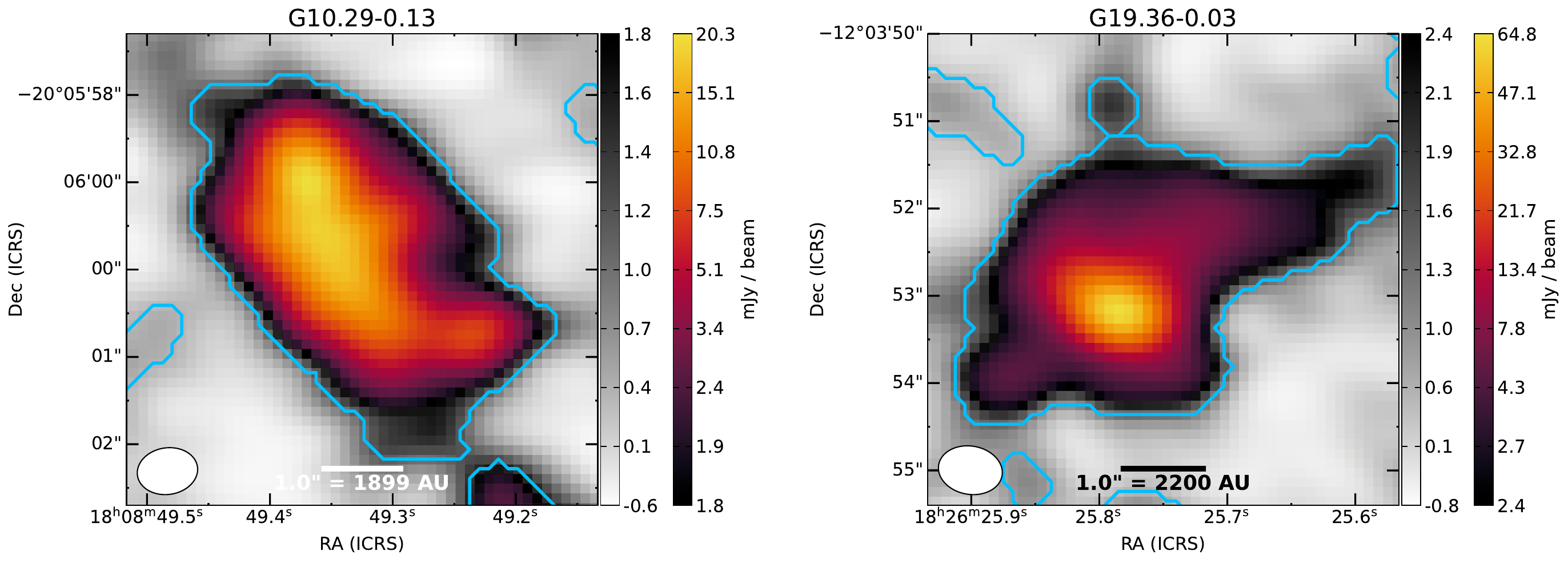}
   \includegraphics[width=\textwidth]{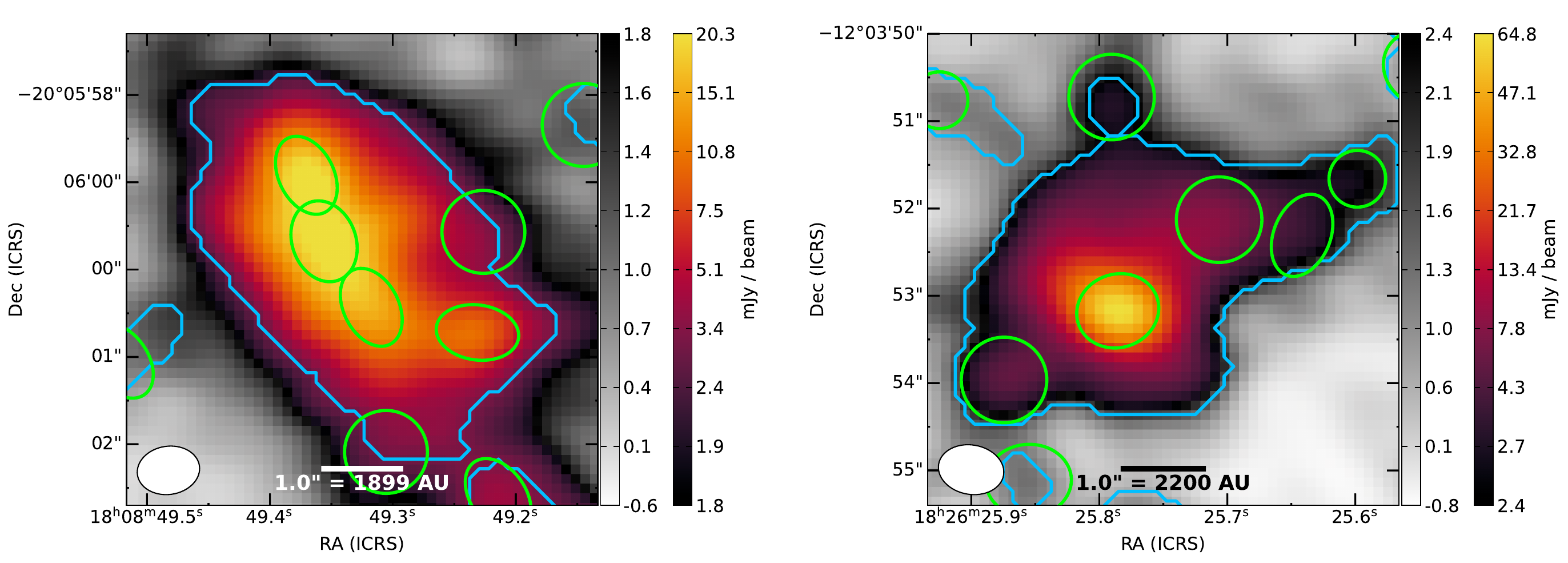}
\caption{Zoom-views of the central regions of G10.29$-$0.13 (\textit{left}) and G19.36$-$0.03 (\textit{right}). Upper panels: ALMA 40\,k$\lambda$ image uncorrected for the primary beam response (colourscale), overlaid with \textit{Astrodendro} leaf boundaries (light blue) identified in this image (Section~\ref{sec:dendro}). Lower panels: ALMA 12\,m-only image (colourscale), overlaid with \textit{Astrodendro} leaf boundaries as above, and FWHM ellipses (green) of the cores identified in the 12\,m-only image with $\textit{Hyper}$ (Section~\ref{sec:hyper}). The colourscale is as defined in Fig.~\ref{fig:mm}, but the 40\,k$\lambda$ $\sigma_\text{rms}$ value (Table~\ref{tab:img}) is used for both panels for each region. In both regions the large, irregular \textit{Astrodendro} leaf is resolved into multiple core-scale Gaussian sources by \textit{Hyper}.} 
\label{fig:dhcomp}
\end{figure*}

After exploring a range of \textit{Astrodendro} input parameter combinations, we adopted \texttt{min$\_$flux} $=$ 4\,$\sigma_{\text{rms}}$, \texttt{min$\_$delta} $=\sigma_{\text{rms}}$ and \texttt{min$\_$npix }$=N_{\text{beam}}$/3 (with $N_{\text{beam}}$ and $\sigma_{\text{rms}}$ measured from the non-primary beam-corrected 40\,k$\lambda$ images, see Table~\ref{tab:img}). We found \texttt{min$\_$npix }$=N_{\text{beam}}$/3 captured some obvious sources for which the size of the emission region above the \texttt{min$\_$flux} threshold was reduced in the 40\,k$\lambda$ images, particularly for weak, isolated sources. To remove spurious sources, we checked the S/N of the resulting leaves in the primary beam-corrected 40\,k$\lambda$ images and discarded leaves with peak intensity $<$5\,$\sigma_\text{rms, pb}$ (see Table~\ref{tab:img}). Overall, we found this procedure improved on our initial \textit{Astrodendro} runs in the 12\,m-only images and gave leaves which reliably corresponded to obvious 1.3\,mm emission features. A persistent issue, however, was the poor separation of blended sources in the crowded central regions of some fields (see Fig.~\ref{fig:dhcomp}). This issue motivated the use of an alternative source extraction algorithm. 

\subsection{Hyper}
\label{sec:hyper}
The `hybrid photometry and extraction routine' \cite[\textit{Hyper};][]{Traficante2015} is a source identification algorithm specifically designed to separate blended sources against complex backgrounds. Originally developed for the \textit{Herschel} Infrared Galactic Plane Survey (Hi-GAL), \textit{Hyper} is now also applied to ALMA 1.3\,mm images \citep{Traficante2023} and synthetic observations \citep{Nucara2025} of massive star-forming regions. 

The code applies a high-pass filter to the input image, amplifying point-like structure and suppressing diffuse emission \citep{Traficante2015}. Source centroids are defined as peaks in the high-pass image at which the intensity in the input image is greater than a user-defined threshold ($\sigma_t$).\footnote{This differs from the initial version described in \citet{Traficante2015}, in which source centroids are defined as positions above a user-defined threshold value in the clean filtered image.}  A 2D Gaussian is then fit at each centroid in the input image, and a source photometric aperture - within which the flux density is integrated - defined as an ellipse with axes 
2 $\times$ the FWHM of the Gaussian fit. If a given source has `companions' (other source centroids within 2 $\times$ the beam FWHM), the source and its companions are fit via multi-Gaussian fitting, and the companion model fits are subtracted prior to flux integration. We did not use \textit{Hyper}'s background subtraction capability, in which the local background of each source is fit with a 2D polynomial and subtracted prior to source photometry. In initial tests, this procedure significantly reduced the measured integrated flux density of bright sources in crowded regions compared to the value in the original image, so we chose to disable it.

Since the high-pass filtering in \textit{Hyper} already suppresses extended emission, we ran the algorithm on the non-primary beam-corrected 12\,m-only images. We used a threshold value $\sigma_t=$ 5\,$\sigma_{\mathrm{rms}}$ in six of the ten regions. In the remaining four (G10.29$-$0.13, G12.91$-$0.03, G18.89$-$0.47 and G28.83$-$0.25), improved \textit{(u, v)}-coverage means the images show more filamentary structure (see Fig.~\ref{fig:mm} and MRS values in Table~\ref{tab:img}), in which some small-scale variations survived the image filtering and were subsequently identified as sources by \textit{Hyper}. We therefore used $\sigma_t=$ 7\,$\sigma_{\mathrm{rms}}$ for these fields, which we found prevented the identification of such features. 

\subsection{Core Catalogue} 
\label{sec:cat}
In general, we find good \textit{Astrodendro}-\textit{Hyper} correspondence, with $\sim$86\% of the \textit{Astrodendro} leaves identified in the 40\,k$\lambda$ images associated with at least one \textit{Hyper} source (i.e. a \textit{Hyper} source centroid lies within the \textit{Astrodendro} leaf boundary). There are however typically more \textit{Hyper} sources than \textit{Astrodendro} leaves in a given field: $\sim$twice as many on average. This is, in part, expected given the desired outcome of \textit{Hyper} separating blended sources comprising single \textit{Astrodendro} leaves. However, an average of 30\% of \textit{Hyper} sources per field are not associated with an \textit{Astrodendro} leaf. Unlike \textit{Astrodendro}, \textit{Hyper} does not require the peak intensity of identified sources to be significant relative to local more extended emission (c.f. \textit{Astrodendro} \texttt{min\_delta} distinguishing leaves from branches and trunks), so may identify sources that are not within the boundary of an \textit{Astrodendro} leaf. Further, whilst limits are placed on the FWHM size of sources identified by \textit{Hyper} (see below), \textit{Hyper} places no limit on the size of the significant ($>\sigma_t$) emission required for source identification (\textit{Astrodendro}: \texttt{min\_npix}). \textit{Hyper} may therefore identify weaker sources missed by \textit{Astrodendro}, or spurious sources in filamentary structures or small-scale noise fluctuations. 

To ensure only the most reliable sources are included in our final core catalogue, we select structures identified by both \textit{Astrodendro} and \textit{Hyper}, i.e. \textit{Hyper} sources with centroids lying within the \textit{Astrodendro} leaves identified in the 40\,k$\lambda$ images. We found this approach gave sources which corresponded well to the core-scale emission in the 1.3\,mm continuum images (with one notable exception, see Appendix~\ref{app:G22}) and generally separated blended sources in the central regions of our fields. This is illustrated in Fig.~\ref{fig:dhcomp}, where large, irregular, central \textit{Astrodendro} leaves ($\sim$10,000\,AU) are separated into multiple core-scale Gaussian sources (FWHM $\approx$ 2000\,AU) by \textit{Hyper}. 

\textit{Hyper} is run twice in a given field. 
First, \textit{Hyper} is run on the non-primary beam-corrected 12\,m-only images to identify source centroids. Then, only those centroids lying within the \textit{Astrodendro} leaves identified in the 40\,k$\lambda$ images are provided to \textit{Hyper} in a second run, for source photometry. In this second run, the source photometry is performed in the 12\,m-only images corrected for the primary beam response. To ensure sensible (multi-)Gaussian fits in this second run, we limited the fit amplitudes to be greater than zero, with an initial estimate equal to the image intensity at the source centroid. We also limited the minimum and maximum allowed size of source FWHMs to within 1 and 1.5 times the FWHM of the image beam, with \textit{Hyper}'s \texttt{aper\_inf} and \texttt{aper\_sup} parameters, respectively. The (multi-)Gaussian fits to all our sources converged, that is, the \textit{Hyper} \texttt{STATUS} output is 0 in all cases \cite[see][]{Traficante2015}.

Our complete source identification procedure is thus:
\vspace{-0.8em}
\begin{itemize}
\item[1.] Run \textit{Astrodendro} on the non-primary beam-corrected 40\,k$\lambda$ image.
\item[2.] Select leaves with a peak intensity $>$ 5\,$\sigma_{\mathrm{rms, pb}}$ in the primary beam-corrected 40\,k$\lambda$ image.
\item[3.] Run \textit{Hyper} on the non-primary beam-corrected 12\,m-only image, with $\sigma_t=$ 5\,$\sigma_{\mathrm{rms}}$ or 7\,$\sigma_{\mathrm{rms}}$, depending on the region.
\item[4.] Select \textit{Hyper} sources with centroids lying within the \textit{Astrodendro} leaves selected in step 2. 
\item[5.] Rerun \textit{Hyper}, providing the source centroids from step 4 for (multi-)Gaussian fitting and aperture photometry, in the primary beam-corrected 12\,m-only image.
\end{itemize}

We find this combined approach leverages the strengths and addresses the shortcomings of using each respective algorithm. In our mosaic images, \textit{Astrodendro} reliably identifies significant emission features, while \textit{Hyper} separates blended sources within these structures. This technique yields a total of 570 sources across the EGO-10 sample (an average of 57 per field), which we take to be the dense cores in these regions. Our core catalogue, presented in Table~\ref{tab:cat}, includes the measured properties and multiwavelength associations (see Section~\ref{sec:cm_assoc}) of these cores. 
Within each field, cores are named in order of decreasing \textit{Hyper} peak intensity, beginning with `MM1' for the brightest core. The number of cores per field ($N_\text{cores}$) is given in Table~\ref{tab:mst}  and ranges from 13 to 135, in G22.04$+$0.22 and G10.29$-$0.13, respectively. 

\begin{landscape}
\begin{table}

  \caption{EGO-10 Core Catalogue: Measured properties and multiwavelength associations of identified cores.}
 \label{tab:cat}
 \begin{tabular}{lccccccccccccccccc}
\hline
\hline
EGO & ID$^\text{a}$ & \multicolumn{2}{c}{Position (ICRS)$^\text{b}$} &  &  & & \multicolumn{3}{c}{Size$^\text{f}$} & P$^\text{g}$ & \multicolumn{2}{c}{6.7\,GHz CH$_3$OH} & \multicolumn{2}{c}{22\,GHz H$_2$O} & \multicolumn{2}{c}{cm-$\lambda$ sources}\\
\cline{3-4}
\cline{8-10}
\cline{12-13}
\cline{14-15}
\cline{16-17}
& & R.A. & Dec. &  $I_\text{1.3mm}$$^\text{c}$ &  $S_\text{1.3mm}$$^\text{d}$ &  $\sigma_{S_\text{1.3mm}}$$^\text{e}$ & $\theta_\text{maj}$  & $\theta_\text{min}$  & P.A. & & Name$^\text{h}$ & $\Delta_\text{off}$$^\text{i}$ & Name$^\text{h}$ & $\Delta_\text{off}$$^\text{i}$ & Name$^\text{h}$ & $\Delta_\text{off}$$^\text{i}$\\
& & (h m s) & ($^\circ$ $^\prime$ $^{\prime\prime}$) & (mJy / beam) & (mJy) & (mJy) & (\arcsec) & (\arcsec) & ($^\circ$) & & & (AU) & & (AU) & & (AU)\\
\hline
G10.29$-$0.13 & MM1 & 18:08:46.589 & -20:05:53.7 & 22.2 & 32.5 & 0.2 & 0.84 & 0.63 & 93.7 & N &  &  &  &  &  &  \\
G10.29$-$0.13 & MM2 & 18:08:49.370 & -20:05:58.9 & 20.4 & 41.2 & 0.2 & 0.95 & 0.63 & 206.9 & Y & CM1-M1 & 210 & CM1-W1 & 210 & CM1 & 310 \\
G10.29$-$0.13 & MM3 & 18:08:49.356 & -20:05:59.7 & 18.1 & 47.4 & 0.2 & 0.95 & 0.73 & 200.2 & N &  &  &  &  &  &  \\
G10.29$-$0.13 & MM4 & 18:08:46.330 & -20:05:59.2 & 17.6 & 22.0 & 0.2 & 0.84 & 0.66 & 119.4 & N &  &  &  &  &  &  \\
G10.29$-$0.13 & MM5 & 18:08:49.318 & -20:06:00.4 & 14.9 & 38.1 & 0.2 & 0.95 & 0.63 & 207.2 & N &  &  &  &  &  &  \\
G10.29$-$0.13 & MM6 & 18:08:49.294 & -20:06:04.5 & 14.1 & 27.6 & 0.2 & 0.95 & 0.68 & 111.4 & N &  &  &  &  &  &  \\
G10.29$-$0.13 & MM7 & 18:08:47.016 & -20:05:42.6 & 14.0 & 19.3 & 0.2 & 0.87 & 0.65 & 128.6 & N &  &  &  &  &  &  \\
G10.29$-$0.13 & MM8 & 18:08:46.661 & -20:05:51.5 & 13.9 & 21.9 & 0.2 & 0.93 & 0.63 & 106.5 & Y &  &  & NC-W1 & 230 &  &  \\
G10.29$-$0.13 & MM9 & 18:08:49.231 & -20:06:00.7 & 10.2 & 21.9 & 0.2 & 0.95 & 0.63 & 262.3 & N &  &  &  &  &  &  \\
G10.29$-$0.13 & MM10 & 18:08:52.462 & -20:06:02.4 & 7.7 & 11.4 & 0.2 & 0.76 & 0.72 & 134.8 & Y &  &  & NC-W4 & 400 &  &  \\
\hline
\end{tabular}
\begin{flushleft}
        \small{$^\text{a}$ Core name, assigned in order of decreasing \textit{Hyper}-reported peak intensity in each region.\\
        $^\text{b}$ ICRS coordinates of the source centroid as reported by \textit{Hyper}. The number of significant figures reflects a one pixel uncertainty (0\farcs11).\\
        $^\text{c}$ \textit{Hyper} source peak intensity (\textit{Hyper}'s \texttt{FLUX\_PEAK\_JY} parameter), which is the image intensity at the source centroid position following companions subtraction, if applicable (see Section~\ref{sec:hyper}).\\
        $^\text{d}$ \textit{Hyper} source integrated flux density (\textit{Hyper}'s \texttt{FLUX} parameter), following companions subtraction, if applicable (see Section~\ref{sec:hyper}).\\
        $^\text{e}$ Uncertainty of the integrated flux density, estimated as $\sigma_{S_\text{1.3mm}}=\sigma_\text{rms, pb}\;\sqrt{N_b}$, where $\sigma_\text{rms, pb}$ is the rms noise of the primary beam-corrected image (Table~\ref{tab:img}) and $N_b$ is the number of beams in the source's photometric aperture \cite[see e.g.][]{Beltran2001, Cantwell2016, Thwala2019}. Note that these uncertainties do not include the absolute flux calibration uncertainty \citep[10\% in Band 6;][]{Braatz2020} and so are dominated by the aperture size.\\
        $^\text{f}$ Source major and minor axes (FWHM) as reported by \textit{Hyper} (\textit{Hyper}'s \texttt{FWHM\_MAX} and \texttt{FWHM\_MIN} parameters). The position angle (P.A., \textit{Hyper}'s \texttt{PA} parameter) lies between 0$^\circ-$ 360$^\circ$ and is measured east of north.\\
        $^\text{g}$ Protostellar association flag: `Y' if the core is associated with maser(s) and/or cm-$\lambda$ continuum source(s), `N' otherwise (see Section~\ref{sec:cm_assoc}).\\
        $^\text{h}$ Name of associated tracer, following the convention described in Section~\ref{sec:cm_assoc}.\\
        $^\text{i}$ Projected physical separation between the tracer position and the centroid of its 1.3\,mm host core (see Section~\ref{sec:cm_assoc}).\\
        }
        (This table is available in its entirety in machine-readable form in the online paper).\\
    \end{flushleft}
\end{table}
\end{landscape}

\section{Protocluster Structure} 
Having catalogued the 1.3\,mm cores in each EGO-10 field, we now examine the structure of these protoclusters, as traced by the spatial distribution of 1.3\,mm cores.
\label{sec:struc}

\begin{figure*}
\includegraphics[width=\textwidth]{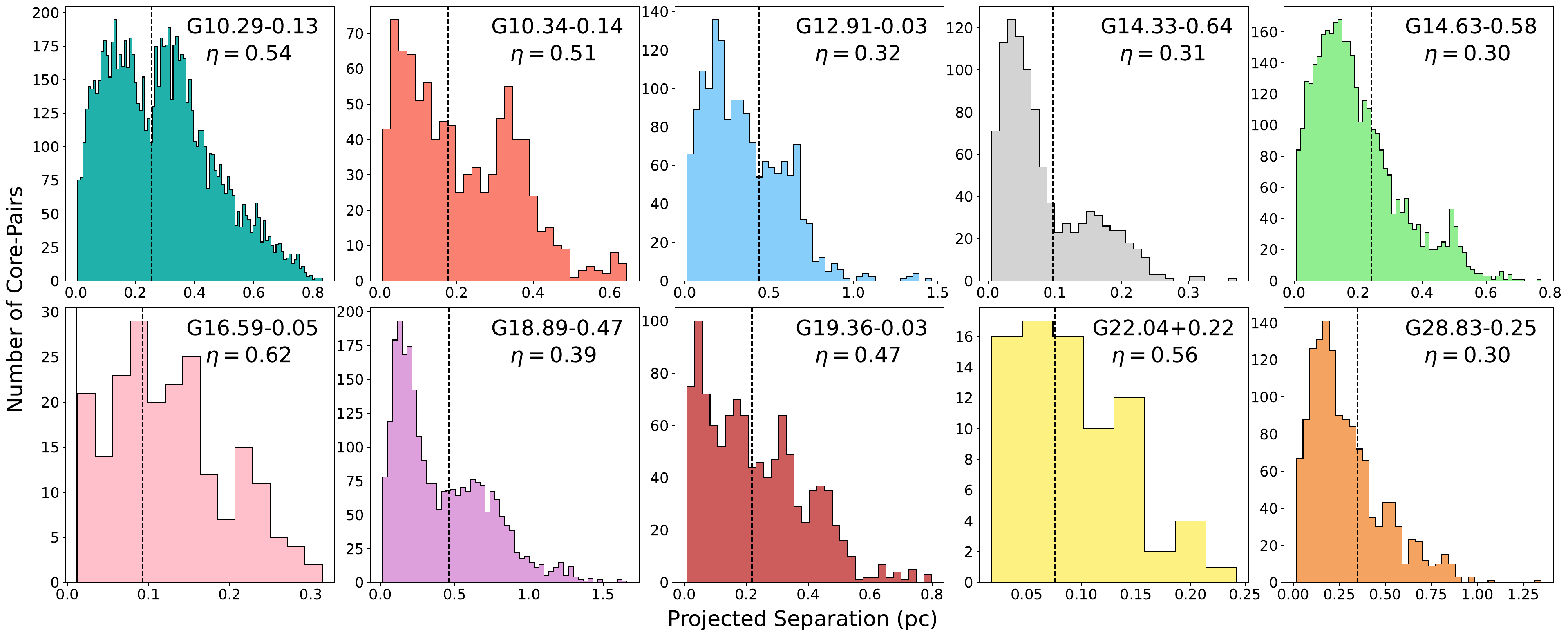}
\caption{Histograms of projected separation between unique core-pairs in each EGO-10 field. The number of bins in each histogram is the square root of the number of unique core-pairs. The value of 0.5\,$R_\text{clus}$ is denoted by the dashed black line. Note the large fraction of the distribution which lies beyond 0.5\,$R_\text{clus}$ in many fields ($\eta$), with some regions exhibiting a clear double-peaked distribution (G10.29$-$0.13, G10.34$-$0.14, G14.33$-$0.64, G14.63$-$0.58, G18.89$-$0.47).}
\label{fig:disthist}
\end{figure*}

\subsection{Distributions of Core Separations} 
\label{sec:hists}
\citet{Cyganowski2017} characterised the structure of the protocluster detected in their ALMA observations of the EGO G11.92$-$0.61 (1.05\,mm, $\sim$0\farcm7 $\times$ 0\farcm7 mosaic). These authors identified a double-peaked profile in the distribution of projected physical separation between unique core-pairs, with a primary peak $\sim$0.1\,pc and a less-pronounced peak $\sim$0.3\,pc (their Figure 5). Fig.~\ref{fig:disthist} shows histograms of this distribution for the EGO-10 regions. 
In all fields, the distribution exhibits a primary peak at a relatively small separation ($\lesssim$0.1\,pc), with 5/10 fields exhibiting a well-defined secondary peak at a larger value\footnote{All primary and secondary peaks occur at values larger than the physical scale of the beam (mean value $\sim$0.01\,pc), so are unlikely to be an effect of resolution limitations.} (G10.29$-$0.13, G10.34$-$0.14, G14.33$-$0.64, G14.63$-$0.58, G18.89$-$0.47). In these regions, the primary and secondary peaks do not necessarily occur at the \citet{Cyganowski2017} values of 0.1\,pc and 0.3\,pc, with the primary peaks occurring at $\sim$0.05--0.1\,pc and the secondary peaks at $\sim$0.15--0.6\,pc. Also plotted in each panel is half the value of the cluster radius $R_\text{clus}$ (see Table~\ref{tab:mst}), calculated as the distance from the mean position of all catalogued cores, which defines the cluster centre, to the catalogued core furthest from this position \citep[][see also Section~\ref{sec:intro}]{Cartwright2004}. Notably, in fields with double-peaked distributions, the primary peak occurs at $\lesssim$0.5\,$R_\text{clus}$ and the secondary peak at $\gtrsim$0.5\,$R_\text{clus}$.

In some fields, the distribution clearly reflects the structure seen in Fig.~\ref{fig:mm}, particularly when considering the relative heights of the primary and secondary peaks. For example, in G10.29$-$0.13 the primary and secondary peaks, at $\sim$0.1\,pc and $\sim$0.3\,pc $\approx$ 0.2\,$R_\text{clus}$ and 0.6\,$R_\text{clus}$, respectively, are of similar height. The number of core-pairs with relatively small separations is therefore similar to that with relatively large separations, reflective of a more extended core spatial distribution. In contrast, in G14.33$-$0.64, in which the primary and secondary peaks occur at 0.05\,pc and 0.15\,pc $\approx$ 0.3\,$R_\text{clus}$ and 0.8\,$R_\text{clus}$, respectively, the secondary peak is less pronounced (only $\sim$1/3
the height of the primary). Such a distribution implies fewer core-pairs with relatively large separations, consistent with a more centrally concentrated core spatial distribution. Table~\ref{tab:mst} reports the fraction of the distribution with separations $>$0.5\,$R_\text{clus}$ in each field ($\eta$; also given in the panels of Fig.~\ref{fig:disthist}).  
 
The clustering of cores at relatively small separations, accompanied by a more extended core distribution, is consistent with simulations demonstrating hierarchical massive star and protocluster formation \cite[e.g.][]{Bonnell2004, Smith2009, Wang2010}. In these simulations, sources cluster close to a central massive source, at relatively small separations ($\lesssim$0.1\,pc), but are also distributed throughout its accretion reservoir ($\gtrsim$few tenths of a pc). We assess the clustering in the environments of the EGO-10 MYSOs in Section~\ref{sec:myso}.

\subsection{Minimum Spanning Trees and the $\mathcal{Q}$-Parameter} 
\label{sec:mst}
\begin{figure*}
   \includegraphics[width=\textwidth]{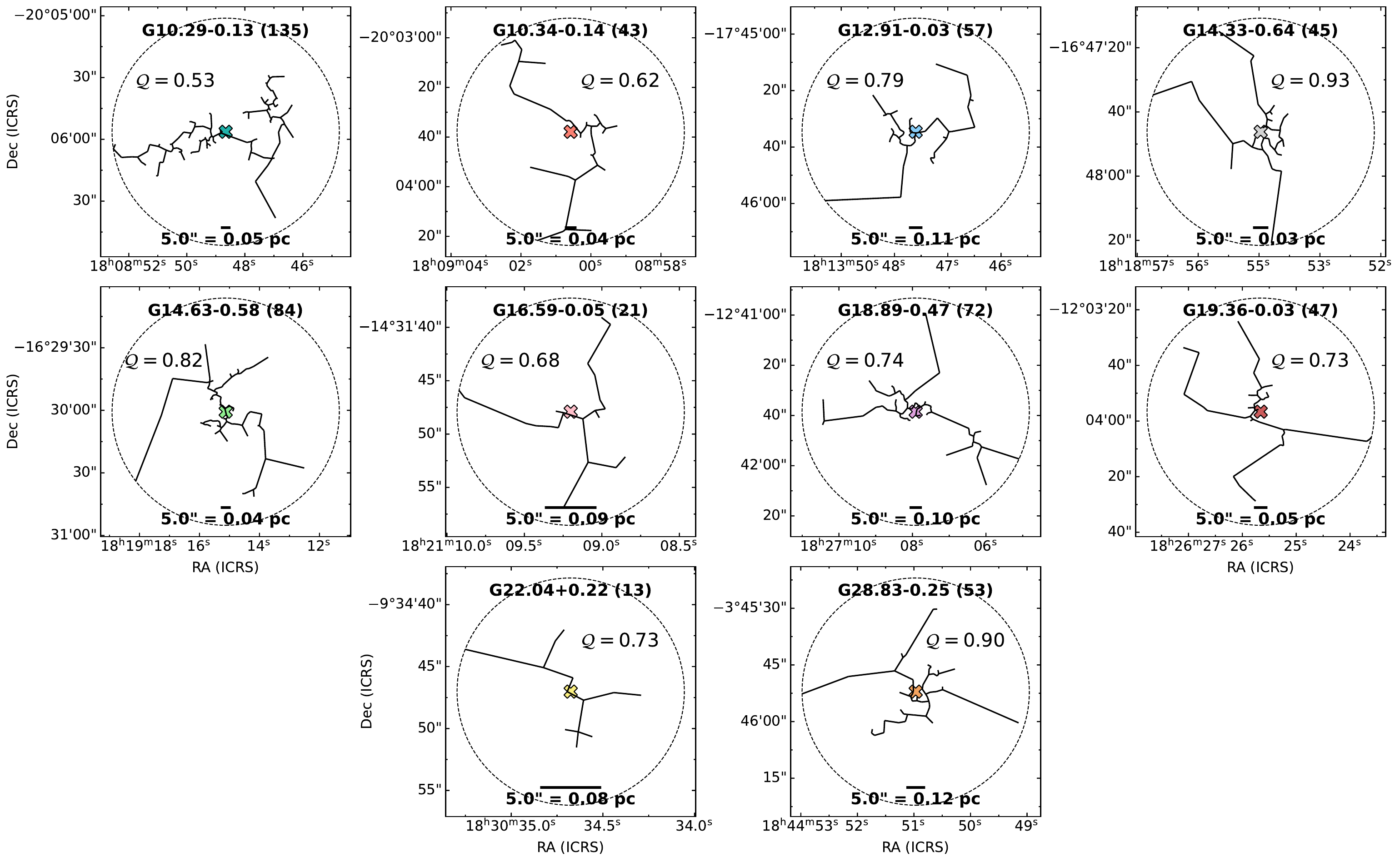} 
\caption{MSTs of the core spatial distributions in the EGO-10 sample. The number of cores identified in each field is given in parentheses in the panel heading. Each panel is centred on the mean position of cores in the field, denoted by a coloured $\times$ (the colour used for each field matches that in Fig.~\ref{fig:disthist}). In each panel, MST edges are shown as black lines, dashed circles indicate the cluster radius ($R_\text{clus}$), and the $\mathcal{Q}$-parameter is given in an in-panel label. Note that each panel is zoomed to $R_\text{clus}$ to best illustrate the structure of the MST in that field, so the panels do not cover the same angular or physical scale.}
\label{fig:mst}
\end{figure*}

To further quantify the structure of the EGO-10 protoclusters, we compute the $\mathcal{Q}$-parameter \citep{Cartwright2004}, derived from the MST of the core spatial distribution. As outlined in Section~\ref{sec:intro}, $\mathcal{Q}<$ 0.8 is taken to indicate a subclustered distribution, with lower values indicating more subclustering. $\mathcal{Q}>$ 0.8 is taken to indicate a centrally-condensed distribution, with higher values indicating increased condensedness\footnote{Since the MST is constructed from the 2D projection of the core spatial distribution, the derived value of the $\mathcal{Q}$-parameter depends on the viewing angle to the protocluster \citep{Cartwright2009, Bastian2009}. Whilst we cannot assess the 3D structure of these protoclusters, we note that \citet{Cartwright2009} find projection effects are negligible when calculating the $\mathcal{Q}$-parameter of individual clusters with aspect ratio $<$3.}. The EGO-10 MSTs, computed with the publicly available \textit{MiSTree} Python package\footnote{\url{https://knaidoo29.github.io/mistreedoc/}} \citep{Naidoo2019}, are shown in Fig.~\ref{fig:mst} and the $\mathcal{Q}$-parameters of the EGO-10 protoclusters are presented in Table~\ref{tab:mst}. To estimate uncertainties on $\mathcal{Q}$, we consider alternative constructions of the core catalogue in each field, testing the effect of mass sensitivity, sampling uncertainty, and an alternative source extraction procedure. As detailed in Appendix~\ref{app:Q_err}, we calculate $\mathcal{Q}$ for each alternative catalogue, and estimate the uncertainty as the standard deviation of the values obtained for each field. The calculated uncertainties, reported in Table~\ref{tab:mst}, are generally small, with a mean relative uncertainty $\sim$5\%. We also explore the potential effect of protocluster distance (Appendix~\ref{sec:app_tests}) and find that protocluster distance does not systematically affect our measured $\mathcal{Q}$-parameters.

The $\mathcal{Q}$-parameter values for the EGO-10 sample span a range 0.53\,$<\mathcal{Q}<$\,0.93. We find subclustered distributions in the majority of regions (7/10), though note that the $^{+}_{-}1\sigma$ uncertainty ranges for G12.91$-$0.03 and G14.63$-$0.58 span  $\mathcal{Q}=$ 0.8, and median and mean values for the sample of $\mathcal{Q}=$ 0.74 and 0.75, respectively. The structural diversity implied by the distributions in Fig.~\ref{fig:disthist} is reflected in the MSTs in Fig.~\ref{fig:mst} and their associated $\mathcal{Q}$-parameters. For example, G10.29$-$0.13 is the most subclustered region with the lowest $\mathcal{Q}=$ 0.53, contrasting with G14.33$-$0.64, which is the most centrally-condensed region with the largest $\mathcal{Q}=$ 0.93. We note, however, that $\mathcal{Q}$ does not always intuitively reflect the structure apparent in the 1.3\,mm continuum images shown in Fig.~\ref{fig:mm}. $\mathcal{Q}$ is independent of $R_\text{clus}$ (Equation~\ref{eq:q}), and so depends only on the relative spatial distribution of sources within $R_\text{clus}$, which is more easily seen in Fig.~\ref{fig:mst}. We also note that the splitting of central \textit{Astrodendro} leaves into multiple \textit{Hyper} cores, as shown in Fig.~\ref{fig:dhcomp}, has only a minor impact on $\mathcal{Q}$ that is captured by the uncertainties presented in Table~\ref{tab:mst}.  For the examples shown in Fig.~\ref{fig:dhcomp}, considering the central source as a single object changes $\mathcal{Q}$ by 0.02, demonstrating that the specific choice of source extraction procedure does not significantly impact these measured $\mathcal{Q}$-parameters (see also Appendix~\ref{app:Q_err}.) 

The $\mathcal{Q}$ values of the EGO-10 sample are similar to those reported in \citet{Cartwright2004} and observed in other massive protoclusters \citep[e.g.][]{Hunter2014, Sanhueza2019, Sadaghiani2020, ONeill2021, Morrii2024, Xu2024, Kinman2025, Schisano2025, Zhang2025}. Notably, the range in $\mathcal{Q}$ spanned by the EGO-10 regions is also similar to that over which $\mathcal{Q}$ evolves in the simulations of \citet{Maschberger2010} and \citet{Laverde-Villarreal2025}. We investigate potential trends between the $\mathcal{Q}$-parameter and indicators of protocluster evolutionary state in Section~\ref{sec:evol}.

We note that the effectiveness of $\mathcal{Q}$ as a diagnostic of protocluster structure may be limited at small source counts, where the source distribution does not sufficiently trace the underlying (proto)cluster structure \cite[e.g.][]{Avison2023, Schisano2025}. The ALMAGAL analysis of \citet{Schisano2025} (their Appendix D) found the $\mathcal{Q}$-parameter distributions of subclustered and centrally-condensed synthetic clusters with \mbox{4 $\leq N_\text{cores} \leq$\,30} can overlap significantly in the range \mbox{0.6 $\lesssim\mathcal{Q}\lesssim$ 0.8}. These authors find increased overlap at smaller $N_\text{cores}$, or when the underlying structure is only weakly subclustered or centrally condensed; in these cases, there is significant overlap even at $N_\text{cores}\approx$ 30 (their Figure D.2). Most EGO-10 fields (8/10) have $N_\text{cores}>$ 40 (Table~\ref{tab:mst}), such that their source distributions and corresponding $\mathcal{Q}$-parameters are more likely to reflect the underlying structure of their host protoclusters. The exceptions are G16.59$-$0.05 and G22.04$+$0.22, with $N_\text{cores}=$\,21 and 13, respectively, both of which also have $\mathcal{Q}$-parameters within the overlap range identified by \citet{Schisano2025}. We therefore consider the $\mathcal{Q}$ values in these regions less reliable than those in the rest of the sample, and exclude them from our statistical analyses in Section~\ref{sec:Q1}. 
These fields have the poorest mass sensitivities in the sample (see Table~\ref{tab:img}), such that deeper observations may yield more sources, and enable a more reliable assessment of their structure.

\begin{table*}
\caption{Structural properties of the dense core spatial distributions in EGO-10.}
\label{tab:mst}
\begin{tabular}{lccccccccc}
\hline
\hline
EGO & $N_\text{cores}$$^\text{a}$ & \multicolumn{2}{c}{$R_\text{clus}$$^\text{b}$} & $\eta$$^\text{c}$ & $m$$^\text{d}$ & $s$$^\text{e}$ & $\bar{m}$$^\text{f}$ & $\bar{s}$$^\text{g}$ & $\mathcal{Q}$$^\text{h}$\\
\cline{3-4}
 & & ($''$) & (pc) & & ($''$) & ($''$) & & \\
\hline
G10.29$-$0.13 & 135 & 55.1 & 0.51 & 0.54 & 2.52 & 31.1 & 0.298 & 0.564 & 0.53 (0.02)\\
G10.34$-$0.14 & 43 & 45.8 & 0.35 & 0.51 & 4.64 & 27.2 & 0.366 & 0.594 & 0.62 (0.01)\\
G12.91$-$0.03 & 57 & 40.2 & 0.88 & 0.32 & 3.05 & 16.1 & 0.318 & 0.400 & 0.79 (0.03)\\
G14.33$-$0.64 & 45 & 35.4 & 0.19 & 0.31 & 3.81 & 15.2 & 0.398 & 0.429 & 0.93 (0.06)\\
G14.63$-$0.58 & 84 & 54.5 & 0.48 & 0.30 & 3.54 & 22.1 & 0.332 & 0.406 & 0.82 (0.06)\\
G16.59$-$0.05 & 21 & 10.6 & 0.18 & 0.62 & 2.01 & 7.3 & 0.465 & 0.682 & 0.68 (0.05)\\
G18.89$-$0.47 & 72 & 45.2 & 0.92 & 0.39 & 3.19 & 20.2 & 0.333 & 0.447 & 0.74 (0.02)\\
G19.36$-$0.03 & 47 & 40.8 & 0.44 & 0.47 & 4.15 & 21.5 & 0.385 & 0.526 & 0.73 (0.02)\\
G22.04$+$0.22 & 13 & 9.2 & 0.15 & 0.56 & 2.17 & 5.6 & 0.445 & 0.607 & 0.73 (0.04)\\
G28.83$-$0.25 & 53 & 30.1 & 0.70 & 0.30 & 2.76 & 12.4 & 0.370 & 0.412 & 0.90 (0.04)\\
\hline
\end{tabular}
\begin{flushleft}
        \small{
        $^\text{a}$ Number of 1.3\,mm cores catalogued in the field using the procedure described in Section~\ref{sec:cat}.\\
        $^\text{b}$ Cluster radius, calculated as the separation from the mean position of all cores to the core furthest from this mean position (Section~\ref{sec:hists}).\\
        $^\text{c}$ Fraction of the distribution of projected separation between core-pairs which is $>$0.5\,$R_\text{clus}$ (see Section~\ref{sec:hists}).\\
        $^\text{d}$ Mean edge length of the MST of the core spatial distribution, computed from the edge lengths returned by MiSTree \citep{Naidoo2019}.\\
        $^\text{e}$ Mean inter-core separation \citep[Section~\ref{sec:intro} and][]{Cartwright2004}.\\
        $^\text{f}$ Normalised mean edge length \cite[Section~\ref{sec:intro} and][]{Cartwright2004}.\\
        $^\text{g}$ Normalised correlation length \cite[Section~\ref{sec:intro} and][]{Cartwright2004}.\\
        $^\text{h}$ Calculated using Equation~\ref{eq:q}. The uncertainty, estimated as the standard deviation of the $\mathcal{Q}$ values listed for each region here and in Table~\ref{tab:Q_err}, is given in parentheses (see Appendix~\ref{app:Q_err}).
    }
    \end{flushleft}
\end{table*}

\section{Association of 1.3\,mm Cores with Centimetre-Wavelength Tracers of Active Star Formation}
\label{sec:cm_assoc}
In this section, we quantify the association of the catalogued ALMA 1.3\,mm cores with known tracers of active star formation and identify cores which are unambiguously protostellar, by leveraging the wealth of high-resolution multiwavelength data available for the EGO-10 sample (see Section~\ref{sec:ego}).
We use three tracers of active star formation: 6.7\,GHz CH$_3$OH masers, 22\,GHz H$_2$O masers and cm-$\lambda$ continuum emission. For the EGO-9 regions (all except G16.59$-$0.05), positions of masers and cm-$\lambda$ continuum sources were taken from \citet{Towner2021} (mean resolution $\sim$0\farcs24 at 1.3\,cm and $\sim$0\farcs35 at 5\,cm). For G16.59$-$0.05, we use the Australia Telescope Compact Array (ATCA) 6.7\,GHz CH$_3$OH maser position from the Methanol Multibeam (MMB) survey \citep[][16.585$-$0.051 in their Table 1]{Green2010} and VLA 22\,GHz H$_2$O maser positions from \citet{Beuther2002} (IRAS 18182$-$1433 in their Table 2). 
The positions of cm-$\lambda$ continuum sources in G16.59$-$0.05 were taken from the \citet{Rosero2016} VLA survey of massive star forming regions (beam $\approx$0\farcs4 at 1.3\,cm and 5\,cm; IRAS 18182$-$1433 in their Table 4). 

In our analysis we consider all star formation tracers within the 10\% response level of the ALMA mosaic of each field (see Fig.~\ref{fig:mm}). For these tracers, we compute the projected separation to the \textit{Hyper} centroid of the nearest 1.3\,mm core ($\Delta_\text{min}$). 
To ensure only robust tracer-core associations are included in our analysis and core catalogue (Table~\ref{tab:cat}), we consider a given maser or cm-$\lambda$ continuum source to be `associated' with a given 1.3\,mm core if the star formation tracer lies within the core's \textit{Hyper} FWHM ellipse (we explore the effect of using alternative, less-restrictive association criteria in Appendix~\ref{app:na_tests}). We classify these cores as protostellar, and label them as such in  Table~\ref{tab:cat}. For tracers associated with 1.3\,mm cores, $\Delta_\text{min}$ is thus the offset between the tracer position and the centroid of its 1.3\,mm host-core, reported as $\Delta_\text{off}$. For protostellar cores, the name of each associated star formation tracer and its $\Delta_\text{off}$ value are given in Table~\ref{tab:cat}. In aggregate, 39/570 ($\sim$7\%) of the ALMA 1.3\,mm cores are associated with either maser or cm-$\lambda$ continuum emission and are thus classified as protostellar. Zoom-views of representative examples are shown in Fig.~\ref{fig:assoc}, with equivalent figures for all protostellar cores available as online Supplementary Material.

We use the naming convention of \citet{Towner2021} to refer to all star formation tracers in the EGO-9 sample. Cm-$\lambda$ continuum sources in a given field are named in decreasing order of integrated flux density at 1.3\,cm. After sources detected at 1.3\,cm are named, sources detected only at 5\,cm are named according to their 5\,cm integrated flux density. Masers are named after their associated cm-$\lambda$ continuum source \cite[][association criterion of angular separation $<$1\arcsec]{Towner2021}, or labelled `NC' if not associated with a cm-$\lambda$ source, followed by a letter indicating the maser species (`M' for 6.7\,GHz CH$_3$OH, `W' for 22\,GHz H$_2$O) and a number for differentiation.
In G16.59$-$0.05, we refer to cm-$\lambda$ sources following \citet{Rosero2016}, in which sources are named alphabetically from east to west, except for their source G16.584$-$0.053 (see Section~\ref{sec:cm}). We approximate the \citet{Towner2021} maser naming convention by naming masers in this field based on their associated cm-$\lambda$ source and the maser species, following the \citet{Towner2021} association criterion. The \citet{Green2010} 6.7\,GHz CH$_3$OH maser, for example, is thus referred to as `C-M1' since it is associated with the \citet{Rosero2016} cm-$\lambda$ source C. We refer to the \citet{Beuther2002} 22\,GHz H$_2$O masers as W1--W5 according to the order in which they appear in their Table 2. 

\subsection{Association Statistics}
\subsubsection{6.7\,GHz CH$_3$OH Masers}
\label{sec:cm_67}
Combining the 11 6.7\,GHz CH$_3$OH masers reported in \citet{Towner2021} with the single \citet{Green2010} detection in G16.59$-$0.05 gives a total of 12 detections across the EGO-10 sample. All masers lie within the 10\% response level of the ALMA mosaics and were therefore included in our analysis. All masers are associated with strong 1.3\,mm continuum emission ($\gtrsim$50--500\,$\sigma_\text{rms}$) and 11/12 (92\%) are associated with 1.3\,mm cores (the sole exception being G18.89$-$0.47 NC-M3). Inversely, of the 570 cores detected across the sample, only $\sim$2\% are associated with 6.7\,GHz masers. 

\begin{figure*}
   \includegraphics[width=\textwidth]{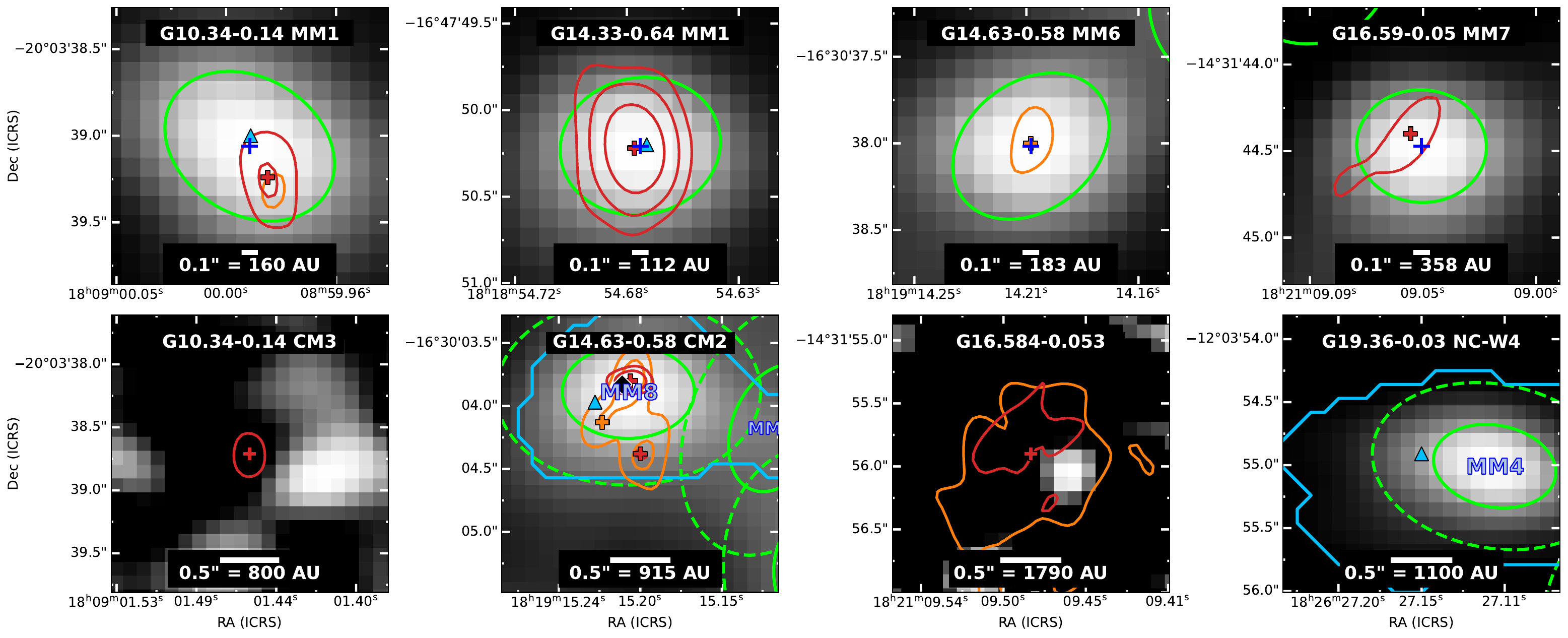}
   
\caption{ALMA 1.3\,mm continuum images (greyscale; corrected for the primary beam response) overlaid with contours of VLA 1.3\,cm (red) and 5\,cm (orange) continuum emission at [4, 8, 16]$\times\sigma_\text{cm}$ (noise from \citealt{Towner2021} or \citealt{Rosero2016}).
Cm-$\lambda$ continuum source positions are marked with thick crosses (\textbf{+}) and
6.7\,GHz CH$_3$OH and 22\,GHz H$_2$O masers with black $\blacklozenge$ and light blue $\blacktriangle$, respectively. 
Core FWHM ellipses are shown in green, with core centroids marked with dark blue crosses (+) in the top row and the core ID (Table~\ref{tab:cat}) in the bottom row. 
\textit{Upper row}: 1\farcs6-FOV zoom-views centred on representative examples of protostellar cores; each panel is labelled with the core ID, and the linear greyscale extends from 5\% to 95\% of the intensity at the core centroid. 
\textit{Lower row}: 2\farcs2-FOV zoom-views centred on representative examples of star formation tracers that are not associated with catalogued 1.3\,mm cores. Each panel is labelled with the tracer name, and the linear greyscale extends from 5\% to 95\% of the peak intensity in the field-of-view. The outer dashed green ellipses are the 2 $\times$ FWHM \textit{Hyper} source photometric apertures and \textit{Astrodendro} leaves identified in the 40\,k$\lambda$ images are outlined in blue (see Sections~\ref{sec:dendro}--\ref{sec:hyper} and Appendix~\ref{app:na_tests}). 
Equivalent figures for all protostellar cores and non-associations are available as online Supplementary Material.}
\label{fig:assoc}
\end{figure*}

\subsubsection{22\,GHz H$_2$O Masers}
\label{sec:cm_22}
Four of the \citet{Towner2021} EGO-9 regions (G10.29$-$0.13, G12.91$-$0.03, G19.36$-$0.03 and G22.04$+$0.22) were observed in two separate epochs (2018.1 and 2019.6). We considered all H$_2$O maser detections across both epochs for these fields, giving a total of 39 detections in the EGO-9 sample. Including the five \citet{Beuther2002} masers detected in G16.59$-$0.05 then gives a total of 44 detections. Where a given maser is detected in both \citet{Towner2021} epochs, we defined its position as that reported in the first (2018.1). One maser - G10.34$-$0.14 NC-W5 - lies outside our ALMA mosaic and thus was not considered in our analysis. 

Of the 43 remaining masers, 28/43 (65\%) are associated with 1.3\,mm cores. Inversely, $\sim$5\% of cores are 22\,GHz H$_2$O maser hosts. It is important to note, however, that given the highly variable and even transient nature of 22\,GHz H$_2$O masers in star-forming regions \citep{Hunter1994, Trinidad2003, Felli2007, Towner2021}, this association rate is likely a lower limit.

\subsubsection{Centimetre Continuum Sources} 
\label{sec:cm}
\citet{Towner2021} report a total of 36 centimetre continuum sources in EGO-9 with well-constrained positions. \citet{Rosero2016} report an additional eight sources in G16.59$-$0.05, but excluded the source G16.584$-$0.053 from their analysis \cite[and from their follow-up characterisation in][]{Rosero2019} as it lies outside the FWHM of the region's \citet{Beuther2002} 1.2\,mm clump. 
As \citet{Towner2021} did not implement a mm-clump-based criterion in constructing their catalogue, we include G16.584$-$0.053 in our analysis for consistency. Combining the two catalogues gives a total of 44 cm-$\lambda$ detections\footnote{The \citet{Towner2021} and \citet{Rosero2016} VLA observations have largest angular scales $\sim$8--9$''$. \citet{Towner2021} report an additional five cm-$\lambda$ continuum sources, all of which are associated with \ion{H}{ii} regions known in the literature (see their Table~5). These extended sources are resolved-out in their observations, such that they are only able to report approximate sizes, flux measurements and positions. We therefore do not consider these sources in our analysis.}. One source - G12.91$-$0.03 CM3 - lies outside our ALMA mosaic, and thus was not considered in our analysis.

\citet{Towner2021} define cm-$\lambda$ source positions at 1.3\,cm, unless a source is only detected at 5\,cm, in which case the position is from the 5\,cm data. 
For sources with a Gaussian morphology,  
\citet{Towner2021} report centroid positions from two-dimensional Gaussian fits, while
for non-Gaussian sources the reported position is that of the peak intensity. \citet{Rosero2016} report the peak intensity position in each of the individual basebands comprising each 1.3\,cm and 5\,cm band. For consistency with \citet{Towner2021}, we define the positions of \citet{Rosero2016} sources at 1.3\,cm. The position of a given \citet{Rosero2016} source was therefore taken as the mean of its reported 20.9\,GHz and 25.5\,GHz positions. If a source was detected in only one 1.3\,cm baseband, we adopt that position; for sources detected in neither 1.3\,cm baseband, the position was taken as the mean of the reported 4.9\,GHz and 7.4\,GHz positions.

Of the 43 cm-$\lambda$ continuum sources, 21/43 (49\%) are associated with 1.3\,mm cores, such that $\sim$4\% of cores are cm-$\lambda$ source hosts. Note that whilst 21 cm-$\lambda$ sources are associated with 1.3\,mm cores, only 20 1.3\,mm cores host cm-$\lambda$ sources, as the 6.7\,GHz CH$_3$OH maser host-core G14.63$-$0.58 MM8\footnote{As G14.63$-$0.58 CM1 lies closer to the core centroid of G14.63$-$0.58 MM8 than G14.63$-$0.58 CM3, the cm-$\lambda$ source name and $\Delta_\text{off}$ reported for this core in Table~\ref{tab:cat} are those of G14.63$-$0.58 CM1. } hosts both G14.63$-$0.58 CM1 and G14.63$-$0.58 CM3 (see Fig.~\ref{fig:MYSO}). This is the only core in the sample associated with more than one tracer of the same type.

\subsubsection{Association Rates and Projected Separations}
\label{sec:seps}
\begin{figure} 
   \includegraphics[width=\columnwidth]{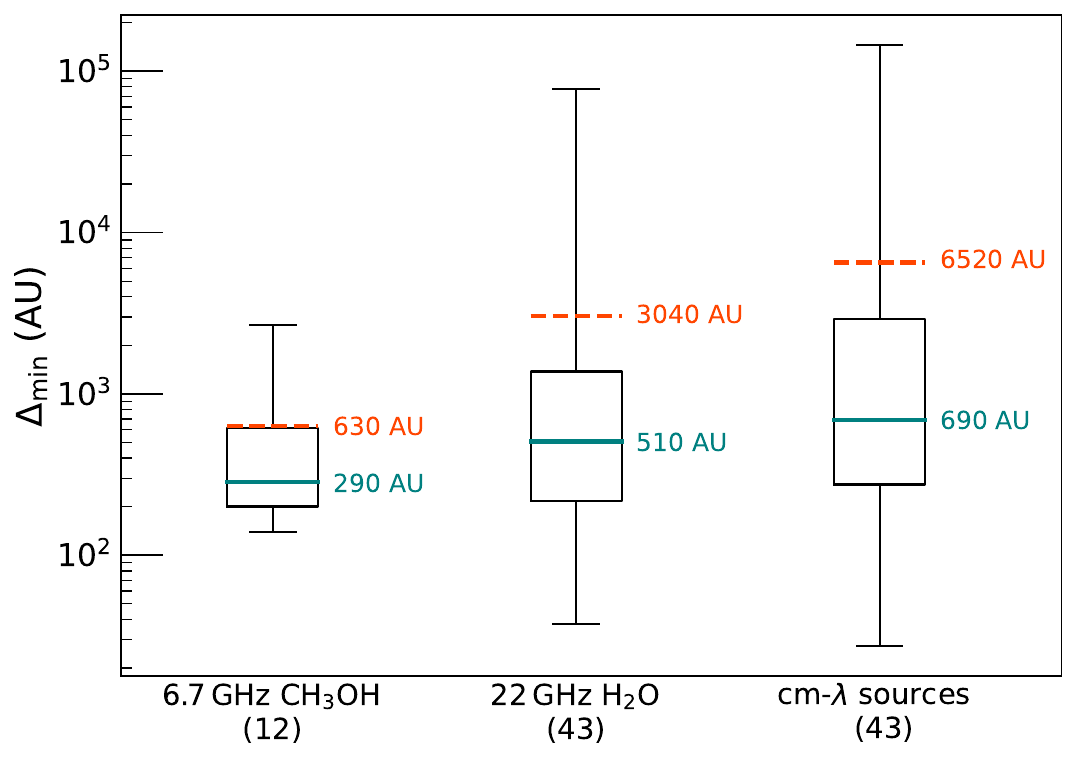}
   \caption{Box plots illustrating the projected physical separation distributions between star formation tracers and their nearest 1.3\,mm core ($\Delta_\text{min}$). The vertical extent of each box denotes the interquartile range, whilst the capped lines show the full extent of each distribution. The solid teal and dashed orange lines denote the median and mean of each distribution, respectively, with their values in AU labelled in the corresponding colour (see also Table~\ref{tab:assocs}). The $x$-axis labels indicate the star formation tracer shown in each box plot and the number of tracers in the corresponding distribution.}
\label{fig:box}
\end{figure}

The association rates and average projected angular and physical $\Delta_\text{min}$ and $\Delta_\text{off}$ for each star formation tracer considered are summarised in Table~\ref{tab:assocs}, along with the association rates obtained with the alternative association criteria described in Appendix~\ref{app:na_tests}; box plots of the $\Delta_\text{min}$ distributions are shown in Fig.~\ref{fig:box}. The 6.7\,GHz CH$_3$OH masers are clearly the star formation tracer most closely associated with 1.3\,mm cores, with the highest association rate and smallest angular and physical $\Delta_\text{min}$. The 6.7\,GHz masers are followed by the 22\,GHz H$_2$O masers, and then the cm-$\lambda$ continuum sources. Cm-$\lambda$ continuum sources are the only tracer with an association rate $<$50\%, and have the largest median and mean $\Delta_\text{min}$. 

The higher association rate and smaller $\Delta_\text{min}$ are expected for the 6.7\,GHz CH$_3$OH masers, which are exclusively pumped in the immediate environment of massive protostars \cite[][and references therein]{Minier2003}. In comparison, both 22\,GHz H$_2$O masers and (weak) cm-$\lambda$ continuum emission can be excited in protostellar outflows \citep[e.g.][cm continuum sensitivities $\lesssim$20$\mu$Jy beam$^{-1}$]{Purser2016, Rosero2016, Sanna2018, Purser2021, Towner2021}, consistent with the lower association rates and wider $\Delta_\text{min}$ distributions we find for these tracers. In particular, 22\,GHz H$_2$O masers with $\Delta_\text{min}\gtrsim$ 1000--10,000\,AU may be pumped in shocks where outflows impact ambient material at relatively high velocity \citep{Moscadelli2020}, though notably \citet{Budaiev2025} find even outflow-associated 22\,GHz masers are confined to $<$2000\,AU of mm continuum sources. Cm-$\lambda$ continuum sources with large $\Delta_\text{min}$ may indicate discrete radio lobes or jet knots, with G16.59$-$0.05 providing a clear example of this in EGO-10. \citet{Rosero2019} identify their sources A, B and C as discrete knots in the well-studied radio jet in this region \cite[IRAS 18182$-$1433 in their work, see also e.g.][and references therein]{Moscadelli2013, Moscadelli2019a}, which is driven by the MYSO associated with our 6.7\,GHz host-core G16.59$-$0.05 MM2. Of these three cm-$\lambda$ continuum sources, only source C is associated with MM2, whilst sources A and B are offset from this core by $\gtrsim$several thousand AU.

\subsubsection{Non-Associations}
\label{sec:nonassoc}
A total of 38/98 (39\%) of star formation tracers are not associated with a 1.3\,mm core, that is, they do not lie within a core FWHM ellipse, as defined by \textit{Hyper}. Zoom-views of representative examples of these non-associated star formation tracers are shown in the lower row of Fig.~\ref{fig:assoc}, with core \textit{Hyper} FWHMs, tracer positions and cm-$\lambda$ continuum contours overlaid. Equivalent figures for all non-associations are available as online Supplementary Material. Some non-associated tracers are genuinely isolated, with no significant nearby 1.3\,mm continuum emission. As already outlined, relatively large projected separations between 22\,GHz H$_2$O masers or cm-$\lambda$ sources and the 1.3\,mm host-cores of their driving (M)YSOs are, in part, expected. Such isolated tracers could also be associated with weak cores below our sensitivity limit. Other non-associated tracers lie in regions of blended 1.3\,mm continuum emission, perhaps associated with as yet unresolved sources or with cores which do not meet the flux or size thresholds required for our source identification. Some unassociated tracers also lie very close to their nearest core, but outside the core's \textit{Hyper} FWHM ellipse (for example, G14.63$-$0.58 CM2 and G19.36$-$0.03 NC-W4 in Fig.~\ref{fig:assoc}). These near-association cases motivated two sets of tests. In \textit{Hyper}, the maximum allowed source FWHM size is a user-defined parameter (Section~\ref{sec:hyper}). We tested whether rerunning \textit{Hyper} with a maximum source size of 2$\times$beam$_{\rm FWHM}$ (compared to 1.5$\times$beam$_{\rm FWHM}$ in Section~\ref{sec:cat}) increased the number of tracer-core associations, but the association statistics were unchanged. We then tested alternative association criteria (Appendix~\ref{app:na_tests}), using the original \textit{Hyper} run described in Section~\ref{sec:cat}. We find that while, as expected, using less stringent association criteria increases the tracer-core association rates (Table~\ref{tab:assocs}), their relative order as discussed in Section~\ref{sec:seps} remains unchanged.

\begin{table*}
\caption{Summary of star formation tracer-core associations.}
\begin{center}
\begin{tabular}{lccccccccccc}
\hline
Tracer &  Total \# of tracer$^\text{a}$ & \# associated with cores$^\text{b}$ & \multicolumn{4}{c}{$\Delta_\text{min}$$^\text{c, e}$} & & \multicolumn{4}{c}{$\Delta_\text{off}$$^\text{d, e}$}\\
\cline{4-7}
\cline{9-12}
&  & & \multicolumn{2}{c}{Median} & \multicolumn{2}{c}{Mean} & & \multicolumn{2}{c}{Median} & \multicolumn{2}{c}{Mean} \\
&  & & ($''$) & (AU) & ($''$) & (AU) & & ($''$) & (AU) & ($''$) & (AU) \\
\hline
6.7\,GHz CH$_3$OH masers & 12 & 11 (92\%, 2\%) & 0.13 & 290 & 0.18 & 630 & & 0.12 & 260 & 0.14 & 440\\
22\,GHz H$_2$O masers & 43 & 28 (65\%, 5\%) & 0.19 & 510 & 1.21 & 3040 & & 0.11 & 270 & 0.14 & 410\\
cm-$\lambda$ sources & 43 & 21 (49\%, 4\%) & 0.42 & 690 & 2.74 & 6520 & & 0.10 & 260 & 0.11 & 240\\
\hline
\hline
(1) \textit{Hyper} 2 $\times$ FWHM associations$^\text{f}$\\
\hline
6.7\,GHz CH$_3$OH masers & 12 & 12 (100\%, 2\%) & -- -- & -- -- & -- -- & -- -- & & -- -- & -- -- & -- -- & -- --\\
22\,GHz H$_2$O masers & 43 & 37 (86\%, 7\%) & -- -- & -- -- & -- -- & -- -- & & -- -- & -- -- & -- -- & -- --\\
cm-$\lambda$ sources & 43 & 29 (67\%, 5\%) & -- -- & -- -- & -- -- & -- -- & & -- -- & -- -- & -- -- & -- --\\
\hline 
\hline
(2) \textit{Astrodendro} associations$^\text{g}$\\
\hline
6.7\,GHz CH$_3$OH masers & 12 & 12 (100\%) & -- -- & -- -- & -- -- & -- -- & & -- -- & -- -- & -- -- & -- --\\
22\,GHz H$_2$O masers & 43 & 35 (81\%) & -- -- & -- -- & -- -- & -- -- & & -- -- & -- -- & -- -- & -- --\\
cm-$\lambda$ sources & 43 & 24 (56\%) & -- -- & -- -- & -- -- & -- -- & & -- -- & -- -- & -- -- & -- --\\
\hline
\end{tabular}
\end{center}
\begin{flushleft}
        \small{$^\text{a}$ Total number of each tracer in the sample considered in our analysis (see Sections~\ref{sec:cm_67}, \ref{sec:cm_22} and \ref{sec:cm}).\\
        $^\text{b}$ Number of each tracer associated with 1.3\,mm cores.  The percentage of the tracer population associated with 1.3\,mm cores (the `tracer-core association rate') is given in parentheses, followed by the percentage of catalogued cores associated with that tracer (the `core-tracer association rate'). \\
        $^\text{c}$ The median and mean of the projected separation between tracers and their nearest 1.3\,mm core ($\Delta_\text{min}$, see Section~\ref{sec:cm_assoc}), in angular and physical units.\\
        $^\text{d}$ For those tracers associated with 1.3\,mm cores, the median and mean of the projected offset ($\Delta_\text{off}$, see Section~\ref{sec:cm_assoc}) between tracers and their host 1.3\,mm core, in angular and physical units.\\
        $^\text{e}$ Angular values do not map directly to the values in AU as the regions of the EGO-10 sample span a range in distance.\\
        $^\text{f}$ Association rates with the association criterion (1) tested in Appendix~\ref{app:na_tests}.\\
        $^\text{g}$ Association rates with the association criterion (2) tested in Appendix~\ref{app:na_tests}. The percentage of cores associated with each tracer is not given because there is not a one-to-one correspondence between \textit{Astrodendro} leaves and our catalogued cores (see Sections~\ref{sec:dendro} and ~\ref{sec:cat} and Figure~\ref{fig:dhcomp}).
      
        }
    \end{flushleft}
\label{tab:assocs}
\end{table*}

\subsection{The EGO-10 MYSOs}
\label{sec:myso}
\begin{figure*}
   \includegraphics[width=\textwidth]{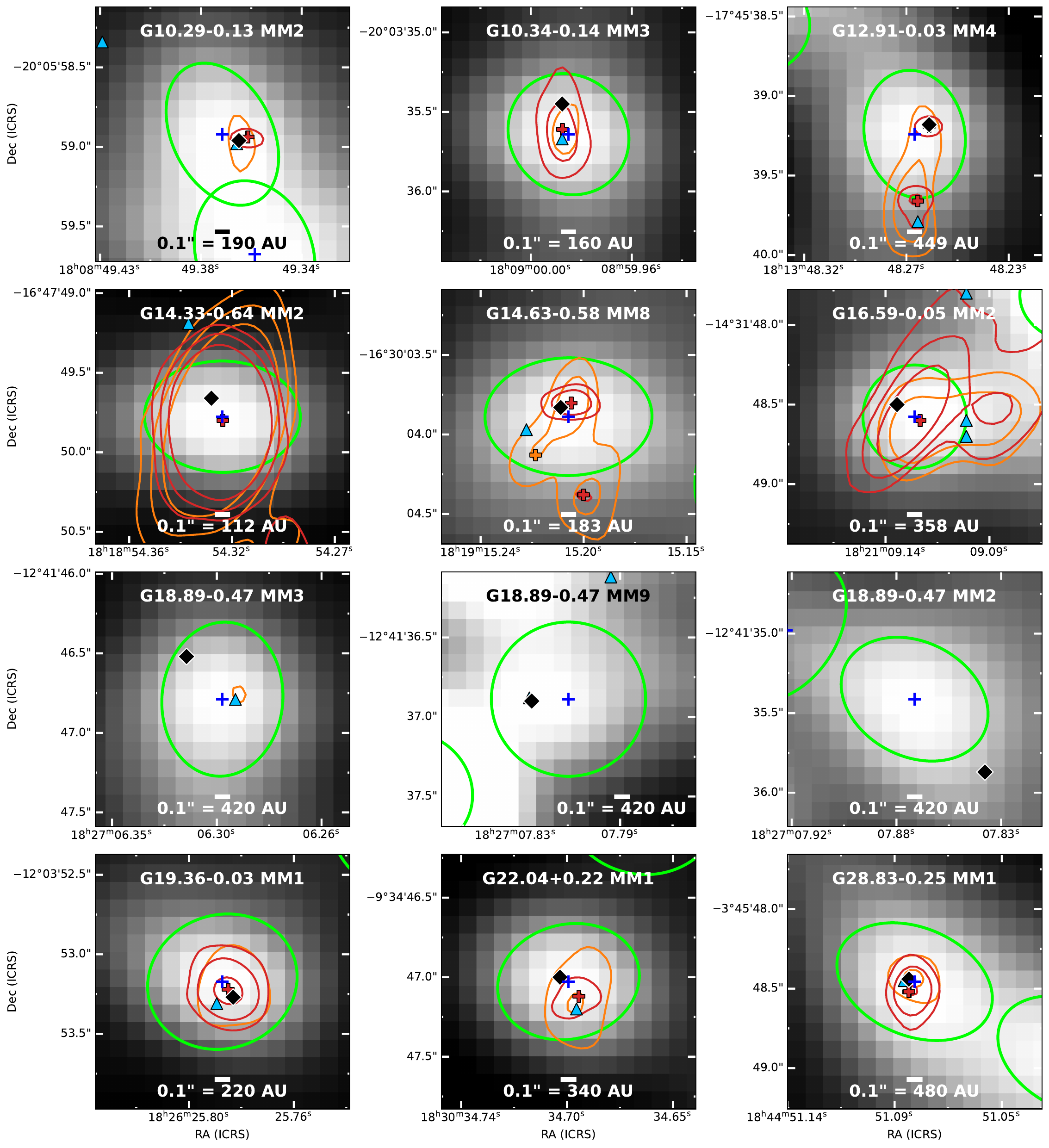}

\caption{1\farcs6-FOV zoom-views centred on the EGO-10 MYSO host-cores - 1.3\,mm cores associated with 6.7\,GHz CH$_3$OH masers (the nearest core to G18.89$-$0.47 NC-M3, G18.89$-$0.47 MM2, is also shown for completeness). Each panel shows the 1.3\,mm ALMA image (12\,m-only, corrected for the primary beam response) in greyscale (linear scale; minimum and maximum corresponding to 5\% to 95\% of the image intensity at the core centroid, respectively) and is labelled with the ID of the centred core, as given in Table~\ref{tab:cat}. Markers and contours are as in the upper panels of Fig.~\ref{fig:assoc}.} 
\label{fig:MYSO}
\end{figure*}

Given 6.7\,GHz CH$_3$OH masers are exclusively pumped in the immediate environment of massive protostars \citep{Minier2003, Breen2013}, their associated 1.3\,mm cores can be unambiguously identified as the host-cores of MYSOs in EGO-10. Zoom-views of these cores, which are listed in Table~\ref{tab:clus67}, are shown in Fig.~\ref{fig:MYSO}, with maser and cm-$\lambda$ continuum source positions and contours overlaid\footnote{The closest 1.3\,mm core to G18.89$-$0.47 NC-M3 - G18.89$-$0.47 MM2 - is also shown for completeness, though we do not consider this maser further in our analysis (see Section~\ref{sec:cm_67})}. Of these MYSO host-cores, 8/11 (73\%) are associated with 22\,GHz H$_2$O masers, 9/11 (82\%) with cm-$\lambda$ continuum sources and 6/11 (55\%) are associated with both 22\,GHz H$_2$O masers and cm-$\lambda$ continuum sources (see Fig.~\ref{fig:MYSO} and Table~\ref{tab:cat}).  

\begin{table*}
\caption{Association between 6.7\,GHz CH$_3$OH masers and 1.3\,mm cores for the 11 masers associated with 1.3\,mm cores.}
\begin{center}
\begin{tabular}{lccccccccccc}
\hline
\hline
\multicolumn{5}{c}{Cluster \& image properties} & & \multicolumn{6}{c}{6.7\,GHz CH$_3$OH maser properties}\\
\cline{1-5}
\cline{7-12}
EGO$^\text{a}$ & Distance$^\text{b}$ & Beam$^\text{c}$ & $N_\text{cores}$$^\text{d}$ & $R_\text{clus}$$^\text{e}$ & & Maser Name$^\text{f}$ & 1.3\,mm Core Name$^\text{g}$ &  $\Delta_\text{off}$$^\text{h}$ & $\Delta_\text{cen}$$^\text{i}$ & $\Delta_\text{nn}$$^\text{j}$ & $N_\text{env}$$^\text{k}$\\
  & (kpc)  & (AU $\times$ AU) & & (pc) & & & & (AU) & (pc) & (AU) & \\
\hline
G14.33$-$0.64 & 1.13 & 930$\times$660 & 45 & 0.19 & & CM1-M1 & MM2 & 150 & 0.035 & 2240 & 28 \\
G10.34$-$0.14 & 1.6 & 1320$\times$930 & 43 & 0.35 & & CM1-M1 & MM3 & 310 & 0.070 & 2630 & 14 \\
G14.63$-$0.58 & 1.83 & 1530$\times$1070 & 84 & 0.48 & & CM1-M1 & MM8 & 140 & 0.030 & 2330 & 14 \\
G10.29$-$0.13 & 1.9 & 1360$\times$1050 & 135 & 0.51 & & CM1-M1 & MM2 & 210 & 0.095 & 1370 & 18 \\
G19.36$-$0.03 & 2.2 & 1660$\times$1240 & 47 & 0.44 & & CM3-M1 & MM1 & 260 & 0.040 & 3380 & 14 \\
G22.04$+$0.22 & 3.4 & 2590$\times$1940 & 13 & 0.15 & & CM1-M1 & MM1 & 200 & 0.005 & 4090 & 3 \\
G16.59$-$0.05 & 3.58 & 2690$\times$2010 & 21 & 0.18 & & C-M1 & MM2 & 480 & 0.020 & 5190 & 7 \\
\hline
G18.89$-$0.47 & 4.2 & 3210$\times$2270 & 72 & 0.92 & & NC-M1 & MM3 & 1470 & 0.505 & 7280 & 2 \\
 &  &  &  &  & & NC-M2 & MM9 & 970 & 0.040 & 3950 & 4 \\
 \hline
G12.91$-$0.03 & 4.5 & 3220$\times$2490 & 57 & 0.88 & & CM2-M1 & MM4 & 490 & 0.140 & 6040 & 3 \\
G28.83$-$0.25 & 4.8 & 3930$\times$2710 & 53 & 0.70 & & CM1-M1 & MM1 & 190 & 0.095 & 5510 & 1 \\
\hline
\end{tabular}
\end{center}
\begin{flushleft}
        \small{
        $^\text{a}$ EGO name, ordered by increasing distance.\\
        $^\text{b}$ EGO distance, as in Table~\ref{tab:tbl1}.\\
        $^\text{c}$ Physical scale of the synthesised beam of the 12\,m-only image (see Table~\ref{tab:img} for values in arcseconds).\\
        $^\text{d}$ Total number of cores catalogued in the field (see Section~\ref{sec:cat}).\\
        $^\text{e}$ Projected physical cluster radius (see Section~\ref{sec:hists}).\\
        $^\text{f}$ 6.7\,GHz CH$_3$OH maser name, following the naming convention described in Section~\ref{sec:cm_assoc}. Note that the full name includes the EGO e.g. G10.29$-$0.13 CM1-M1.\\
        $^\text{g}$ ID of the 6.7\,GHz CH$_3$OH maser's associated 1.3\,mm core, as in Table~\ref{tab:cat}. Note that the full name includes the EGO e.g. G10.29$-$0.13 MM1.\\
        $^\text{h}$ Projected physical separation between the maser and the centroid of its 1.3\,mm host-core (see Section~\ref{sec:cm_67}).\\             
        $^\text{i}$ Projected physical separation between the maser and the cluster centre (see Sections~\ref{sec:hists} and ~\ref{sec:myso_clust}), rounded to the nearest 0.005\,pc.\\
        $^\text{j}$ Projected physical separation between the maser and the centroid of the nearest-neighbouring core (see Section~\ref{sec:myso_clust}).\\
        $^\text{k}$ The number of 1.3\,mm core centroids within a projected radius of 10,000\,AU from the maser position (not including the maser's host-core, see Section~\ref{sec:myso_clust}).
       
       }
    \end{flushleft}
\label{tab:clus67}
\end{table*}

\begin{figure*}
   \includegraphics[width=0.47\textwidth]{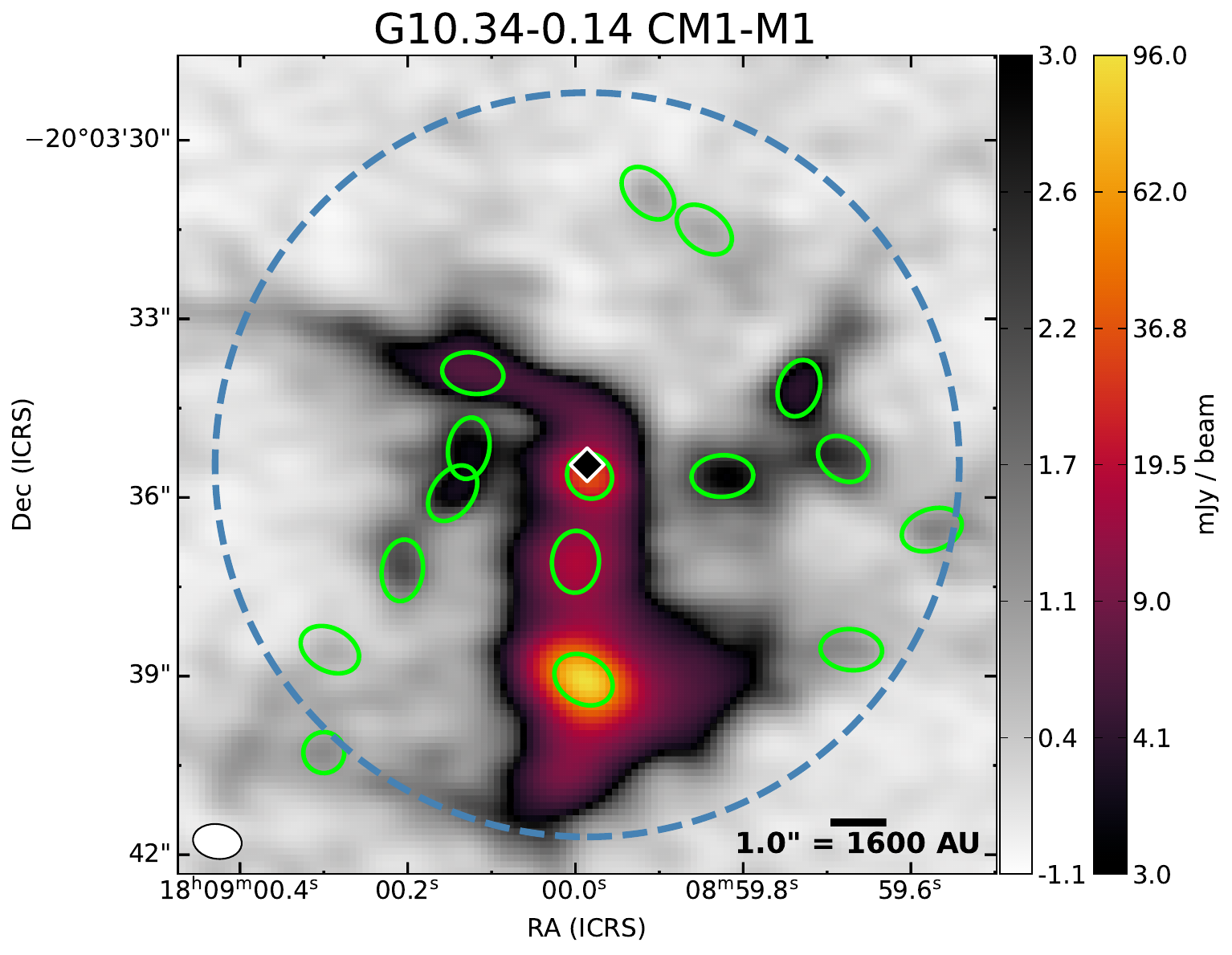}
   \includegraphics[width=0.47\textwidth]{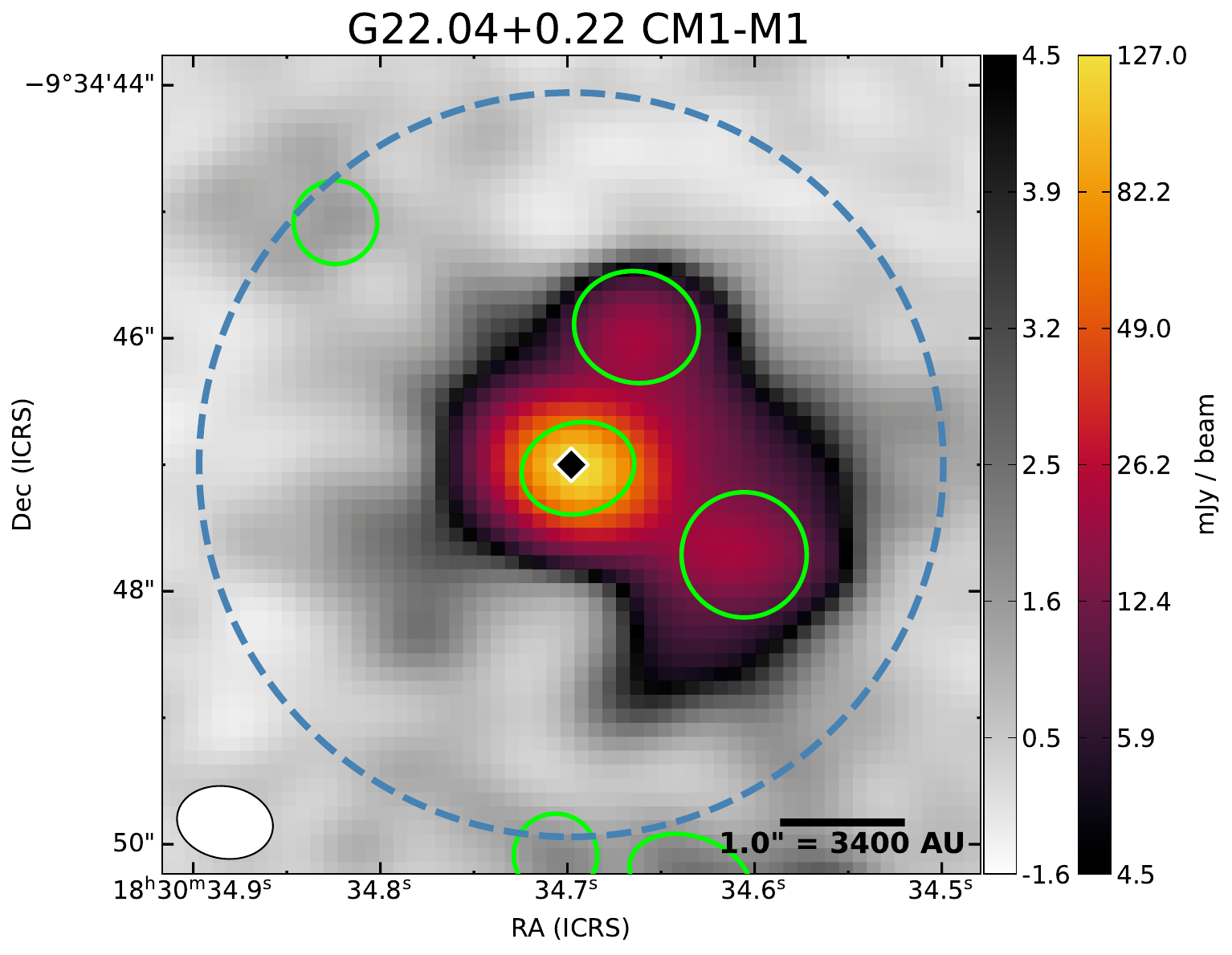}
\caption{Representative examples of the ALMA 1.3\,mm continuum emission (images and colourscale as in Fig.~\ref{fig:mm}) and catalogued cores within a projected radius of 10,000\,AU (dashed blue circle) of MYSOs traced by 6.7\,GHz CH$_3$OH masers.  \textit{Left}: G10.34-0.14 CM1-M1 ($d=$ 1.6\,kpc, $N_\text{env}=$ 14).  \textit{Right}: G22.04+0.22 CM1-M1 ($d=$ 3.4\,kpc, $N_\text{env}=$ 3; $N_\text{env}$ is the number of 1.3\,mm cores within a projected physical separation of 10,000\,AU of the 6.7\,GHz maser position). 6.7\,GHz CH$_3$OH maser positions ($\blacklozenge$) and FWHM ellipses of catalogued 1.3\,mm cores (green) are overlaid in each panel, and the synthesized beam is shown at lower left. Panel titles give the name of the centred maser, following the naming convention in Section~\ref{sec:cm_assoc}. Equivalent figures for the other MYSOs discussed in Section~\ref{sec:myso_clust} are available as online Supplementary Material.}
\label{fig:AUzoom}
\end{figure*}

\subsubsection{Spatial Distribution}
\label{sec:myso_spat}
Taking the 6.7\,GHz CH$_3$OH maser positions to mark the locations of MYSOs, we can investigate the spatial distribution of MYSOs within their respective protoclusters. We compute the projected physical separation between each MYSO and its respective protocluster centre (defined as the mean position of all cores in a given field, see Section~\ref{sec:hists}), given in Table~\ref{tab:clus67} as $\Delta_\text{cen}$.
We find MYSOs are generally located near protocluster centres, with $\Delta_\text{cen}<$ 0.2\,pc for 10/11 (91\%) of the 6.7\,GHz masers considered\footnote{The exception is G18.89$-$0.47 NC-M1, associated with the 
1.3\,mm 
core G18.89$-$0.47 MM3, which lies $\sim$0.5\,pc from the protocluster centre.} and a median and mean $\Delta_\text{cen}$ of 0.045\,pc and 0.095\,pc, respectively. $\Delta_\text{cen}$ is generally a small fraction ($<$20\%) of the projected physical $R_\text{clus}$ value (see Table~\ref{tab:clus67}), which has a median and mean of 0.46\,pc and 0.48\,pc, respectively. 
Indeed a directional Mann-Whitney U-test\footnote{With \texttt{scipy.stats}' \texttt{mannwhitneyu}. This is a non-parametric statistical test of the null hypothesis that two samples have been drawn from the same underlying distribution.} reveals the MYSO host-cores, as a population, lie closer to cluster centres than the population of cores which do not host MYSOs, to a statistically significant degree (\mbox{$p$-value $\approx$ 0.002}). The preferential location of MYSOs at protocluster centres is consistent with simulations demonstrating hierarchical massive star and protocluster formation \cite[e.g.][]{Bonnell2004, Smith2009, Wang2010}. Whilst we do not, in this work, calculate core masses and thus cannot further quantify any mass segregation that may be present, this result is at least 
consistent with primordial mass segregation \cite[e.g.][and references therein]{Xu2024}. Interestingly, in the \citet{Morii2023} analysis of the full ASHES sample, the population of their most massive cores - comprised of the most massive core in each clump, with masses determined from 1.3\,mm continuum flux densities - was not preferentially located nearer clump centres compared to the rest of the core population. This held whether the clump centre in a given field was taken as the position of the ATLASGAL peak intensity or the mean position of all cores in each clump (as in this work). This difference could indicate an evolutionary effect, given the (70\,$\mu$m-dark) regions of the ASHES sample generally occupy earlier evolutionary stages than those of EGO-10, which \citet{Towner2019} find to be in transition from the IR-quiet to IR-bright phases. 

The location of 6.7\,GHz masers near protocluster centres is also consistent with these masers' close association with millimetre source peaks on clump-scales \cite[e.g.][]{Beuther2002, Urquhart2013, Billington2019, Billington2020}. Notably, however, our high-resolution interferometric observations reveal 6.7\,GHz masers are not necessarily associated with the brightest 1.3\,mm core in a given protocluster. In fact, this is only the case in 3/10 regions\footnote{The host-core of the 6.7\,GHz maser G10.29$-$0.13 CM1-M1 would be the brightest core in this field based on its peak intensity in the ALMA image, but due to the overlap of its photometric ellipse with a companion source, \textit{Hyper} performs multi-Gaussian fitting and companions subtraction, reducing the reported peak intensity. The 6.7\,GHz host-core is therefore the second-brightest core in this field (MM2) and the brightest core (MM1), which has no companions, is located near the edge of the mosaic in the western ATLASGAL clump AGAL010.284$-$00.114.}: G19.36$-$0.03, G22.04$+$0.22 and G28.83$-$0.25. This contrasts with the cm-$\lambda$ case, where 7/11 6.7\,GHz masers (64\%) are associated with the brightest cm-$\lambda$ source in a given field (largest integrated flux density at 1.3\,cm). These include six of the \citet{Towner2021} 6.7\,GHz masers and the \citet{Green2010} maser in G16.59$-$0.05 associated with \citet{Rosero2016} source C. 

\subsubsection{Clustered Massive Star Formation}
\label{sec:myso_clust}
We quantify the degree of clustering in the immediate environment of the MYSOs in our sample with two additional parameters, reported in Table~\ref{tab:clus67}. $\Delta_\text{nn}$ is the projected physical separation between a 6.7\,GHz CH$_3$OH maser and its nearest-neighbouring core. $N_\text{env}$ is the number of cores with projected physical separation $<$10,000\,AU from the maser position \cite[$\sim$0.05\,pc, within the size scale of massive prestellar cores in monolithic collapse models of massive star formation, e.g.][]{McKee2002, McKee2003, Krumholz2009, Meyers2013}. In computing these parameters, the maser's host-core was not considered, as we assume this core hosts the same MYSO driving the maser emission. In 5/11 cases we find highly clustered MYSO environs, with $\Delta_\text{nn}\lesssim$ 4000\,AU and $N_\text{env}>$ 10 (corresponding to a core number density $\approx$2$\times$10$^4$\,pc$^{-3}$, assuming all sources lie within a spherical volume). Notably, for all five MYSOs at $d<$ 3\,kpc (corresponding to a beam spatial scale $\lesssim$ 1700 $\times$ 1200\,AU), $\Delta_\text{nn}\lesssim$ 4000\,AU and $N_\text{env}>$ 10, suggesting MYSOs in the more distant regions in our sample would likely exhibit a similarly high degree of clustering if observed at higher spatial resolution. 
Fig.~\ref{fig:AUzoom} shows representative examples of the 1.3\,mm continuum emission within 10,000\,AU of MYSOs, in G10.34$-$0.14 ($d=$ 1.6\,kpc) and G22.04$+$0.22 ($d=$ 3.4\,kpc). Equivalent figures for the other MYSOs discussed in this section are available as online Supplementary Material.

The presence of multiple 1.3\,mm cores within 10,000\,AU of the MYSOs in our sample 
is inconsistent with massive star formation by monolithic collapse, in which a $\sim$0.05--0.1\,pc-radius core collapses to form a single star or small-$N$ multiple system \cite[e.g.][]{McKee2002, McKee2003, Krumholz2009, Meyers2013}. Clustering to this degree and on these scales is, however, inherent to hierarchical formation scenarios such as competitive accretion \citep{Bonnell2001, Bonnell2004, Bonnell2006, Smith2009} and global hierarchical collapse \citep{Vazquez-Semadeni2017, Vazquez-Semadeni2019}. Simulations have demonstrated such clustering as a result of turbulent fragmentation, thus enabling the formation of stellar multiple systems with initial separations $\lesssim$10,000\,AU \cite[e.g.][and references therein]{Guszejnov2023, Kuruwita2023, Offner2023}. 

The impact of such densely clustered environs on the massive star formation process has not yet been fully characterised. Simulations have shown a surrounding cluster of cores can intercept inter-core material which would otherwise reach the central MYSO, impeding accretion and limiting the final stellar mass \cite[the `accretion-shielding' in fragmentation-induced starvation, e.g.][]{Peters2010a, Peters2010b, Girichidis2012}. Similarly, outflows driven by surrounding YSOs can disrupt large-scale infall or filamentary accretion flows toward an MYSO \citep{Wang2010, Lebreuilly2024}. Notably, 9/11 (82\%) of the MYSOs considered here have at least one 22\,GHz H$_2$O maser or cm-$\lambda$ continuum source within a projected separation of 10,000\,AU that is not associated with the MYSO host-core. In 6/9 (67\%) of these cases, these additional tracer(s) are associated with 1.3\,mm core(s). Both tracers can be excited in protostellar outflows (see Section~\ref{sec:seps}), indicating the presence of additional outflow-driving sources in the immediate environment of these MYSOs.

\section{Protocluster Structure and Evolutionary State} 
\label{sec:evol}
\begin{figure}
\includegraphics[width=\columnwidth]{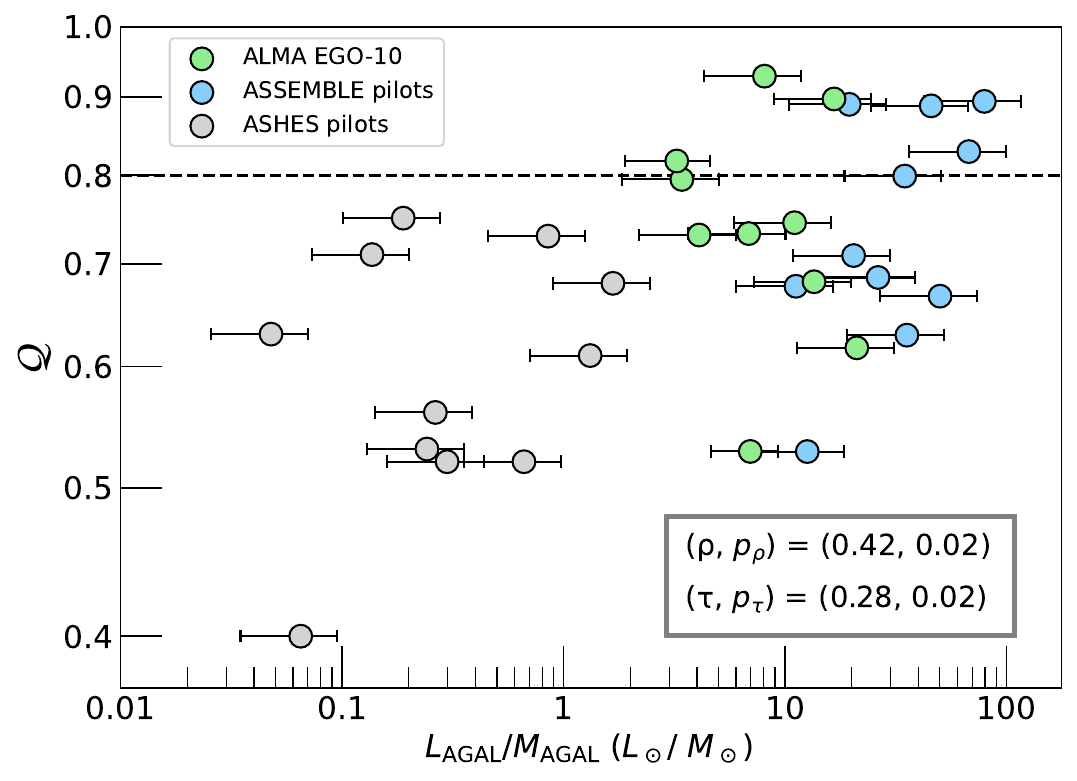} 
\caption{The \citet{Cartwright2004} $\mathcal{Q}$-parameter as a function of the \citet{Urquhart2018} ATLASGAL luminosity-to-mass ratio ($L_\text{AGAL}/M_\text{AGAL}$) for the ASSEMBLE \citep[blue;][]{Xu2024} and ASHES \citep[grey;][]{Sanhueza2019} pilot samples, as in Figure~11 of \citet{Xu2024}.  Green points show the ALMA EGO-10 protoclusters. The dashed horizontal line denotes the $\mathcal{Q}=$ 0.8 value which distinguishes subclustered ($\mathcal{Q}<$ 0.8) from centrally-condensed ($\mathcal{Q}>$ 0.8) (proto)cluster structure. The label lists the Spearman ($\rho$) and Kendall ($\tau$) correlation coefficients and their associated $p$-values, evaluated for the combined sample.}
\label{fig:LM_ALL}
\end{figure}

Fig.~\ref{fig:LM_ALL} shows the EGO-10 protoclusters alongside the ASHES and ASSEMBLE regions in the parameter space of \citet{Xu2024} Figure~11, i.e. $\mathcal{Q}$ v.\ the \citet{Urquhart2018} ATLASGAL luminosity-to-mass ratio ($L_\text{AGAL}/M_\text{AGAL}$, see Tables~\ref{tab:tbl1} and \ref{tab:Lcm} for EGO-10). The statistically significant positive correlation between $\mathcal{Q}$ and $L_\text{AGAL}/M_\text{AGAL}$ identified by \citet{Xu2024} in the combined ASHES and ASSEMBLE data persists upon the addition of the EGO-10 sample (see in-panel label of Fig.~\ref{fig:LM_ALL}). 
Taking $L_\text{AGAL}/M_\text{AGAL}$ as an evolutionary indicator \citep[e.g.][]{Molinari2008, Molinari2016}, this is consistent with an increase in $\mathcal{Q}$ with evolutionary state, suggesting the evolution of protocluster structure from subclustered to centrally condensed, as discussed in Section~\ref{sec:intro}. 

As noted in Section~\ref{sec:intro}, \citet{Xu2024} also found a positive correlation between $\mathcal{Q}$ and $L_\text{AGAL}/M_\text{AGAL}$ within the ASSEMBLE pilot sample itself, consisting of 11 regions (which was then strengthened by their inclusion of the ASHES pilot sample).
We do not find any statistically significant relationship between $\mathcal{Q}$ and $L_\text{AGAL}/M_\text{AGAL}$ within the EGO-10 sample. This is not especially surprising however, as the EGO-10 targets were specifically selected to occupy a narrow range in evolutionary state based on their MIR morphology. The $L_\text{AGAL}/M_\text{AGAL}$ range of the EGO-10 sample is relatively narrow as a result, as is evident in Fig.~\ref{fig:LM_ALL}: \mbox{3 $<L_\text{AGAL}/M_\text{AGAL}<$ 21\,$L_\odot/M_\odot$}, compared to \mbox{11 $<L_\text{AGAL}/M_\text{AGAL}<$ 79\,$L_\odot/M_\odot$} for the ASSEMBLE pilot sources. \citet{Urquhart2018} also estimate large relative uncertainties of \mbox{$\Delta L_\text{AGAL}\approx$ 40\%} and \mbox{$\Delta M_\text{AGAL}\approx$ 20\%}, such that \mbox{$\Delta$($L_\text{AGAL}/M_\text{AGAL}$)$\approx$ 50\%} (see Table~\ref{tab:Lcm}), which may contribute scatter to the relation. 
Notably, the \citet{Schisano2025} analysis of ALMAGAL clumps, a much larger sample covering a wider range in $L/M$ (see Table~\ref{tab:summ}; $L/M$ values from \citealt{Molinari2025}), yielded only a weak positive correlation between $\mathcal{Q}$ and $L/M$ compared to that in \citet{Xu2024}. The quantification of protocluster structure in ALMAGAL is however limited by it being a single-pointing survey; as noted by \citet{Schisano2025}, additional cores may lie outside the field-of-view, particularly in regions at $d<$ 4\,kpc.

We note that we find no statistically significant relationship between $\mathcal{Q}$ and $L_\text{AGAL}/M_\text{AGAL}$ both for the full EGO-10 sample and when excluding G16.59$-$0.05 and G22.04$+$0.22, for which we consider our measured $\mathcal{Q}$-parameters less reliable due to small $N_\text{cores}$ (Section~\ref{sec:mst}). We exclude these regions from our analyses in the following sections, and note that the eight regions in the remaining (sub)sample are all members of the EGO-9 sample of \citet{Towner2021} (see Section~\ref{sec:ego}).

\begin{table*}
\caption{Summary of data used to investigate correlations with $\mathcal{Q}$.}

\begin{tabular}{lccccccccc}
\hline
\hline
EGO$^\text{a}$ &  AGAL$^\text{a}$ & $L_\text{AGAL}$$^\text{b}$ & $M_\text{AGAL}$$^\text{c}$ & $L_\text{AGAL}/M_\text{AGAL}$$^\text{d}$ & $L_\text{1.3cm}$$^\text{e}$ & $L_\text{5cm}$$^\text{e}$ & & $L_\text{1.3cm}/M_\text{AGAL}$$^\text{f}$ & $L_\text{5cm}/M_\text{AGAL}$$^\text{f}$\\
\cline{6-7}
\cline{9-10}\noalign{\vspace{0.5ex}}
 & & (10$^3L_\odot$) & (10$^3M_\odot$) & ($L_\odot/M_\odot$) & \multicolumn{2}{c}{(mJy kpc$^2$)} & & \multicolumn{2}{c}{(10$^{-3}$mJy kpc$^2/M_\odot$)}\\
\hline
G10.29$-$0.13 & \makecell[c]{010.288-00.124\\010.284-00.114} & 15 (4) & 2.2 (0.3) & 7 (2) & 0.3 (0.1) & 0.15 (0.05) & & 0.14 (0.05) & 0.07 (0.03)\\
\hline
G10.34$-$0.14 & 010.342-00.142 & 18 (7) & 0.8 (0.17) & 20 (10) & 1.3 (0.2) & 0.17 (0.04) & & 1.5 (0.4) & 0.21 (0.06)\\
G12.91$-$0.03 & 012.904-00.031 & 7 (3) & 2 (0.4) & 3 (1.6) & 4 (1) & 4 (1) & & 1.9 (0.6) & 1.9 (0.6)\\
G14.33$-$0.64 & 014.331-00.644 & 6 (2) & 0.7 (0.14) & 8 (4) & 6.1 (0.8) & 4.8 (0.7) & & 9 (2) & 7 (2)\\
\hline
G14.63$-$0.58 & \makecell[c]{014.632-00.577\\014.621-00.579} & 1.8 (0.7) & 0.55 (0.09) & 3 (1.4) & 1.6 (0.2) & 0.81 (0.06) & & 3 (0.6) & 1.5 (0.3)\\
\hline
G16.59$-$0.05 & 016.586-00.051 & 18 (8) & 1.3 (0.3) & 14 (6) & 20 (1.9) & 14 (1.7) & & 15 (3) & 10 (2)\\
G18.89$-$0.47 & 018.888-00.474 & 50 (20) & 5 (1) & 11 (5) & 5 (1) & 1.6 (0.5) & & 1 (0.3) & 0.3 (0.1)\\
G19.36$-$0.03 & 019.362-00.031 & 3 (1.2) & 0.7 (0.14) & 4 (1.9) & 17 (4) & 9 (1.7) & & 24 (7) & 13 (3)\\
 &  & & & & 4.4 (0.9)$^\text{g}$  & 4 (1)$^\text{g}$ & & 6 (2)$^\text{g}$ & 6 (2)$^\text{g}$\\
G22.04+0.22 & 022.038+00.222 & 5 (1.9) & 0.7 (0.13) & 7 (3) & 2.1 (0.7) & 0.8 (0.3) & & 3 (1) & 1.2 (0.5)\\
G28.83$-$0.25 & 028.831-00.252 & 40 (15) & 2.2 (0.4) & 17 (8) & 10 (2) & 9 (2) & & 5 (1) & 4 (1)\\
\hline
\end{tabular}
\begin{flushleft}
        \small{
        $^\text{a}$ EGO name and associated ATLASGAL clump designation(s), as in Table~\ref{tab:tbl1}. G16.59$-$0.05 and G22.04$+$0.22 are included for completeness (see Section~\ref{sec:evol}).\\
        $^\text{b}$ Luminosity of the ATLASGAL clump(s) calculated from the log($L_\text{AGAL}$) reported by \citet{Urquhart2018}. For G10.29$-$0.13 and G14.63$-$0.58 this is the sum of the luminosities of both clumps with peak positions within the ALMA mosaic. The \citet{Urquhart2018} estimated uncertainty is given in parentheses. \\
        $^\text{c}$ Mass of the ATLASGAL clump(s) calculated from the log($M_\text{AGAL}$) reported by \citet{Urquhart2018}. For G10.29$-$0.13 and G14.63$-$0.58 this is the sum of the masses of both ATLASGAL clumps with peak positions within the ALMA mosaic. The \citet{Urquhart2018} estimated uncertainty is given in parentheses. \\
        $^\text{d}$ The ATLASGAL luminosity-to-mass ratio, rounded to reflect the propagated uncertainty, which is given in parentheses (see also Table~\ref{tab:tbl1}).\\
        $^\text{e}$ Total $L_\text{cm}$ of the field, calculated using the integrated flux densities ($S_\text{cm}$) of cm-$\lambda$ sources reported in \citet{Towner2021} (for the EGO-9 regions) and \citet{Rosero2016} (for G16.59$-$0.05), and the distances in Table~\ref{tab:tbl1} (see Equation~\ref{eq:cm}). The uncertainty is given in parentheses, calculated by propagating the uncertainties in $S_\text{cm}$ and distance, using the mean uncertainty for distances with asymmetric uncertainties. For sources in the EGO-9, we adopted the $S_\text{cm}$ uncertainties reported by \citet{Towner2021}. For sources in G16.59$-$0.05, we used the mean of the values reported by \citet{Rosero2016} in the individual basebands comprising each of the 1.3 and 5\,cm observing bands to evaluate $S_\text{cm}$ and followed these authors by assuming an uncertainty of 10\% in both the integrated flux density and the absolute flux calibration. \\  
        $^\text{f}$ The ratio of the total $L_\text{cm}$ of the protocluster to the mass of its associated ATLASGAL clump(s). Uncertainties, calculated by propagating the uncertainties on $L_\text{cm}$ and $M_\text{AGAL}$, are given in parentheses.\\
        $^\text{g}$ Values used in the EGO-10$^\prime$ subsample in which the cm-$\lambda$ source G19.36$-$0.03 CM1 is excluded from the calculation of $L_\text{cm}$ (see Section~\ref{sec:samps}).
        
        }
    \end{flushleft}
\label{tab:Lcm}
\end{table*}

 \begin{figure*}
   \includegraphics[width=\textwidth]{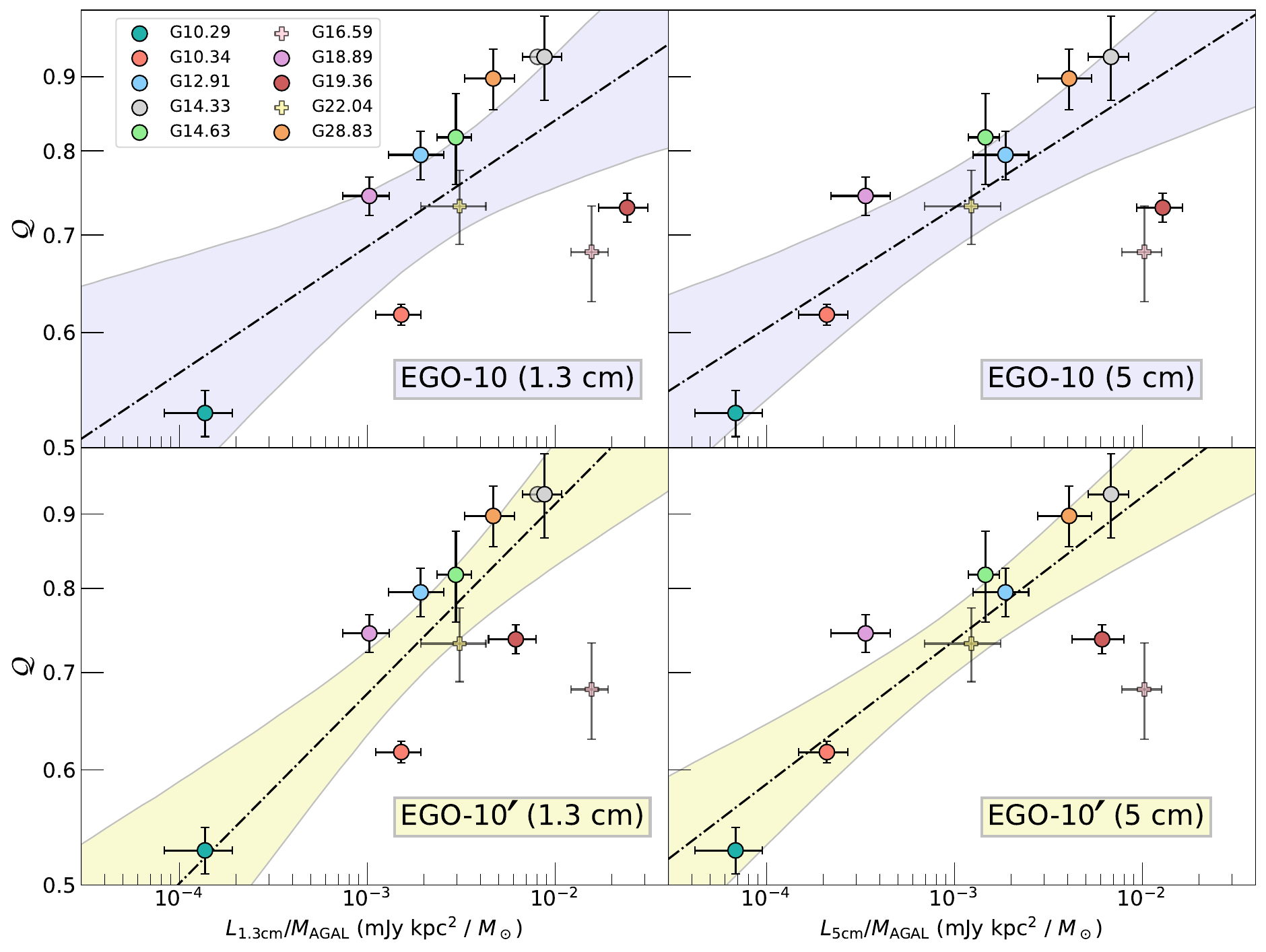} 
   \caption{$\mathcal{Q}$ plotted as a function of $L_\text{cm}/M_\text{AGAL}$ for the EGO-10 (upper panels) and EGO-10$^\prime$ (lower panels) subsamples (see Section~\ref{sec:samps}). In-panel labels give the subsample name and the wavelength for $L_\text{cm}$. In each panel, the dot-dashed line and shaded area show the LinMix fit and its 1-$\sigma$ uncertainty envelope, respectively. Correlation statistics and LinMix fit results are reported in Table~\ref{tab:corr1}. G16.59$-$0.05 and G22.04$+$0.22, which are not considered in the statistical analysis or LinMix fit due to their small $N_{\rm cores}$ and so less reliable $\mathcal{Q}$-parameters (see Section~\ref{sec:mst} and Section~\ref{sec:evol}), are shown as crosses. The transparent grey point in the 1.3\,cm panels denotes the location of G14.33$-$0.64 if the cm-$\lambda$ continuum source G14.33$-$0.64 CM4 is excluded from the calculation of $L_\text{1.3cm}$ (see Appendix~\ref{app:alpha}).}
\label{fig:QvLM}
\end{figure*}

\subsection{\texorpdfstring{$L_\text{cm}/M_\text{AGAL}$}{L\_cm/M\_AGAL} - An Alternative Evolutionary Indicator}
\label{sec:Q1}
The high-resolution multiwavelength data available for the EGO-10 sample provides access to other evolutionary indicators, specifically the cm-$\lambda$ continuum source detections in \citet{Towner2021}. As outlined in Section~\ref{sec:ego}, the EGO-10 regions are in an early evolutionary state, in which their MYSO(s) are just beginning to excite cm-$\lambda$ continuum emission. \citet{Towner2021} find a strong correlation between radio and bolometric luminosity in EGO-9, and conclude the observed cm-$\lambda$ emission arises primarily from thermal free-free emission in ionised jets. Relatively weak cm-$\lambda$ jet emission is typically observed prior to the emergence of radiatively ionised regions \citep[e.g.][]{Guzman2010, Guzman2012}, with stronger emission expected as an MYSO evolves and forms an \ion{H}{ii} region \citep[and references therein]{Hoare2007a, Hoare2007b, Anglada2018, Purser2021}. 
The observed cm-$\lambda$ luminosity of nascent \ion{H}{ii} regions is also expected to increase as they evolve \citep{Hoare2007a, Hoare2007b, Galvn-Madrid2011, Tanaka2017}, albeit not monotonically, with cm-$\lambda$ \ion{H}{ii} region variability characterised in both simulations \cite[][]{Peters2010a, Galvn-Madrid2011} and observations \cite[e.g.][and references therein]{Depree2014, Yang2025}.  Overall, however, the total cm-$\lambda$ continuum luminosity of the (M)YSO population in a given protocluster would be expected to generally increase with time, including as additional (M)YSO(s) reach the stage of exciting cm-$\lambda$ continuum emission.

We therefore propose $L_\text{cm}/M_\text{AGAL}$ - the ratio of a protocluster's total cm-$\lambda$ continuum distance-luminosity ($L_\text{cm}$; see Equation~\ref{eq:cm}) to its ATLASGAL clump mass ($M_\text{AGAL}$) - as an indicator of protocluster evolutionary state, and investigate trends with $\mathcal{Q}$. To our knowledge, this is the first use of this specific quantity as an indicator of protocluster evolutionary state.

\subsubsection{Evaluating $L_\text{cm}/M_\text{AGAL}$ and Subsample Selection}
\label{sec:samps}
To evaluate $L_\text{cm}$ in each protocluster, we consider 
all \citet{Towner2021} detections within the 10\% response level of the ALMA mosaic in each field (see also  Section~\ref{sec:cm}) and compute 

\begin{equation}
L_\text{cm} = \sum{S_\text{cm}\,d^2} 
	\label{eq:cm}
\end{equation}

\noindent where $S_\text{cm}$ is the integrated flux density of a source at a given wavelength, and $d$ the protocluster distance (see Table~\ref{tab:tbl1}). We use the $S_\text{cm}$ values reported by \citet{Towner2021}, which they measured with CASA's \texttt{imfit} for Gaussian-like sources, and with aperture photometry for irregular sources. $L_\text{cm}/M_\text{AGAL}$ was then evaluated at each cm wavelength using the \citet{Urquhart2018} $M_\text{AGAL}$ values (see Table~\ref{tab:Lcm}). The uncertainty in $L_\text{cm}/M_\text{AGAL}$ was evaluated by propagating the integrated flux density uncertainties reported by \citet{Towner2021}, the protocluster distance uncertainties (see Table~\ref{tab:tbl1}), and the \citet{Urquhart2018} estimated $M_\text{AGAL}$ uncertainty.

We also construct an additional subsample, designated \mbox{EGO-10$^\prime$}, in which we exclude the cm-$\lambda$ continuum source G19.36$-$0.03 CM1 from the calculation of $L_\text{cm}$ for the protocluster G19.36$-$0.03. G19.36$-$0.03 CM1 was identified as a potentially more evolved object by  \citet{Cyganowski2011_VLA} (their source F G19.36$-$0.03-CM2), consistent with our spectral index analysis in Appendix~\ref{app:alpha}. This source dominates the cm-$\lambda$ continuum flux density of this region (74\% contribution at 1.3\,cm; 52\% contribution at 5\,cm) and could therefore significantly influence correlations and fitting results, motivating the construction of EGO-10$^\prime$. G19.36$-$0.03 CM1 is associated with a 1.3\,mm source (G19.36$-$0.03 MM10). For consistency, in our analyses of the EGO-10$^\prime$ subsample, we recompute the MST and $\mathcal{Q}$-parameter of G19.36$-$0.03 without G19.36$-$0.03 MM10, which changes the $\mathcal{Q}$ value only marginally, from 0.73 to 0.74.

\begin{figure*}
   \includegraphics[width=\textwidth]{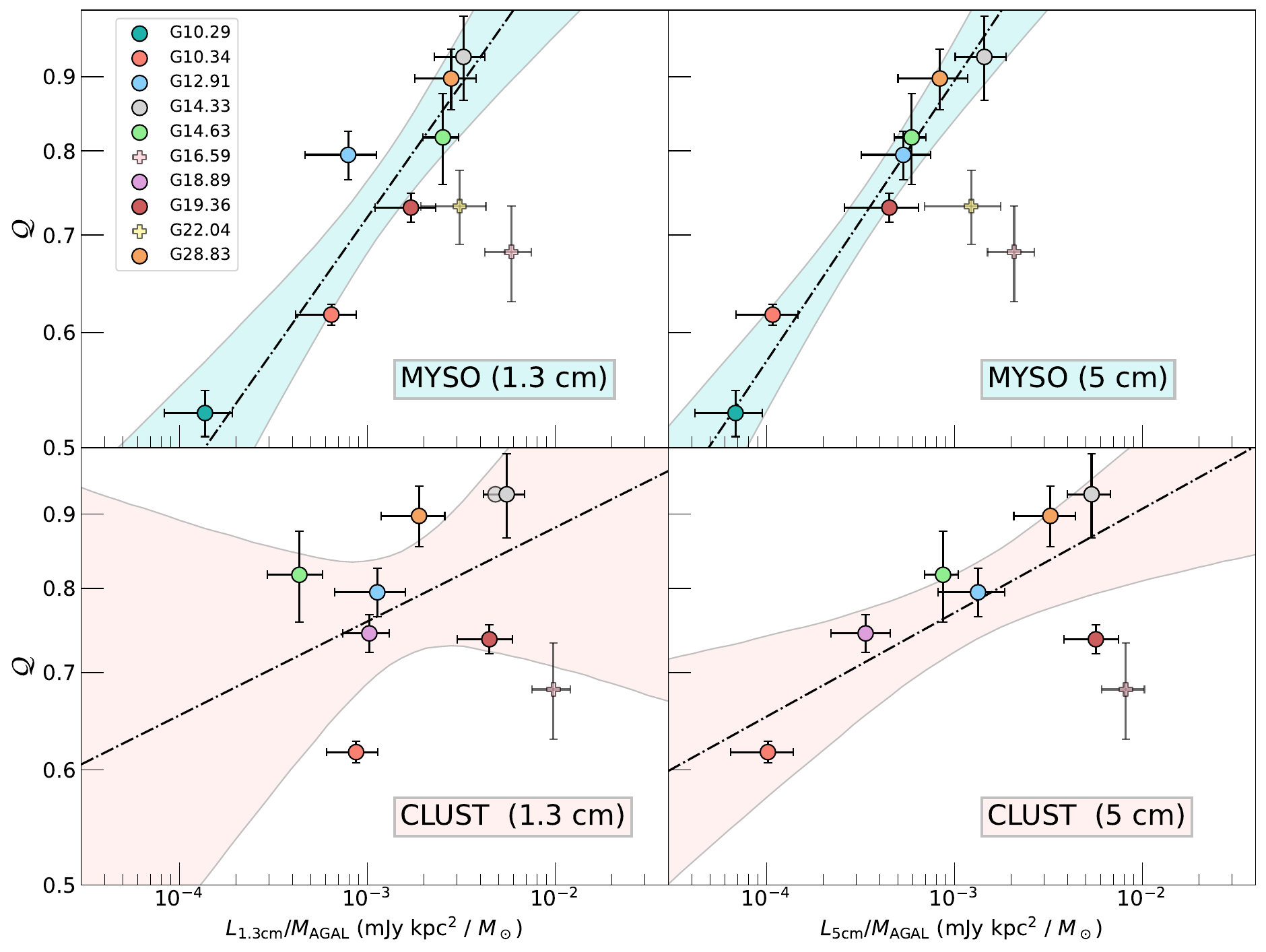} 
   \caption{$\mathcal{Q}$ plotted as a function of $L_\text{cm}/M_\text{AGAL}$ for the MYSO (upper panels) and CLUST (lower panels) subsamples (see Section~\ref{sec:cm_67}). 
   As in Fig.~\ref{fig:QvLM}, in-panel labels give the subsample name and the wavelength for $L_\text{cm}$, and the dot-dashed line and shaded area show the LinMix fit and its 1-$\sigma$ uncertainty envelope. Correlation statistics and LinMix fit results are reported in Table~\ref{tab:corr1}. Markers and $x$- and $y$-axis ranges are the same as in Fig.~\ref{fig:QvLM}. G16.59$-$0.05 and G22.04$+$0.22, which are not considered in the statistical analysis or LinMix fit due to their small $N_{\rm cores}$ and so less reliable $\mathcal{Q}$-parameters (Section~\ref{sec:mst} and Section~\ref{sec:evol}), are shown as crosses, and the transparent grey point denotes the position of G14.33$-$0.64 without the contribution from G14.33$-$0.64 CM4 (see Appendix~\ref{app:alpha}).}
\label{fig:QvLM_2}
\end{figure*}

\begin{table*}
 \caption{$\mathcal{Q}$ vs. $L_\text{cm}/M_\text{AGAL}$ correlation coefficients and fitting results for subsamples described in Section~\ref{sec:samps} and Section~\ref{sec:QvLM_2}.}
 \begin{tabular}{lccccccccccccc}
 \hline
 \hline
   & & \multicolumn{4}{c}{Spearman} & & \multicolumn{4}{c}{Kendall} & & \multicolumn{2}{c}{LinMix Fit$^\text{e}$}\\
\cline{3-6}
\cline{8-11}
\cline{13-14}
\noalign{\vskip 2pt}
\textit{Subsample (1.3 cm)} & & $\rho$$^\text{a}$ & $p_\rho$$^\text{b}$ & $\overline{\rho_{_{MC}}}$$^\text{c}$ & \%$_\rho^{MC}$$^\text{d}$ & & $\tau$$^\text{a}$ & $p_\tau$$^\text{b}$ & $\overline{\tau_{_{MC}}}$$^\text{c}$ & \%$_\tau^{MC}$$^\text{d}$ & & $\alpha$ & $\beta$\\
\hline
EGO-10 & & 0.60 & 0.13 & 0.58(0.11) & 10 & & 0.57 & 0.06 & 0.49(0.1) & 11 & & 0.1(0.17) & 0.09(0.06)\\
EGO-10$'$ & & 0.71 & 0.06 & 0.69(0.12) & 40 & & 0.64 & 0.03 & 0.57(0.12) & 39 & & 0.22(0.16) & 0.13(0.06)\\
MYSO & & 0.96 & 2.4e-03 & 0.79(0.14) & 62 & & 0.9 & 2.4e-03 & 0.67(0.15) & 54 & & 0.41(0.25) & 0.18(0.08)\\
CLUST & & 0.39 & 0.4 & 0.29(0.24) & 0.8 & & 0.33 & 0.38 & 0.22(0.19) & 0.7 & & 0.07(0.51) & 0.06(0.18)\\
\hline
\hline

\textit{Subsample (5 cm)} & \\
\hline
EGO-10 & & 0.62 & 0.12 & 0.64(0.09) & 21 & & 0.57 & 0.06 & 0.55(0.1) & 30 & & 0.12(0.14) & 0.08(0.05)\\
EGO-10$'$ & & 0.74 & 0.04 & 0.72(0.11) & 52 & & 0.64 & 0.03 & 0.62(0.11) & 55 & & 0.16(0.14) & 0.1(0.04)\\
MYSO & & 1.0 & 4.0e-04 & 0.84(0.13) & 77 & & 1.0 & 4.0e-04 & 0.73(0.16) & 70 & & 0.53(0.21) & 0.19(0.06)\\
CLUST & & 0.43 & 0.36 & 0.56(0.16) & 10 & & 0.43 & 0.24 & 0.48(0.14) & 8 & & 0.1(0.19) & 0.07(0.06)\\
\hline
\end{tabular}
\begin{flushleft}
        \small{
        $^\text{a}$ Correlation coefficient computed from the exact ($\mathcal{Q}$, $L_\text{cm}/M_\text{AGAL}$) measurements. \\
        $^\text{b}$ $p$-value of the correlation coefficient, evaluated empirically using Monte Carlo realisations of the null distribution (see Section~\ref{sec:QvLM}). The combination of small sample size (7) and large correlation coefficients in the MYSO subsample is such that $p_\rho$ and $p_\tau$ are identical; an equal number of null-realisations have Spearman and Kendall correlation coefficients as extreme as $\rho$ and $\tau$.\\
        $^\text{c}$ Mean of the distribution of the correlation coefficient obtained from Monte Carlo simulation. The uncertainty, estimated as the standard deviation of the corresponding distribution, is given in parentheses (see Section~\ref{sec:QvLM}).\\
        $^\text{d}$ Percentage of Monte Carlo realisations in which the correlation is statistically significant ($p<$ 0.05).\\
        $^\text{e}$ The intercept ($\alpha$) and gradient ($\beta$) of the linear fit to the log-log data ($\text{log}_{10}(\mathcal{Q}) = \alpha + \beta\;\text{log}_{10}(L_\text{cm}/M_\text{AGAL})$) computed by LinMix (see Section~\ref{sec:QvLM}). Both are evaluated as the mean of the $\alpha$ and $\beta$ distributions output by LinMix. Uncertainties, estimated as the standard deviations of their corresponding distributions, are given in parentheses.
        }
    \end{flushleft}
\label{tab:corr1}
\end{table*}

\subsubsection{$\mathcal{Q}$ and $L_\text{cm}/M_\text{AGAL}$}
\label{sec:QvLM}

We quantify the $\mathcal{Q}$--$L_\text{cm}/M_\text{AGAL}$ relationship in each subsample with the Spearman ($\rho$) and Kendall ($\tau$) correlation coefficients - computed with \texttt{scipy.stats}' \texttt{spearmanr} and \texttt{kendalltau}, respectively -  and their associated $p$-values ($p_\rho$, $p_\tau$). We compute the $p$-values empirically, from $n=$\,10,000 Monte Carlo realisations of the null distribution, made by sampling each ($\mathcal{Q}$, $L_\text{cm}/M_\text{AGAL}$) measurement from Gaussian distributions with standard deviations equal to their associated uncertainties and then randomly permuting the resulting pairs. The $p$-value is then computed as $p=(b+1)/(n+1)$\footnote{This formula prevents $p=$ 0, since this is, in general, not possible to establish via Monte Carlo simulation given the finite number of null-realisations.} where $b$ is the number of null-realisations with correlation coefficient at least as extreme as measured \cite[see e.g.][and references therein]{Davison1997, Phipson2016}. To assess the robustness of the $\mathcal{Q}$--$L_\text{cm}/M_\text{AGAL}$ relationship to uncertainty, we compute the mean of the $\rho$ and $\tau$ distributions of the 10,000 ($\mathcal{Q}$, $L_\text{cm}/M_\text{AGAL}$) Monte Carlo realisations (prior to random permutation) \mbox{($\overline{\rho_{_{MC}}}$, $\overline{\tau_{_{MC}}}$)} and estimate their associated uncertainties as the standard deviation. We also compute the percentage of Monte Carlo realisations in which the correlation is statistically significant, with \mbox{$p<$ 0.05} (\mbox{\%$_\rho^{MC}$, \%$_\tau^{MC}$)}. We compute linear fits in log-log space, of the form $\text{log}_{10}(\mathcal{Q}) = \alpha + \beta\;\text{log}_{10}(L_\text{cm}/M_\text{AGAL})$, with the Python implementation\footnote{\url{https://linmix.readthedocs.io/en/latest/src/linmix.html}} of the LinMix algorithm \citep{Kelly2007}. When provided with the uncertainties in $\mathcal{Q}$ and $L_\text{cm}/M_\text{AGAL}$, LinMix uses a Bayesian technique to perform the linear regression and produce a distribution of fit parameters. The ($\alpha$, $\beta$) values and their associated uncertainties are estimated as the mean and standard deviation of the resulting distributions, respectively. All statistics and fitting results are reported in Table~\ref{tab:corr1}.

We find positive correlations between $\mathcal{Q}$ and $L_\text{cm}/M_\text{AGAL}$ in both EGO-10 and EGO-10$'$, at 1.3\,cm and 5\,cm, which persist in  distance-limited subsamples (see Appendix~\ref{sec:app_tests}). The relationship is evident in Fig.~\ref{fig:QvLM}, where $\mathcal{Q}$ is plotted against $L_\text{cm}/M_\text{AGAL}$, with the LinMix fits and their 1-$\sigma$ uncertainty envelopes also indicated. The positive correlation becomes stronger, and statistically significant, in EGO-10$'$ (Table~\ref{tab:corr1}). The relationship is also more robust in this subsample, with smaller relative uncertainties on $\overline{\rho_{_{MC}}}$ and $\overline{\tau_{_{MC}}}$, and larger values of \%$_\rho^{MC}$ and \%$_\tau^{MC}$: $\sim$40\% and $\sim$50\% at 1.3 and 5\,cm, respectively, compared to $\sim$10\% and $\lesssim$30\% in the EGO-10 subsample. These results are also reflected in the slightly better-constrained linear fit for EGO-10$'$ (see Fig.~\ref{fig:QvLM} and Table~\ref{tab:corr1}). Notably, the larger 5\,cm (\%$_\rho^{MC}$, \%$_\tau^{MC}$) values in both subsamples suggest the relationship is more robust at 5\,cm than at 1.3\,cm, which may reflect potential contributions to the observed 1.3\,cm flux densities from warm dust emission \citep{Towner2021}.

\subsubsection{MYSO and Cluster-Scale Relationships}
\label{sec:QvLM_2}
As noted by \citet{Towner2021}, the 6.7\,GHz CH$_3$OH masers are typically associated with the brightest cm-$\lambda$ continuum source in a given field (see also Section~\ref{sec:myso_spat}). Having identified the 1.3\,mm MYSO host-cores via their association with these masers (see Section~\ref{sec:myso}), we can investigate to what degree the correlations identified in Section~\ref{sec:QvLM} are driven solely by the cm-$\lambda$ continuum source(s) associated with the EGO-10 MYSOs. To do this, we constructed two further subsamples:

\begin{itemize}
    \item[--] MYSO: Only the integrated flux density of the cm-$\lambda$ source(s) associated with the MYSO host-core(s) contributes to $L_\text{cm}$ in a given field.
    
    \item[--] CLUST: Only the integrated flux densities of cm-$\lambda$ source(s) not associated with the MYSO host-core(s) contribute to $L_\text{cm}$ in a given field. 
\end{itemize}

\noindent We note that, by construction, not all protoclusters are included in each subsample. G18.89$-$0.47 is excluded from the MYSO subsample, since its two MYSO host-cores do not host any cm-$\lambda$ continuum sources. G10.29$-$0.13 is excluded from the CLUST subsample since the sole cm-$\lambda$ continuum source in this field is associated with the MYSO host-core. The MYSO and CLUST subsamples are therefore each comprised of seven regions, since as in Section~\ref{sec:QvLM}, we do not include G16.59$-$0.05 and G22.04$+$0.22 in this analysis. All regions in the MYSO subsample contain a single MYSO host-core, which is associated with a single cm-$\lambda$ continuum source detected at both 1.3 and 5\,cm in 6/7 cases (see Fig.~\ref{fig:MYSO}). The exception is the core G14.63$-$0.58 MM8, which is associated with two 5\,cm sources - G14.63$-$0.58 CM1 and CM3 (see Section~\ref{sec:cm} and Fig.~\ref{fig:MYSO}) - both of which contribute to $L_\text{5cm}$.
In evaluating $L_\text{cm}/M_\text{AGAL}$ in the CLUST subsample, we excluded G19.36$-$0.03 CM1 and its associated 1.3\,mm core (G19.36$-$0.03 MM10), and used the corresponding \mbox{$\mathcal{Q}=$ 0.74}, as in the EGO-10$^\prime$ subsample (Section~\ref{sec:QvLM}, see also Section~\ref{sec:samps}). All correlation statistics and LinMix fit results are computed as in Section~\ref{sec:QvLM}, and reported in Table~\ref{tab:corr1}.

Fig.~\ref{fig:QvLM_2} shows $\mathcal{Q}$ plotted against $L_\text{cm}/M_\text{AGAL}$ for the MYSO and CLUST subsamples at both 1.3\,cm and 5\,cm. The relationship in the MYSO subsample is clearly stronger. This is evident in Table~\ref{tab:corr1}, with strong ($\rho, \tau \gtrsim$ 0.9), statistically significant ($p_\rho, p_\tau\lesssim$ 0.0025) correlations in this subsample at both 1.3 and 5\,cm. The relationship is also robust, with relatively small uncertainties on $\overline{\rho_{_{MC}}}$ and $\overline{\tau_{_{MC}}}$ and with (\%$_\rho^{MC}$, \%$_\tau^{MC}$)$\gtrsim$ 50\%. As in EGO-10 and EGO-10$'$, above, the relationship is stronger at 5\,cm than at 1.3\,cm, with larger correlation coefficients with smaller $p$-values, and larger (\%$_\rho^{MC}$, \mbox{\%$_\tau^{MC}$)$\gtrsim$ 70\%}. In contrast, at both 1.3 and 5\,cm, correlations in the CLUST subsample do not meet the threshold for statistical significance. The 1.3\,cm relationship is particularly weak, with (\%$_\rho^{MC}$, \%$_\tau^{MC}$)$<$ 1\% and a LinMix slope consistent with no trend, within uncertainty.

The strong, statistically significant relationship in the MYSO subsample, contrasting with the lack of significance in the CLUST subsample, suggests the $\mathcal{Q}$--$L_\text{cm}/M_\text{AGAL}$ relationship identified in Section~\ref{sec:QvLM} is driven by the MYSOs in these protoclusters. This further suggests that the proposed metric of $L_\text{cm}/M_\text{AGAL}$  is indeed a valid tracer of the evolutionary state of the MYSOs in these fields.

\subsection{Global Collapse in Massive Protoclusters}
The statistically significant positive correlations we identify in Section~\ref{sec:QvLM} and Section~\ref{sec:QvLM_2} provide observational evidence of dynamic massive protocluster structure, consistent with observations \cite[e.g.][]{Traficante2023} and simulations \cite[e.g.][]{Bonnell2003, Ballone2020, Guszejnov2022}. More specifically, the positive correlation between the central-condensedness of massive protocluster structure (as traced by the $\mathcal{Q}$-parameter of the core spatial distribution) and protocluster evolutionary state (as traced by $L_\text{cm}/M_\text{AGAL}$) suggests the evolution of structure from initially subclustered to centrally condensed, consistent with the massive protocluster simulation analyses of \citet{Maschberger2010}, \citet{Ballone2020} and \citet{Laverde-Villarreal2025}. As outlined in Section~\ref{sec:intro}, in these simulations, the global gravitational contraction of the star-forming cloud drives gas and cores inward, such that the $\mathcal{Q}$-parameter of the sink particle distribution increases over time. Our results are also in accordance with the global collapse inherent to hierarchical, clump-fed scenarios of massive star and protocluster formation, such as competitive accretion \citep{Bonnell2001, Bonnell2004, Smith2009} and global hierarchical collapse \citep{Vazquez-Semadeni2017, Vazquez-Semadeni2019}. Our results are broadly consistent with those of the ASSEMBLE pilot survey \cite[which used $L_\text{AGAL}/M_\text{AGAL}$;][]{Xu2024}. Notably, however, the EGO-10 sample occupies a narrower range in evolutionary state than the ASSEMBLE sample (see Table~\ref{tab:summ}). Our $\mathcal{Q}$--$L_\text{cm}/M_\text{AGAL}$ relationship therefore suggests the evolution of massive protocluster structure - from subclustered to centrally condensed - occurs in the narrowly defined, early evolutionary stage(s) of massive star and protocluster formation signposted by EGOs.  Further, our finding that the $\mathcal{Q}$--$L_\text{cm}/M_\text{AGAL}$ relationship is driven by the MYSOs in the sample suggests the formation and evolution of massive (proto)stars is coupled with the global evolution of the structure of their host protoclusters. 

Observational evidence of global collapse in a given protocluster requires the detection of clump-scale infalling velocity flows. Indeed, the regions of the \citet{Xu2024} ASSEMBLE pilot sample were preselected for optically thick blue-asymmetric profiles, a signature of isotropic collapse \cite[e.g.][]{Evans1999, Smith2012}. 
Attempts to detect similar blue-asymmetric profiles toward the EGO-10 regions have been made. \citet{Cyganowski2009} used the James Clerk Maxwell Telescope (JCMT) to observe the HCO$^+$(3-2) line toward 19 EGOs, including six EGO-10 regions, reporting blue-asymmetric profiles in three EGO-10 regions (G10.34$-$0.14, G18.89$-$0.47 and G19.36$-$0.03). \citet{Chen2010} used the Purple Mountain Observatory 13.7\,m telescope to observe the HCO$^+$(1-0) line toward 88 EGOs, including all EGO-10 regions, and found a blue-asymmetric profile in only one EGO-10 region - G16.59$-$0.05. These non-detections do not necessarily indicate an absence of global collapse, however, particularly given the limitations of these single-dish, single-pointing surveys \cite[see][and references therein]{Chen2010, Xu2023, Xu2023_SDC}. Results from numerical simulations and synthetic observations have also demonstrated the limited effectiveness of the blue-asymmetric profile as an infall diagnostic, given the complex line-of-sight velocity structure produced by the anisotropic, filamentary global collapse of massive star-forming regions (\citealp{Smith2013, Juvela2022}; see also \citealp{Bonnell2006}). Notably, the EGO-10 region G22.04$+$0.22, which did not present a blue-asymmetric profile in \citet{Cyganowski2009} or \citet{Chen2010}, has since been revealed as a hub-filament system by \citet{Yuan2018}, through their analysis of archival interferometric observations and single-dish maps of the region in a variety of molecular tracers. This work revealed $\sim$several pc longitudinal velocity flows along filaments converging at the EGO-hosting clump, as well as clump- and core-scale blue-asymmetric profiles. In a future work, we will combine the 12\,m ALMA EGO-10 data with the 7\,m ACA observations, to perform a kinematic analysis in a number of molecular tracers covered by our 1.3\,mm tuning, to assess whether such systems are common across the sample. 

\section{Conclusions}
\label{sec:concl}
In this work, we present an analysis of the 1.3\,mm continuum observations from the ALMA EGO-10 survey - a sensitive, high-resolution imaging survey of the EGO-10 sample of ten GLIMPSE Extended Green Objects (EGOs) with ALMA. Our key findings are as follows:

\begin{itemize}

    \item[1.] The EGO-10 fields are revealed as massive protoclusters, hosting morphologically diverse 1.3\,mm continuum emission, comprised of large-scale filamentary structures and numerous compact sources. In all fields, the target EGO is associated with significant 1.3\,mm continuum emission.
    
    \item[2.] Combining two source identification algorithms - \textit{Astrodendro} and \textit{Hyper} - we identify a total of 570 millimetre cores across the EGO-10 sample (an average of 57 cores per field), ranging from 13 to 135 per field. The fundamental properties of these cores are presented in a core catalogue.

    \item[3.] The distribution of projected separation between unique core-pairs exhibits an obvious double-peaked profile in 5/10 regions, with a primary peak at $\sim$0.05--0.1\,pc and a secondary peak at $\sim$0.15--0.6\,pc.
    
    \item[4.] We quantify protocluster structure with the \citet{Cartwright2004} $\mathcal{Q}$-parameter of the core spatial distribution. The mean $\mathcal{Q}$ of the sample is 0.75 and the majority of regions (7/10) are subclustered, with $\mathcal{Q}<$ 0.8. The $\mathcal{Q}$-parameter spans a range \mbox{0.53 $<\mathcal{Q}<$ 0.93}, indicating structural diversity within the sample. 

    \item[5.] Correlating the catalogued 1.3\,mm cores with published high-resolution observations of tracers of active star formation - 6.7\,GHz CH$_3$OH masers, 22\,GHz H$_2$O masers and cm-$\lambda$ continuum sources - we classify 39 cores as protostellar based on their association with maser and/or cm-$\lambda$ continuum emission. We find that 92\% of 6.7\,GHz CH$_3$OH masers, 65\% of 22\,GHz H$_2$O masers and 49\% of the cm-$\lambda$ continuum sources in the published samples are associated with catalogued 1.3\,mm cores. Conversely, the percentage of 1.3\,mm cores associated with star formation tracers is small: 2\%, 5\% and 4\% of cores with 6.7\,GHz masers, 22\,GHz masers and cm-$\lambda$ continuum sources, respectively. 

    \item[6.] We classify the eleven 1.3\,mm cores that are associated with 6.7\,GHz CH$_3$OH masers as MYSO host-cores.
    Only in 3/10 fields is the brightest 1.3\,mm core an MYSO host-core. We find MYSO host-cores are located close to the centre of their host protocluster - 10/11 within a projected separation $\lesssim$0.2\,pc - closer than cores which do not host MYSOs, to a statistically significant degree. All of the MYSO host-cores in regions at $d<$ 3\,kpc are found in highly clustered environments, with the nearest-neighbouring 1.3\,mm core within a projected separation $\lesssim$4000\,AU of the 6.7 GHz maser, and with $>$10 cores within 10,000\,AU of the maser position.

    \item[7.] Using the sensitive, high-resolution cm-$\lambda$ continuum observations available for the EGO-10 sample, we construct a new indicator of protocluster evolutionary state: $L_\text{cm}/M_\text{AGAL}$ - the ratio of a protocluster's total cm-$\lambda$ continuum luminosity to the mass of its associated ATLASGAL clump(s). We identify a statistically significant positive correlation between the protocluster $\mathcal{Q}$-parameter and $L_\text{cm}/M_\text{AGAL}$, and find this correlation is primarily driven by the cm-$\lambda$ continuum emission directly associated with the EGO-10 MYSOs. These results suggest dynamic massive protocluster structure, evolving from subclustered to centrally condensed in the early stage(s) of massive star formation signposted by EGOs, and that the formation and evolution of massive (proto)stars is coupled with that of their host protoclusters. 
\end{itemize}

These results are, in general, consistent with the predictions of hierarchical, clump-fed massive star formation, particularly the positive correlation between $\mathcal{Q}$ and $L_\text{cm}/M_\text{AGAL}$. Notably, the relationship between protocluster structure and evolutionary state is only recovered within the EGO-10 sample by taking $L_\text{cm}/M_\text{AGAL}$ as the evolutionary indicator, as opposed to the clump-scale luminosity-to-mass ratio used in previous works with samples covering a wider range in evolutionary state. This suggests $L_\text{cm}/M_\text{AGAL}$ may be a more powerful diagnostic of evolutionary state in the early stages of massive star formation, in which MYSO(s) are just beginning to excite cm-$\lambda$ continuum emission. Our results therefore highlight the power of the complementary high-resolution multiwavelength observations available for the EGO-10 sample to reveal new insights into the early stages of the evolution of massive stellar protoclusters. 

\section*{Acknowledgements} 
MCL acknowledges support from a UK Science and Technologies Facilities Council (STFC) PhD studentship, under training grant ST/W507817/1 (project reference 2748654).  CJC gratefully acknowledges support from the STFC (grant ST/Y002229/1).  MCL thanks G.M. Williams for helpful guidance with \textit{Astrodendro}, J. Hernandez Santisteban for support with LinMix and other statistical techniques, F. Xu for providing ASSEMBLE and ASHES data and S.N. Longmore for helpful discussions. 
This paper makes use of the following ALMA data: ADS/JAO.ALMA\#2016.1.00747.S, ADS/JAO.ALMA\#2017.1.00983.S. ALMA is a partnership of ESO (representing its member states), NSF (USA) and NINS (Japan), together with NRC (Canada), NSTC and ASIAA (Taiwan), and KASI (Republic of Korea), in cooperation with the Republic of Chile. The Joint ALMA Observatory is operated by ESO, AUI/NRAO and NAOJ. The National Radio Astronomy Observatory is a facility of the National Science Foundation operated under cooperative agreement by Associated Universities. 
This research has made use of the following Python packages: Numpy \citep{Harris2020}, Astropy \citep{astropy:2013,astropy:2018,astropy:2022}, Scipy \citep{Virtanen2020}, AplPy \citep{aplpy2012}, Matplotlib \citep{Hunter2007}, CMasher \citep{vanderVelden2020}.

\section*{Data Availability}
The observational ALMA data is publicly available via the ALMA Archive under project codes 2016.1.00747.S and 2017.1.00983.S (PI C.\ Brogan). The ALMA EGO-10 1.3\,mm continuum 12\,m-only and 40\,k$\lambda$ images, and the IDL \textit{Hyper} source extraction and photometry code used, are available at doi:10.5281/zenodo.20156717.  



\bibliographystyle{mnras}
\bibliography{lib} 

\section*{Supporting Information}
Supplementary data are available at MNRAS online. These include the complete core catalogue, provided as comma-separated-value (CSV) text, and three figure sets.  The figure sets are provided as zip files of individual pdf figures.  Note the figure set zip files do not include the representative examples already shown in the main text.\\

\noindent Supplementary files provided:\\
\noindent \textbf{CORECAT.csv} The complete ALMA EGO-10 1.3\,mm core catalogue, see Table~\ref{tab:cat}.\\
\noindent \textbf{PROTO.zip} Figures as in Fig.~\ref{fig:assoc} for 1.3\,mm cores classified as protostellar.\\   
\noindent \textbf{NA.zip} Figures as in Fig.~\ref{fig:assoc} for star formation tracers that are not associated with 1.3\,mm cores.\\ 
\noindent \textbf{MYSO.zip} Figures as in Fig.~\ref{fig:AUzoom} for 1.3\,mm emission and identified cores within 10,000\,AU of 1.3\,mm cores classified as MYSO hosts.\\ 

\noindent Please note: Oxford University Press is not responsible for the
content or functionality of any supporting materials supplied by
the authors. Any queries (other than missing material) should be
directed to the corresponding author for the article.



\appendix
\section{Manual Addition of a Source in G22.04+0.22}
\label{app:G22}

\begin{figure}
   \includegraphics[width=\columnwidth]{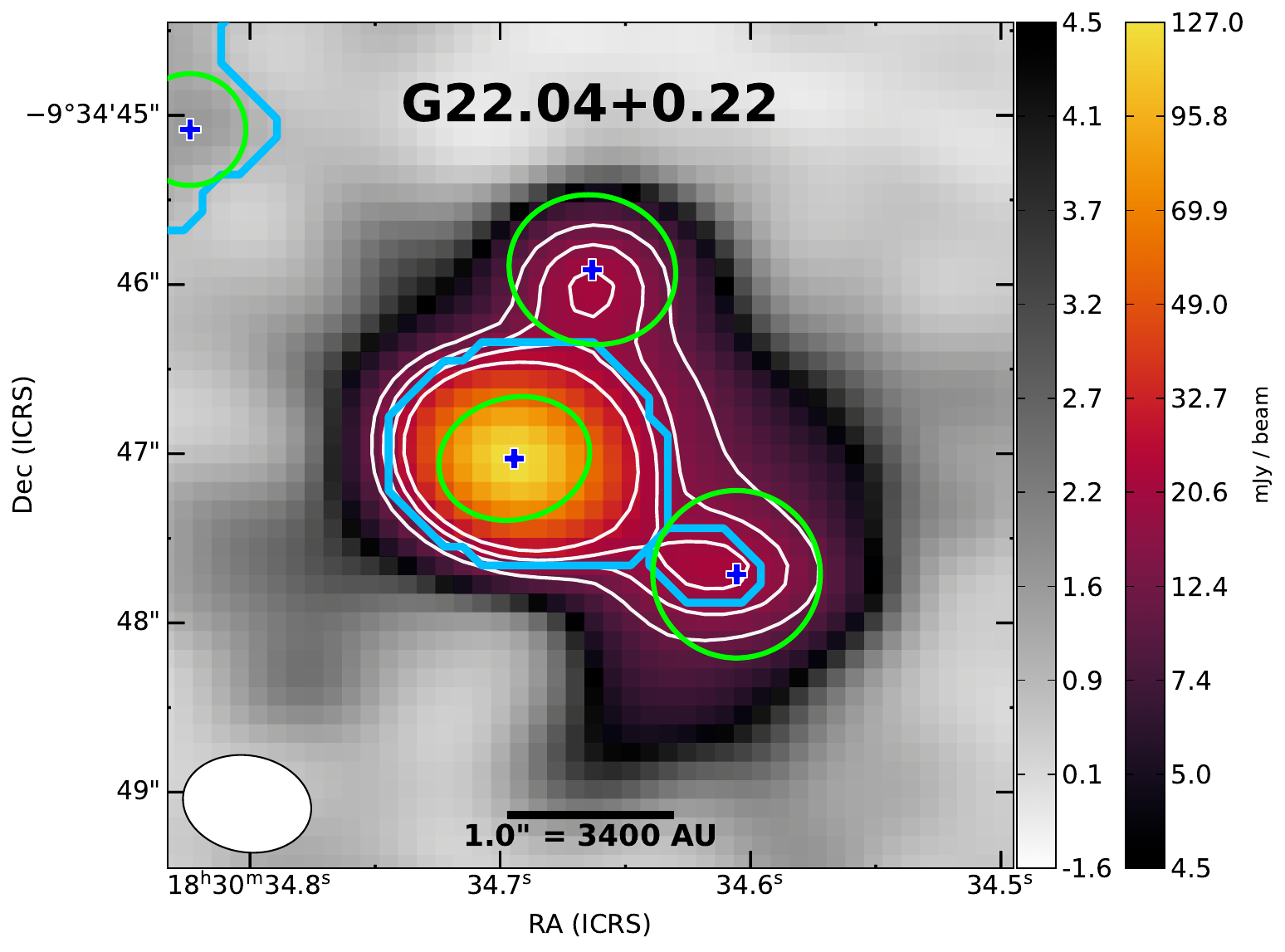} 
   \caption{The central 1.3\,mm clump of G22.04$+$0.22. The background colourscale is identical to that of Fig.~\ref{fig:mm}, with white contours at [50, 70, 90, 120] $\times\sigma_\text{rms}$. The synthesised beam is shown bottom left. The \textit{Astrodendro} leaves extracted in the non-primary beam-corrected 40\,k$\lambda$ image are outlined in light blue and the centroids and FWHMs of \textit{Hyper} sources are shown as blue crosses and green ellipses, respectively.}
\label{fig:G22_app}
\end{figure}

In Section~\ref{sec:source} we sought to apply a uniform source extraction technique to identify 1.3\,mm cores in the EGO-10 sample.
As noted in Section~\ref{sec:cat}, there is one notable case in G22.04$+$0.22 where this approach does not capture the core-scale 1.3\,mm emission.
The 1.3\,mm continuum in this field is simple in comparison to most other EGO-10 fields, covering only a small ($\sim$10\arcsec) portion of the mosaic at the pointing centre (see Fig.~\ref{fig:mm}). The strongest emission occurs in a central clump comprised of three point-like sources, shown in the zoom-view in Fig.~\ref{fig:G22_app}, in which these sources are outlined by the overlaid emission contours. The central source peaks $\gtrsim$500\,$\sigma_\text{rms}$, with the northern and southwestern sources both peaking $\sim$100\,$\sigma_\text{rms}$. Also shown in Fig.~\ref{fig:G22_app} are the \textit{Astrodendro} leaves identified in the non-primary beam-corrected 40\,k$\lambda$ image in Section~\ref{sec:dendro}. The central and southwestern sources are separated, but no leaf is assigned to the northern source. In fact, this source is only recovered with \texttt{min\_npix} $=N_\text{beam}$/4. This value is however atypically small, and implementation in other fields predictably gives many weak, spurious sources, precluding application across the sample.

The \textit{Hyper} source centroids are also overlaid in Fig.~\ref{fig:G22_app}, showing all three sources are identified. We therefore manually added the northern source to our core catalogue, despite it having no associated \textit{Astrodendro} leaf. As with all sources, its fundamental properties were obtained by providing \textit{Hyper} with the centroid identified in the non-primary beam-corrected image, for Gaussian fitting in the primary beam-corrected image. The resulting source FWHM ellipses are shown in Fig.~\ref{fig:G22_app} and the source properties are reported in the core catalogue in Table~\ref{tab:cat}. Following the naming convention in Section~\ref{sec:cat}, 
the central, southwestern and northern sources are G22.04$+$0.22 MM1, MM2 and MM3, respectively. 

\section{Estimating the Uncertainty on the $\mathcal{Q}$-parameter}
\label{app:Q_err}
As outlined in Section~\ref{sec:mst}, we estimate uncertainties on the $\mathcal{Q}$-parameters as the standard deviation of the $\mathcal{Q}$ values calculated using alternative constructions of the core catalogue in each field, testing the effect of mass sensitivity, sampling uncertainty, and an alternative source extraction procedure.

To test the effect of mass sensitivity in the sample, we estimate the masses of the cores in Table~\ref{tab:cat} using Equation 1 of \citet{Cyganowski2017} \cite[using the clump-scale temperatures of][reported in Table~\ref{tab:tbl1} - as for the $M_{5\sigma}$ values in Table~\ref{tab:img}]{Cyganowski2013}. We then remove cores in each field with masses less than the largest 5$\sigma_\text{rms, pb}$ point-mass sensitivity in the sample - 0.26\,$M_\odot$ in G22.04$+$0.22. This only affects the catalogues in the five regions at $d<$\,3\,kpc - G10.29$-$0.13, G10.34$-$0.14, G14.33$-$0.64, G14.63$-$0.58 and G19.36$-$0.03 - removing seven, four, five, five and one core(s) from each, respectively. The resulting recalculated $\mathcal{Q}$-parameters for these regions are reported as $\mathcal{Q}_M$ in Table~\ref{tab:Q_err}. 

To test the effect of sampling uncertainty, we assumed Poisson statistics and randomly removed $\sqrt{N_\text{cores}}$ of the cores catalogued in each field 100 times, recalculating $\mathcal{Q}$ for each realisation. The mean of the resulting $\mathcal{Q}$ distribution was taken as an additional estimate of $\mathcal{Q}$ for each protocluster, reported as $\mathcal{Q}_N$ in Table~\ref{tab:Q_err} \cite[similar to][]{Sadaghiani2020}.

\begin{table}
\centering
\caption{$\mathcal{Q}$-parameter values for alternative constructions of the core catalogue in each field.}
\begin{tabular}{lccccc}
\hline
\hline
EGO & $\mathcal{Q}_M$$^\text{a}$ & $\mathcal{Q}_N$$^\text{a}$ & $\mathcal{Q}_\text{12m}$$^\text{b}$ & $\mathcal{Q}_{\text{40k}\lambda}$$^\text{b}$ & $\sigma_\mathcal{Q}$$^\text{c}$\\
\hline
G10.29$-$0.13 & 0.528 & 0.53 & 0.542 & 0.579 & 0.0194 \\
G10.34$-$0.14 & 0.638 & 0.609 & 0.623 & 0.611 & 0.0105 \\
G12.91$-$0.03 & -- -- & 0.834 & 0.749 & 0.829 & 0.0305 \\
G14.33$-$0.64 & 0.981 & 1.025 & 0.843 & 0.915 & 0.0618 \\
G14.63$-$0.58 & 0.713 & 0.779 & 0.844 & 0.886 & 0.0588 \\
G16.59$-$0.05 & -- -- & 0.686 & 0.774 & 0.800 & 0.0517 \\
G18.89$-$0.47 & -- -- & 0.733 & 0.686 & 0.713 & 0.0223 \\
G19.36$-$0.03 & 0.727 & 0.729 & 0.699 & 0.752 & 0.0169 \\
G22.04$+$0.22 & -- -- & 0.644 & 0.640 & 0.72 & 0.0429 \\
G28.83$-$0.25 & -- -- & 0.825 & 0.915 & 0.959 & 0.0430 \\ 
\hline
\end{tabular}
\begin{flushleft}
        \small{
        $^\text{a}$ Calculated using a subset of the cores catalogued in each field in Section~\ref{sec:cat}. For $\mathcal{Q}_M$, cores with estimated masses less than the worst mass sensitivity limit in the sample are removed. $\mathcal{Q}_N$ is the mean value from 100 Monte Carlo realisations of the catalogue in which $\sqrt{N_\text{cores}}$ cores are removed each time (see $N_\text{cores}$ values in Table~\ref{tab:mst}).\\
        $^\text{b}$ Calculated from \textit{Astrodendro} catalogues; $\mathcal{Q}_\text{12m}$ and $\mathcal{Q}_{\text{40k}\lambda}$ from leaves identified in the 12\,m-only and 40\,k$\lambda$ images in Section~\ref{sec:dendro}, respectively.\\
        $^\text{c}$ Standard deviation of calculated $\mathcal{Q}$-parameters tabulated here and in Table~\ref{tab:mst}, adopted as the uncertainty on $\mathcal{Q}$ (see Section~\ref{sec:mst} and Section~\ref{sec:evol}).
        }
        
    \end{flushleft}
    \label{tab:Q_err}
\end{table}

To test the effect of using an alternative source extraction procedure, we recalculated $\mathcal{Q}$ using the $>$5\,$\sigma_\text{rms, pb}$ \textit{Astrodendro} leaves identified in our initial runs in the 12\,m-only images, and again separately using those identified in the 40\,k$\lambda$ images (see Section~\ref{sec:dendro}). These values are reported in Table~\ref{tab:Q_err} as $\mathcal{Q}_\text{12m}$ and $\mathcal{Q}_{\text{40k}\lambda}$, respectively.

\begin{figure}
\includegraphics[width=\columnwidth]{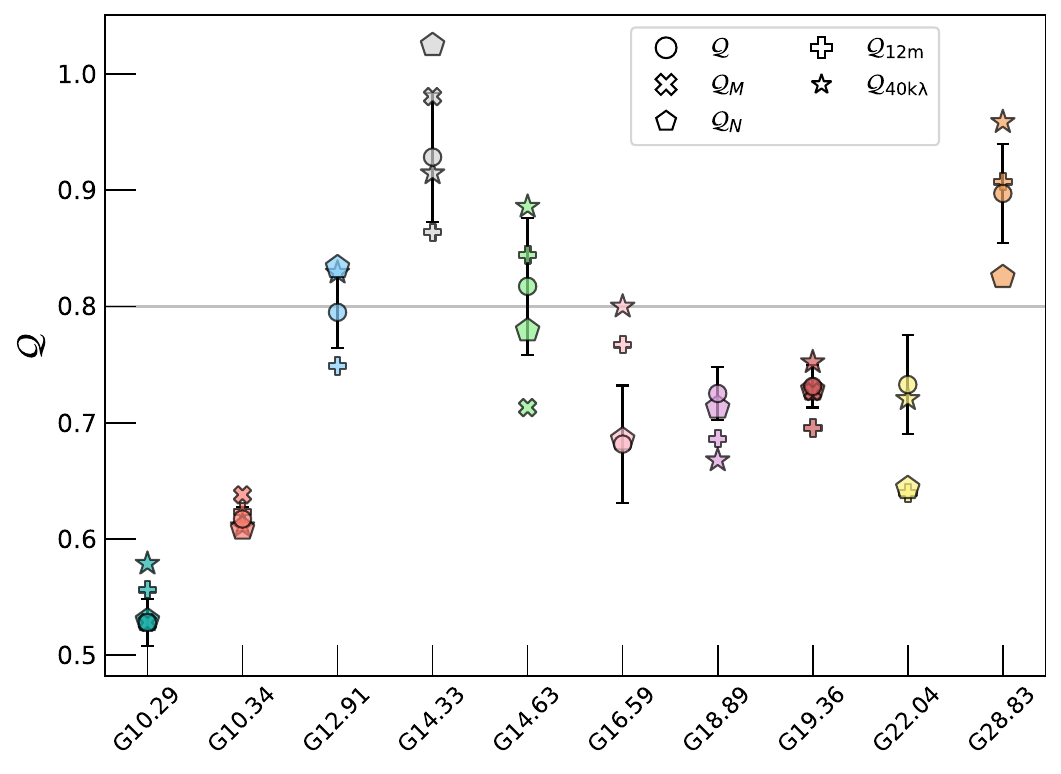}
\caption{The variation of the $\mathcal{Q}$-parameter in EGO-10 using alternative constructions of the core catalogue in each field. Circles denote the $\mathcal{Q}$ derived and reported in Section~\ref{sec:mst}. Different symbols denote the different tests described in Appendix~\ref{app:Q_err}: $\times$ for the mass sensitivity test ($\mathcal{Q}_{M}$); pentagons for the Poisson sampling statistics test ($\mathcal{Q}_{N}$); $+$ for the \textit{Astrodendro} 12\,m-only image test ($\mathcal{Q}_\text{12m}$); and stars for the \textit{Astrodendro} 40\,k$\lambda$ image test ($\mathcal{Q}_{\text{40k}\lambda}$). Error bars are centred on the $\mathcal{Q}$-parameters reported in Section~\ref{sec:mst}, with the error equal to the standard deviation of all $\mathcal{Q}$ values in a given field.}
\label{fig:Q_err}
\end{figure}

Fig.~\ref{fig:Q_err} shows all $\mathcal{Q}$ values calculated for each protocluster, including those calculated in Section~\ref{sec:mst}. The error bars indicate the standard deviation of the $\mathcal{Q}$ values obtained in each field (i.e. the 4--5 values in Tables~\ref{tab:mst} and \ref{tab:Q_err}), which are the uncertainties presented in Table~\ref{tab:mst} and used in the statistical analyses in Section~\ref{sec:evol}.

\section{Effect of Distance on Protocluster Structure and Relationships}
\label{sec:app_tests}
\begin{figure*}
\includegraphics[width=\textwidth]{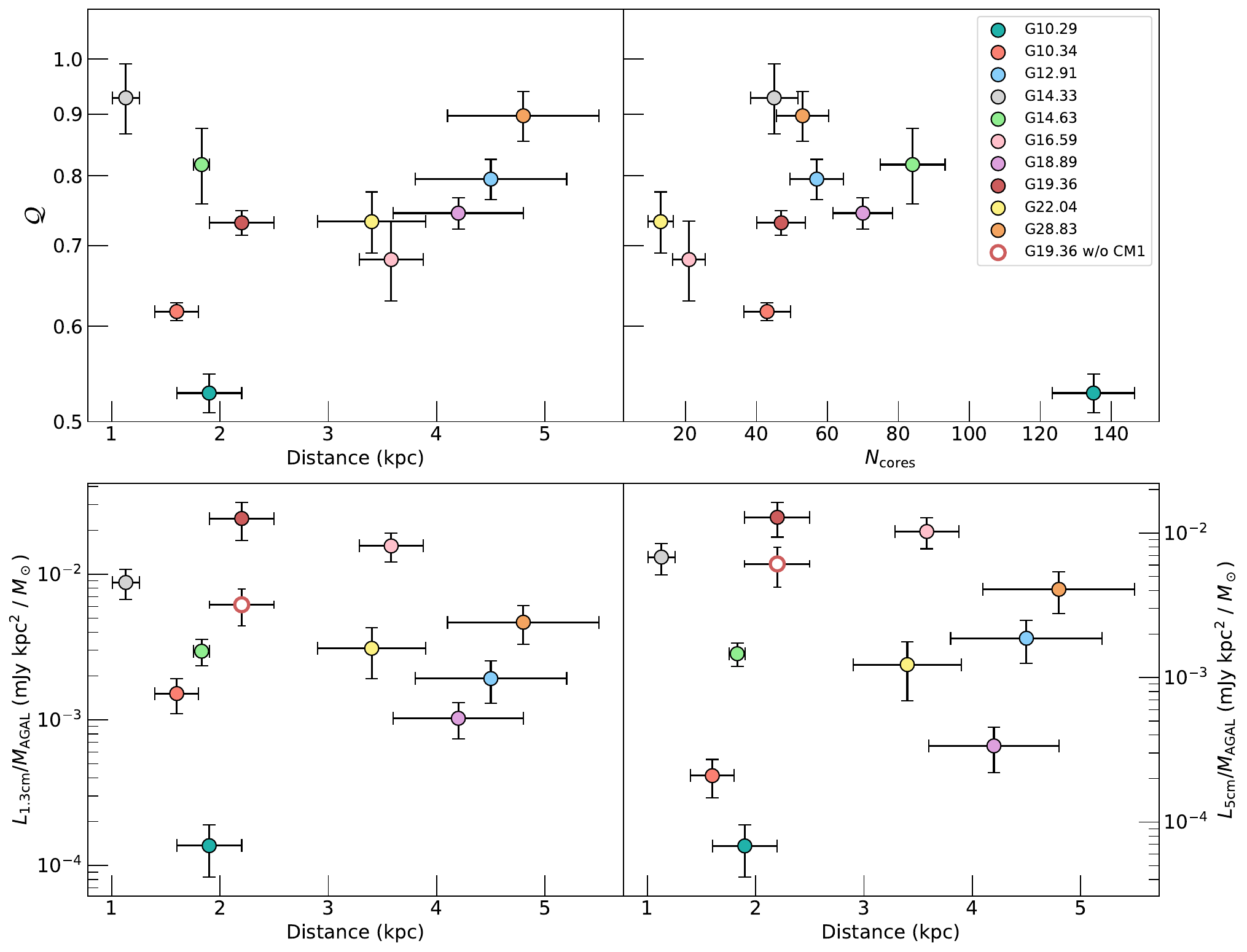}
\caption{Upper panels: The $\mathcal{Q}$-parameter is plotted as a  function of: \textit{left}: protocluster distance and \textit{right}: number of catalogued cores per field ($N_\text{cores}$). Lower panels: the evolutionary indicator $L_\text{cm}/M_\text{AGAL}$ is plotted as a function of protocluster distance at: \textit{left}: 1.3\,cm and \textit{right}: 5\,cm. The open red circle in the lower panels denotes the position of G19.36$-$0.03 if the cm-$\lambda$ continuum source G19.36$-$0.03 CM1 does not contribute to $L_\text{cm}$. Correlation coefficients and associated $p$-values are given in Table~\ref{tab:tests}.}
\label{fig:app_tests_multi}
\end{figure*}

\begin{figure*}
\includegraphics[width=\textwidth]{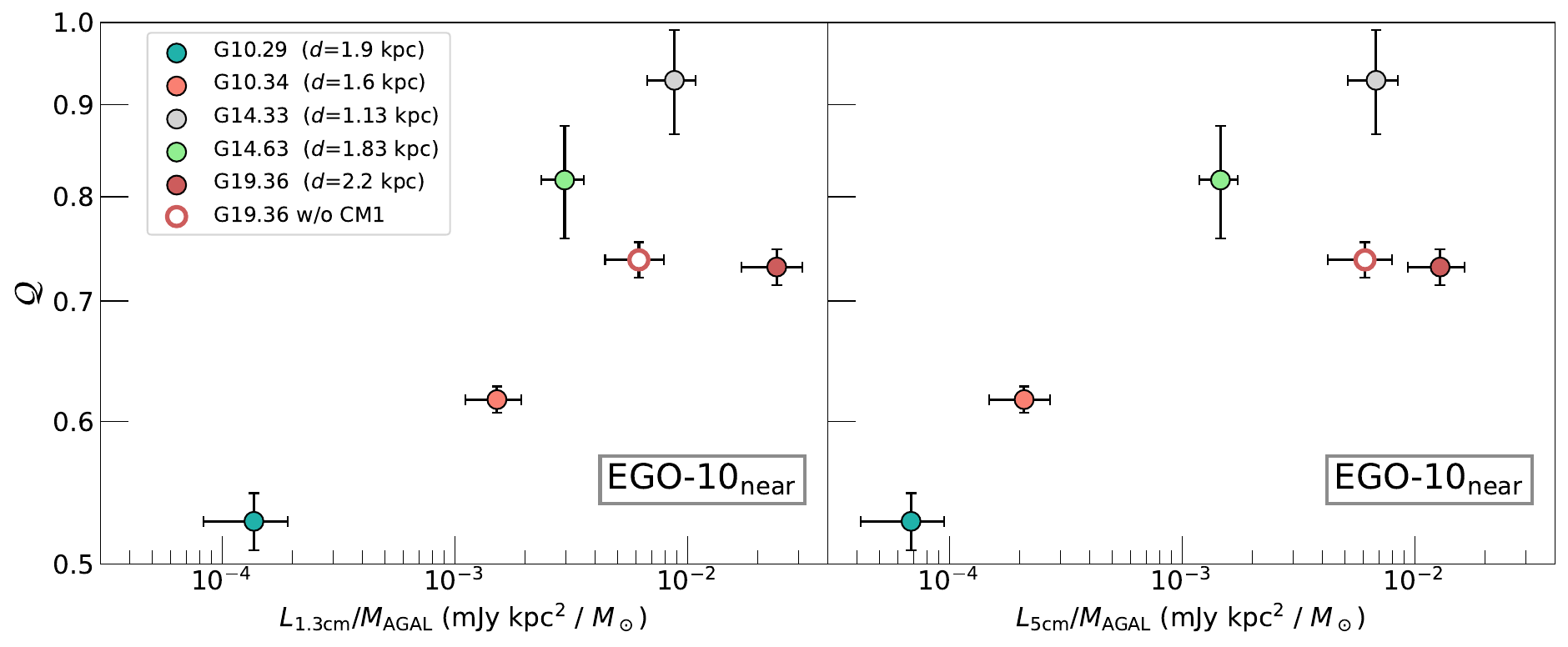}
\includegraphics[width=\textwidth]{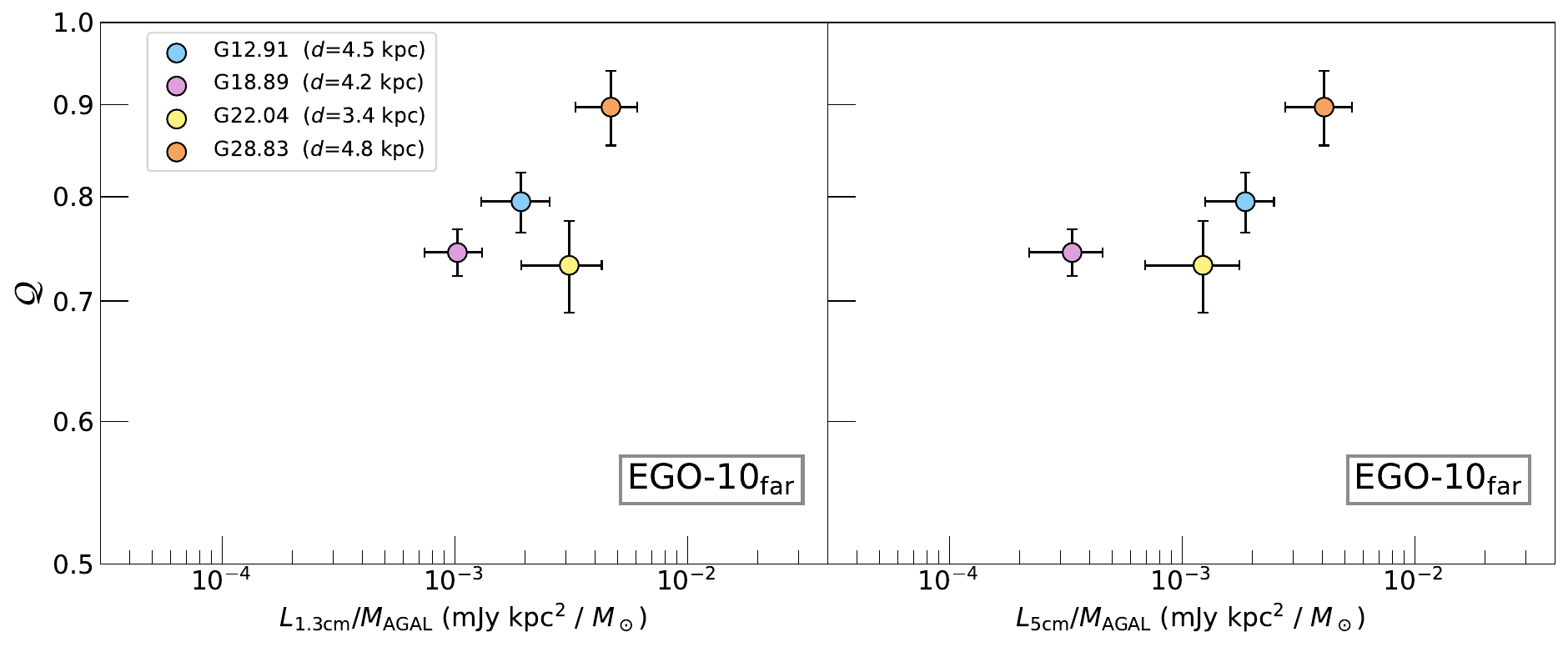}
\caption{$\mathcal{Q}$ plotted as a function of $L_\text{cm}/M_\text{AGAL}$ for the EGO-10$_\text{near}$ (upper panels) and EGO-10$_\text{far}$  (lower panels) subsamples. The EGO-10$_\text{near}$ subsample consists of the five regions at $d<$ 3\,kpc. The EGO-10$_\text{far}$ subsample consists of regions at $d>$ 3\,kpc. Marker colours and $x$- and $y$-axis ranges are the same as in Fig.~\ref{fig:QvLM}. Correlation coefficients and associated $p$-values are given in Table~\ref{tab:tests}.}
\label{fig:app_tests}
\end{figure*}

The regions of the EGO-10 sample span a distance range \mbox{1$\lesssim d \lesssim$ 5\,kpc} (see values in Table~\ref{tab:tbl1}). Here, we present the results of tests to determine whether the inhomogeneous protocluster distance within the EGO-10 sample significantly impacts our results. In Fig.~\ref{fig:app_tests_multi}, we plot the $\mathcal{Q}$-parameter (Table~\ref{tab:mst}) as a function of protocluster distance (Table~\ref{tab:tbl1}). Given the distance will also affect the number of cores detected per region ($N_\text{cores}$, see values in Table~\ref{tab:mst}) via the mass sensitivity limit and beam spatial scale, Fig.~\ref{fig:app_tests} also shows $\mathcal{Q}$ as a function of $N_\text{cores}$. In the lower panels, we plot our evolutionary indicator - $L_\text{cm}/M_\text{AGAL}$ - as a function of distance at both 1.3\,cm ($L_\text{1.3cm}/M_\text{AGAL}$) and 5\,cm ($L_\text{5cm}/M_\text{AGAL}$). Correlation coefficients and $p$-values  are given in Table~\ref{tab:tests}. We find no statistically-significant correlations. These tests therefore suggest the inhomogeneous protocluster distance within the EGO-10 sample does not systematically impact our determination of protocluster structure or evolutionary state.

As noted in Section~\ref{sec:myso_clust}, the EGO-10 regions can be separated by distance, with half of the regions at $d<$ 3\,kpc. To test whether the positive correlation between $\mathcal{Q}$ and $L_\text{cm}/M_\text{AGAL}$ found in Section~\ref{sec:Q1} is affected by protocluster distance, we examined the relationship separately within these distance-limited subsamples: EGO-10$_\text{near}$ (G10.29$-$0.13, G10.34$-$0.14, G14.33$-$0.64, G14.63$-$0.58 and G19.36$-$0.03) and EGO-10$_\text{far}$ (G12.91$-$0.03, G16.59$-$0.05, G18.89$-$0.47, G22.04+0.22 and G28.83-0.25). Fig.~\ref{fig:app_tests} shows the $\mathcal{Q}$-parameter as a function of $L_\text{cm}/M_\text{AGAL}$ within these subsamples, at both 1.3\,cm and 5\,cm, with correlation coefficients and $p-$values reported in Table~\ref{tab:tests}. In EGO-10$_\text{near}$, correlation coefficients remain positive at both wavelengths, though the small sample size is such that these are not statistically significant. Correlation coefficients in the EGO-10$_\text{far}$ subsample indicate no trend, due to the influence of the outlier region G16.59$-$0.05 (see also Fig.~\ref{fig:QvLM}). Positive correlation coefficients do return however, upon removal of this source from the subsample (reported under EGO-9$_\text{far}$ in Table~\ref{tab:tests}). Overall these tests suggest the positive correlation between protocluster structure ($\mathcal{Q}$) and evolutionary state ($L_\text{cm}/M_\text{AGAL}$) persists, even within distance-limited subsamples of EGO-10, and that the relationship is not purely a result of distance effects.

\begin{table}
\centering
 \caption{Correlation coefficients and associated $p$-values of distance tests.}
 \begin{tabular}{lcccccc}
 \hline
 \hline
   & & \multicolumn{2}{c}{Spearman} & & \multicolumn{2}{c}{Kendall}\\

\cline{3-4}
\cline{6-7}

Test & & $\rho$ & $p_\rho$ & & $\tau$ & $p_\tau$\\
\hline
EGO-10$_\text{near}$$^\text{a}$\\
\hline
($\mathcal{Q}$, $L_\text{1.3cm}/M_\text{AGAL}$) & & 0.7 & 0.235 & & 0.6 & 0.235\\
($\mathcal{Q}$, $L_\text{5cm}/M_\text{AGAL}$) & & 0.7 & 0.231 & & 0.6 & 0.231\\
\hline
\hline
EGO-10$_\text{far}$$^\text{a}$\\
\hline
($\mathcal{Q}$, $L_\text{1.3cm}/M_\text{AGAL}$) & & -0.3 & 0.687 & & -0.2 & 0.819\\
($\mathcal{Q}$, $L_\text{5cm}/M_\text{AGAL}$) & & -0.1 & 0.951 & & 0.0 & 1.0 \\
\hline
\hline
EGO-9$_\text{far}$$^\text{a}$ & & \\
\hline
($\mathcal{Q}$, $L_\text{1.3cm}/M_\text{AGAL}$) & & 0.4 & 0.746 & & 0.33 & 0.746\\
($\mathcal{Q}$, $L_\text{5cm}/M_\text{AGAL}$) & &  0.8 & 0.337 & & 0.67 & 0.337\\
\hline
\hline
($\mathcal{Q}$, $d$)$^\text{b}$ & & 0.14 & 0.7 & & 0.20 & 0.48\\
($\mathcal{Q}$, $N_\text{cores}$)$^\text{b}$ & & 0.09 & 0.8 & & 0.02 & 1.0\\
($L_\text{1.3cm}/M_\text{AGAL}$, $d$)$^\text{b}$ & & 0.02 & 0.96 & & 0.02 & 1.0\\
($L_\text{5cm}/M_\text{AGAL}$, $d$)$^\text{b}$ & & 0.16 & 0.65 & & 0.11 & 0.73\\
\hline
\end{tabular}
\begin{flushleft}
        \small{
        $^\text{a}$ Correlation coefficients and $p$-values were evaluated as in Section~\ref{sec:QvLM}. \\
        $^\text{b}$ Correlation coefficients and $p$-values from \texttt{scipy.stats}.\\

        }
    \end{flushleft}
\label{tab:tests}
\end{table}

\section{Tests of Alternative Association Criteria}
\label{app:na_tests}
In general, defining a given star formation tracer to be `associated' with a given 1.3\,mm core depends on the core identification algorithm and corresponding input parameters used, and the specific association criteria chosen. We tested the effect of two alternative criteria for the association of 1.3\,mm cores with tracers of active star formation on the association statistics presented in Section~\ref{sec:cm_assoc}: (1) a given star formation tracer is `associated' with a given 1.3\,mm core if it lies within the core's \textit{Hyper} photometric aperture (which is 2 $\times$ the source FWHM ellipse, see Section~\ref{sec:hyper}), and (2) a given star formation tracer is `associated' with a 1.3\,mm core if it lies within the boundary of an \textit{Astrodendro} leaf identified in the 40\,k$\lambda$ image that meets the peak intensity criterion described in Section~\ref{sec:dendro}. 
For both of these cases, the tracer-core association rates increase (see Table~\ref{tab:assocs}), though their relative order remains unchanged (see Section~\ref{sec:seps}). 

As already outlined, a goal of this multiwavelength comparison is to identify which of the ALMA 1.3\,mm cores catalogued in Section~\ref{sec:cat} are unambiguously protostellar, based on their association with previously detected masers and/or cm-$\lambda$ continuum sources. The less-restrictive association criteria tested above increase tracer-core association rates, but also yield ambiguous associations in some cases. For example, 
some tracers lie within an \textit{Astrodendro} leaf which has been resolved into multiple core-scale sources with \textit{Hyper} (see non-association figure set in  online Supplementary Material). In other cases, a tracer lies within a core's \textit{Hyper} photometric aperture, but outside the \textit{Astrodendro} leaf, or vice versa. Misleading or false associations are also more likely with less-restrictive association criteria. For example, as discussed in Section~\ref{sec:seps}, the \citet{Rosero2019} cm-$\lambda$ continuum source G16.59$-$0.05 A is likely a discrete knot in the radio jet driven by the MYSO associated with our 1.3\,mm core G16.59$-$0.05 MM2. However, the offset between G16.59$-$0.05 A and G16.59$-$0.05 MM2 is such that the cm-$\lambda$ source lies, in projection, within the 2 $\times$ FWHM photometric aperture of another 1.3\,mm core, G16.59$-$0.05 MM9. 
To avoid ambiguity and minimise false associations, the association statistics and $\Delta_\text{off}$ values reported in Sections~\ref{sec:cm_67}, \ref{sec:cm_22} and \ref{sec:cm}, and in the core catalogue in Table~\ref{tab:cat}, correspond to the more robust associations found with the criterion used in Section~\ref{sec:cm_assoc}: a given star formation tracer is considered `associated' with a given 1.3\,mm core only if it lies within the core's \textit{Hyper} FWHM ellipse.

\section{Spectral Index Analysis}
\label{app:alpha}
\citet{Towner2021} and \citet{Rosero2016} report the spectral indices ($\alpha$) of their cm-$\lambda$ continuum sources between 1.3 and 5\,cm. Of the 21 cm-$\lambda$ continuum sources we find in association with 1.3\,mm cores, the reported $\alpha$ value is constrained in 12 cases, spanning a range 0.13 $<\alpha<$ 1.58, consistent with expected values from free-free emission in ionised jets and radiatively-ionised (\ion{H}{ii}) regions \citep[][and references therein]{Towner2021}. We use $\alpha$ to extrapolate the contribution of these cm-$\lambda$ continuum sources to the 1.3\,mm integrated flux density of their host-cores, as measured by \textit{Hyper} (see Section~\ref{sec:cat} and Table~\ref{tab:cat}). In the majority of cases (10/12), the extrapolated 1.3\,mm integrated flux density accounts for $<$10\% of the measured value, indicating the 1.3\,mm emission of these cores is dominated by the thermal dust contribution. The two exceptions are G14.63$-$0.58 MM8 (hosting G14.63$-$0.58 CM1 and CM3) and G19.36$-$0.03 MM10 (hosting G19.36$-$0.03 CM1). 

The extrapolated 1.3\,mm integrated flux density of G14.63$-$0.58 CM1 is $\sim$46\% of the measured 1.3\,mm flux density of G14.63$-$0.58 MM8\footnote{This does not include the contribution from G14.63$-$0.58 CM3 (see Section~\ref{sec:cm}), for which \citet{Towner2021} report $\alpha$ as an upper limit. If extrapolated, this source would contribute only marginally ($\sim$1\%) to the 1.3\,mm flux of G14.63$-$0.58 MM8.}. \citet{Towner2021} report $\alpha=$ 1.58 for G14.63$-$0.58 CM1. This value is consistent with thermal free-free emission \cite[][and references therein]{Towner2021}. From our observations we calculate $\alpha_{\rm 1.3cm\_1.3mm}=$ 1.91, considering the measured integrated flux densities of MM8 and CM1, respectively, with this steeper value approaching the $\alpha$ expected for thermal dust emission.

G19.36$-$0.03 CM1 is the only cm-$\lambda$ continuum source in the sample for which the extrapolated 1.3\,mm integrated flux density is larger than the measured value of its associated 1.3\,mm core - G19.36$-$0.03 MM10 - 15\,mJy and 5\,mJy, respectively. Additionally from our observations we calculate $\alpha_{\rm 1.3cm\_1.3mm}=$ 0.28 between 1.3\,cm and 1.3\,mm. This is inconsistent with the 1.3\,mm continuum flux arising from thermal dust emission. The source is associated with bright, point-like multiband IRAC emission, and was identified as a potentially more evolved object by \citet{Cyganowski2011_VLA} due to a lack of 24 $\mu$m emission. This source also has no 19.7\,$\mu$m or 37.1\,$\mu$m emission in the \citet{Towner2019} EGO-12 SOFIA observations, and the cm position from \citet{Towner2021} lies just outside the 0.25\,$I_\text{peak}$ ATLASGAL 870\,$\mu$m contour \cite[see Figure~1 of][]{Towner2021}. Taken together, its observational characteristics suggest this source may not be a young object associated with the 1.3\,mm protocluster.

Of the 1.3\,mm cores associated with cm-$\lambda$ continuum sources for which \citet{Towner2021} and \citet{Rosero2016} report $\alpha$ as an upper or lower limit, one source, G14.33$-$0.64 CM4, has $\alpha>$ 2, which is inconsistent with thermal free-free emission \mbox{($-$0.1$<\alpha<$ 2)}. \citet{Towner2021} report $\alpha>$ 2.25 for this source, consistent with the $\alpha$ expected from thermal dust emission in dense cores. Using our measured 1.3\,mm integrated flux density, we calculate $\alpha_{\rm 1.3cm\_1.3mm}=$ 2.95, consistent with thermal dust emission. This source is associated with the 1.3\,mm core G14.33$-$0.64 MM1, shown in Fig.~\ref{fig:assoc}.


\bsp	
\label{lastpage}
\end{document}